\documentclass[prx, reprint,superscriptaddress]{revtex4-2}
\usepackage[utf8]{inputenc}

\usepackage{wrapfig, graphicx, tikz, xcolor, siunitx, hyperref} 
\usepackage{amsmath, amssymb, bbm, bm, tikz-cd}
\usepackage{amsfonts}
\usepackage{layouts}
\usepackage{comment}
\usepackage[]{algorithm2e}

\usepackage[export]{adjustbox}

\usepackage{verbatim}
\usepackage{l3keys2e}

\newcommand{\bmh}[1]{\bm{\hat #1}}

\newcommand{\rev}[1]{{#1}}

\begin{document}

\title{More is less in unpercolated active solids}

\author{Jack Binysh}
\email{j.a.c.binysh@uva.nl}
\affiliation{Institute of Physics, Universiteit van Amsterdam, 1098 XH Amsterdam, The Netherlands}
\author{Guido Baardink}
\affiliation{Department of Physics, University of Bath, Claverton Down, Bath BA2 7AY, UK}
\author{Jonas Veenstra}
\affiliation{Institute of Physics, Universiteit van Amsterdam, 1098 XH Amsterdam, The Netherlands}
\author{Corentin Coulais}
\email{coulais@uva.nl}
\affiliation{Institute of Physics, Universiteit van Amsterdam, 1098 XH Amsterdam, The Netherlands}
\author{Anton Souslov}
\email{as3546@cam.ac.uk}
\affiliation{
TCM Group, Cavendish Laboratory, JJ Thomson Avenue, Cambridge, CB3 0US, United Kingdom}
\date{\today}

\begin{abstract}
{A remarkable feat of active matter physics is that systems as diverse as collections of self-propelled particles, nematics mixed with molecular motors, and interacting robots can all be described by symmetry-based continuum theories. These descriptions rely on reducing complex effects of individual motors to a few key active parameters, which increase with activity. Here we observe a striking anomaly in the continuum description of non-reciprocal active solids, a ubiquitous class of active materials. Using a combination of  metamaterial experiments and coarse-graining theory we find that as microscopic activity increases, macroscale active response can vanish: more is less. In this highly active regime, non-affine and localized modes prevail and destroy the large-scale signature of microscopic activity. These modes exist in any dilute periodic structure and emerge in random lattices below a percolation transition. Our results unveil a counterintuitive facet of active matter, offering new principles for engineering materials far from equilibrium.} 
\end{abstract}

\maketitle

The stiffer the steel---the stiffer the bridge. This principle broadly applies to mechanical structures in equilibrium: By rigidifying even a single component, the entire structure becomes more rigid. Beyond architecture, this idea connects structure with function in any material,
from polymers and glasses to nano-- and meta--materials. This one-to-one connection derives from the same logic that guarantees stability in mechanical equilibrium, and is more generally known as Le Chatelier's principle. The principle dictates that near equilibrium, a system will counteract an external stimulus. Related concepts also apply to fluidic, electric, and chemical networks, and form the foundation of  materials mechanics.

Active materials---composed of units that use ambient energy to exert local forces, pushing the system far from equilibrium~\cite{marchettiHydrodynamicsSoftActive2013,gompper2025MotileActive2025}---routinely challenge the principles of thermodynamics. These materials unlock properties that are traditionally forbidden, from long-range orientational order in 2D and phase separation in the absence of attraction~\cite{catesMotilityInducedPhaseSeparation2015} to time-crystalline states~\cite{brunoImpossibilitySpontaneouslyRotating2013,watanabeAbsenceQuantumTime2015,
fruchartNonreciprocalPhaseTransitions2021,liuPhotonicMetamaterialAnalogue2023}. Active solids~\cite{volpeRoadmapAnimateMatter2025,tanOddDynamicsLiving2022,baconnierSelectiveCollectiveActuation2022a,zhangPulsatingActiveMatter2023,xuAutonomousWavesGlobal2023,armonModelingEpithelialTissues2021,perez-verdugoExcitableDynamicsDriven2024} combine this far-from-equilibrium character with a stable reference state, in sharp contrast to the instabilities and nonlinearities that plague many active systems~\cite{alertActiveTurbulence2022}.
Significantly, even away from the stable state, these materials exhibit well-controlled dynamics~\cite{veenstraAdaptiveLocomotionActive2025}, which can be tuned via external stimuli~\cite{saintyvesSelforganizingRoboticAggregate2024,devlinMateriallikeRoboticCollectives2025}. These features make active solids especially attractive for functional materials, but being far from equilibrium presents a challenge in designing their linear response. Here we ask a basic question: when does Le Chatelier's principle, including the monotonicity of macroscale rigidity with microscale stiffness, break down in active solids?

To address this question, we focus on non-reciprocal active matter, which is found in contexts as diverse as spins~\cite{hanaiNonreciprocalFrustrationTime2024,
 avniDynamicalPhaseTransitions2025, loosLongRangeOrderDirectional2023}, optomechanical devices~\cite{raskatlaContinuousSpaceTimeCrystal2024,liskaPTlikePhaseTransition2024,reisenbauerNonHermitianDynamicsNonreciprocity2024,ruesinkNonreciprocityMagneticfreeIsolation2016}, optics~\cite{weidemannTopologicalFunnelingLight2020}, acoustics~\cite{fleurySoundIsolationGiant2014,limAcousticManipulationMultibody2024}, hydrodynamics~\cite{beatusPhononsOnedimensionalMicrofluidic2006,bililignMotileDislocationsKnead2022,guEmergenceCollectiveOscillations2025}, reaction--diffusion chemistry and colloidal physics~\cite{youNonreciprocityGenericRoute2020,braunsNonreciprocalPatternFormation2024,sahaScalarActiveMixtures2020, dinelliNonreciprocityScalesActive2023,osatNonreciprocalMultifariousSelforganization2023,meredithPredatorPreyInteractions2020}. Crucially, non-reciprocity can also be defined at the level of mechanics~\cite{scheibnerOddElasticity2020,fruchartOddViscosityOdd2023}, where the non-equilibrium nature of linear response manifests itself via amplified waves and linear work cycles. However, in all of those studies, Le Chatelier's principle is respected: the macroscopic response is monotonic in the microscopic activity.

Here we create materials in which increasing the microscopic activity weakens linear response, violating Le Chatelier's principle. Combining experiments on robotic metamaterials with analytical coarse graining, we discover that this anomalous behaviour is rooted in a percolation transition within the active parts of the elastic solid. Below percolation, active floppy modes emerge, which kill the macroscopic manifestation of microscale activity. At the same time that continuum response fails to capture the signatures of non-reciprocity, we discover 
a proliferation of active higher-frequency optical modes.
Our work establishes the percolation of active units as a general principle for understanding active solids, with applications ranging from the design of self-learning networks~\cite{goodrichPrincipleIndependentBondLevel2015, rocksDesigningAllosteryinspiredResponse2017, yanArchitectureCoevolutionAllosteric2017,sternSupervisedLearningPhysical2021} and mechanical metamaterials~\cite{saintyvesSelforganizingRoboticAggregate2024, devlinMateriallikeRoboticCollectives2025,duMetamaterialsThatLearn2025} to the dynamics of biological lattices~\cite{armonModelingEpithelialTissues2021,perez-verdugoExcitableDynamicsDriven2024, roncerayStressdependentAmplificationActive2019, duqueRuptureStrengthLiving2024, chaoSelectiveExcitationWorkgenerating2026}.

\begin{figure*}[t!]
\centering
\includegraphics[width=\linewidth]{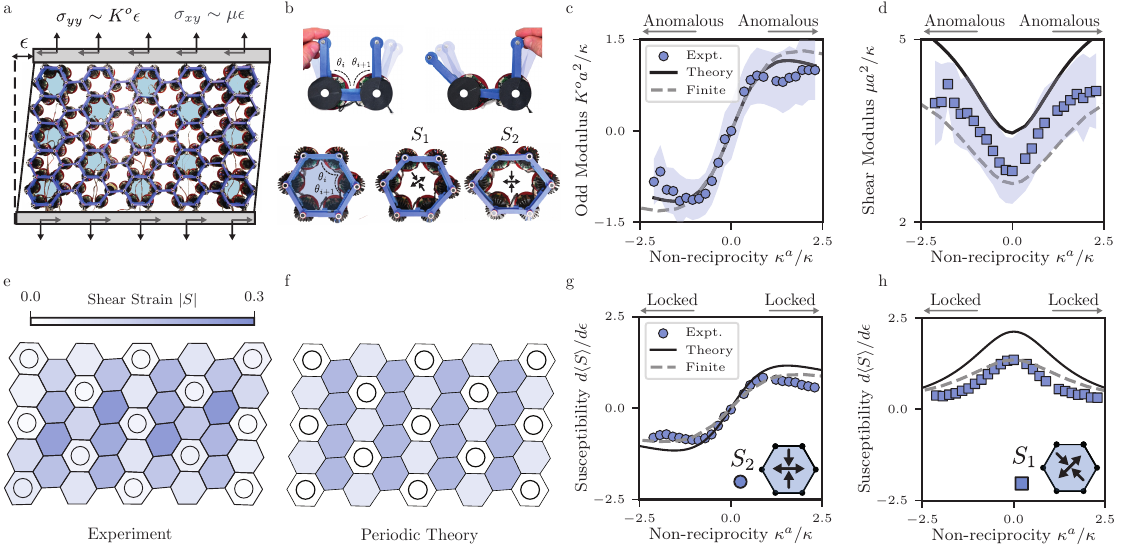}
 \caption{{\bf Anomalous odd response in non-reciprocal materials.} \rev{Transversely shearing an isotropic lattice built from non-reciprocal plaquettes generates normal stresses, the hallmark of odd elasticity. These odd moduli break equilibrium conditions of monotonic response, and can anomalously vanish as microscopic activity increases. (a-b) A robotic metamaterial built from a mixture of active (blue highlights) and passive hexagonal plaquettes. Each active plaquette is made from three bar linkages (side length $a$). Motors on each linkage vertex exert torques according to a non-reciprocal torque-angle relation $\tau_i = \kappa^a(\delta \theta_{i+1}-\delta \theta_{i-1})$, alongside a passive elastic response $\tau_i = -\kappa \delta \theta_{i}$. The lowest deformation modes of these plaquettes are the transverse $S_1$ and vertical $S_2$ shears. (c,d) Shearing the lattice reveals an anomalous regime where the odd modulus $K^o$ decreases as microscale non-reciprocity $\kappa^a$ diverges, with crossover occurring at $\kappa^a/\kappa \sim 1$. Concomitantly, the shear modulus $\mu$ actively stiffens. Points correspond to experimental runs, with error bars quantifying the spread of force sensor measurements on either side of the metamaterial. Solid curves indicates coarse-grained periodic theory. Dashed grey curves indicate ball-spring simulations that account for finite-size effects. (e,f) Visualising the strain field within the lattice in experiment and periodic theory reveals a non-affine checkerboard pattern in which active plaquettes lock, admitting no shear deformation at all. Plaquette colors show hexagonal strain $|S|=\sqrt{S_1^2 + S_2^2}$, with active plaquettes indicated by inset circles. (g,h) To track the plaquette-level response to an applied shear $\epsilon$, we average the strain over all active plaquettes, $\langle S \rangle$, and then compute the susceptibility $ {d \langle S \rangle}/{d \epsilon}$ as a function of activity $\kappa^a$. We find that the anomalous plaquette-locked regime emerges at high activity, with crossover in the shear modes occurring concurrently with the peak and decay of the odd modulus.}}
\label{fig:Phenomenology}
\end{figure*}

We first illustrate how macroscale non-reciprocal response can vanish even as microscale non-reciprocity increases. Figure~\ref{fig:Phenomenology}(a) shows a robotic metamaterial built from hexagonal plaquettes of torque motors. When the motors are off, each hinge in the lattice has a passive torque-angle relation $\tau_i=-\kappa \delta \theta_i$ with stiffness $\kappa$. Within each active plaquette [Fig.~\ref{fig:Phenomenology}(a), blue highlights] the $i^\mathrm{th}$ motor additionally exerts an active torque $\tau_i=\kappa^a(\delta \theta_{i+1}-\delta \theta_{i-1})$ via an antisymmetric combination of neighboring angular deflections $\delta \theta_{i+1}, \delta \theta_{i-1}$ [Fig.~\ref{fig:Phenomenology}(b)] (Methods \S\ref{sec:Construction} and Supplemental movie S1~\cite{SupplementaryMaterial}). This microscale antisymmetry, quantified by $\kappa^a$, breaks Maxwell-Betti reciprocity and generates an antisymmetric coupling between the horizontal $S_1$ and vertical $S_2$ shear modes of the lattice. 

\rev{We apply a macroscopic transverse strain $\epsilon \sim S_1$ to the lattice and measure both normal and tangential shear stresses, $\sigma_{yy}$ and $\sigma_{xy}$ [Fig.~\ref{fig:Phenomenology}(a)]---in Methods~\ref{sec:ExperimentalProtocol} and Fig.~\ref{fig:DataProcessing} we detail our experimental protocol and show full stress-strain sweeps. Large-scale non-reciprocal response is quantified by the odd modulus, $K^o=d\sigma_{yy}/ d\epsilon$ at $\epsilon=0$. This linear modulus vanishes for passive materials. Previous measurements of odd elasticity have been inferred indirectly from displacement data~\cite{veenstraAdaptiveLocomotionActive2025} in highly dynamic, fluctuating active solids~\cite{tanOddDynamicsLiving2022, bililignMotileDislocationsKnead2022}, which can be complicated by noise~\cite{choiNoisedrivenOddElastic2024} or additional degrees of freedom~\cite{chaoSelectiveExcitationWorkgenerating2026}. Here our robotic platform and experimental setup allow a direct measurement of $K^o$ from stress-strain data in the static regime. }

\rev{For weak non-reciprocity $\kappa^a$, we measure a monotonic response, $K^o \sim \kappa^a$. Yet on further increase of $\kappa^a$ we observe a peak and subsequent decay of $K^o$ [Fig.~\ref{fig:Phenomenology}(c)], accompanied by an active stiffening of the linear shear modulus $\mu= d\sigma_{xy}/ d\epsilon$ [Fig.~\ref{fig:Phenomenology}(d)]. The anomalous response in $K^o$ vs. $\kappa^a$ stands in stark contrast to the scaling of elastic moduli with the stiffness of microscopic building blocks in passive isotropic materials, which must be monotonic (Methods \S\ref{sec:Monotonic}).}

\begin{figure*}[t!]
\centering
\includegraphics[width=\linewidth]{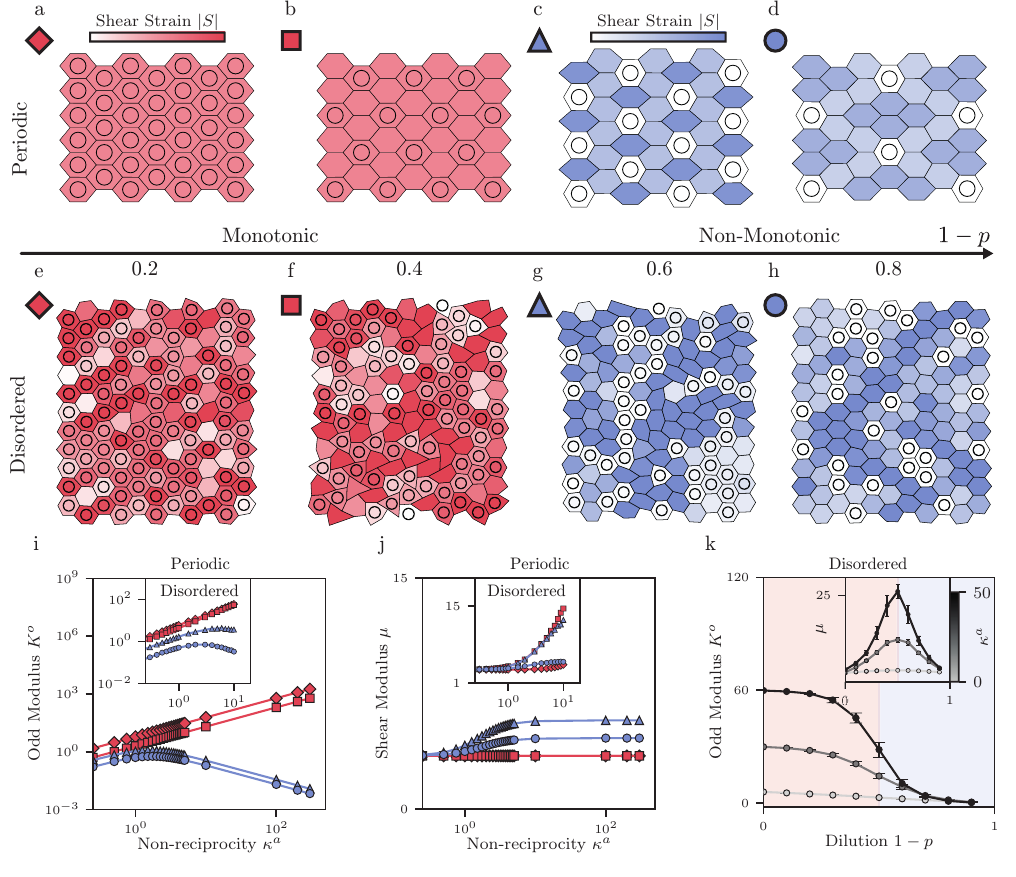}
  \caption{{\bf Odd response vanishes via a percolation transition in active degrees of freedom}. (a-h) We remove active plaquettes from \rev{ball-spring models of} a fully tiled honeycomb lattice, either through isotropy-preserving periodic dilutions (a-d) or through random removals with probability $1-p$ (e-h). For large $p$, active plaquettes deform in system-spanning clusters, generating odd response. For small $p$, these plaquettes decouple from one another, and the strain in the active units vanishes. Colors show strain magnitude $|S|$ per plaquette under a vertical pure shear strain $\epsilon=0.05$. Red/blue dichotomy indicates lattices with monotonic/non-monotonic odd response at large activity. Panels (e-h) show typical random realizations at $1-p=0.2,0.4,0.6,0.8$. (i, j) Dilutions either exhibit either monotonic or anomalous $K^o(\kappa^a)$, with accompanying active stiffening of $\mu(\kappa^a)$, as the active plaquette fraction $p$ decreases and active force chains are destroyed. Markers in main panels show results for the structured lattices in (a-d). Markers in insets correspond to averages over random lattices (e-h). 
  (k) Varying $p$ continuously in random lattices (e-h), we find a step jump in $K^o(p)$ and a spike in $\mu(p)$ as $p$ crosses the percolation threshold $p=p_c=1/2$. Error bars denote standard deviations of the fluctuating quantities $K^o$ and $\mu$. Throughout the figure $\kappa=a=1$.
  }
\label{fig:DesignerDilution}
\end{figure*}

\rev{To unravel the cause of the anomalous decay of $K^o$ and active stiffening of $\mu$, we develop and coarse-grain a periodic model of our metamaterial to analytically predict its elastic moduli. In Methods \S\ref{sec:CoarseGraining}--\ref{sec:Moduli} we give compact expressions for the moduli and internal deformations of a generic non-reciprocal lattice built from non-pairwise interactions, such as those shown in Fig.~\ref{fig:Phenomenology}. Our theory predicts that}
\begin{align}
\begin{split}
 K^o &= \frac{3}{2}\,\frac{\kappa^a}{a^2} \,\frac{1}{1 + c_1(\kappa^a/\kappa)^2}, \\ 
 \mu &= 2\sqrt{3}\,\frac{\kappa}{a^2} \,\frac{1+ c_2 (\kappa^a/\kappa)^2}{1  + c_1 (\kappa^a/\kappa)^2}, 
 \label{eq:TheoryKek20}
\end{split}
\end{align}
\rev{where $a$ is the hexagonal side length of a single linkage, and $c_2=45/64 > c_1=27/64$ are positive numerical constants that reflect the geometry of the lattice. We predict that $K^o$ peaks when $\kappa^a/\kappa \sim 1$, before decaying as $K^o\sim \kappa^2/\kappa^a$ in the limit of large activity. In this same limit, we also predict that the shear modulus $\mu$ plateaus at $\mu = 2\sqrt{3} \kappa c_2/c_1$, a rescaled value above the passive $\kappa^a=0$ limit. This peak in $K^o$ and active stiffening of $\mu$ are both in good agreement with our experiments [Fig.~\ref{fig:Phenomenology}(c,d)].}

\rev{Our periodic theory has a systematic overshoot relative to experiment, particularly in the shear modulus $\mu$ [Fig.~\ref{fig:Phenomenology}(d)]. By developing ball-spring simulations of our metamaterials, we find that the dominant cause of this discrepancy is finite-size effects, which are caused by the relatively small number of unit cells in our experiment (Methods \S\ref{sec:Deviation}], Figs.~\ref{fig:FiniteSizeScaling1},~\ref{fig:FiniteSizeScaling2}). However, the twin signatures that our theory predicts---a peak in $K^o$ and a stiffening $\mu$---are robust to these experimental imperfections. What is the mechanism behind these phenomena?}

\rev{ We characterize the micromechanics underpinning these effects by projecting the deformations of every plaquette onto the shear modes of a single hexagon [Fig.~\ref{fig:Phenomenology}(e,f)]. At large activity we observe a highly non-affine checkerboard of deformations: In both experiment and periodic theory the magnitude of the strain $\langle|S|\rangle= \langle \sqrt{S_1^2 + S_2^2} \rangle$ in the active plaquettes vanishes [Figs.~\ref{fig:Phenomenology}(e,f)]. To better quantify this observation, we first fix $\kappa^a$, and measure sweeps of applied strain $\epsilon$ against plaquette-level shear $S$ (Fig.~\ref{fig:DataProcessing}). The slopes of these relations, $d S /d \epsilon$ for $S_1$ and $S_2$, measure how susceptible a given hexagon is to the externally applied strain, and into which mode it will shear.}

\rev{We now vary $\kappa^a$ and track the susceptibilities $d \langle S \rangle /d \epsilon$ averaged across active plaquettes [Fig.~\ref{fig:Phenomenology}(g,h)]. When $\kappa^a=0$ all hexagons are passive and the shear in each plaquette follows the applied strain field: $S_1 \sim \epsilon $ dominates with $S_2 \approx 0$. As $\kappa^a$ increases, $d \langle S_2 \rangle /d \epsilon$ peaks as the coupling between shear modes generates a normal shear strain $S_2 \sim K^o S_1$. However as activity further increases $S_2$ and $S_1$ both decay: the active plaquettes lock, and the passive degrees of freedom shoulder the entire strain field applied to the sample boundary. Our systematic sweep of $d \langle S \rangle /d \epsilon$ against $\kappa^a$ [Fig.~\ref{fig:Phenomenology}(g,h)] demonstrates that the locking behaviour underpinning the checkerboard in Figs.~\ref{fig:Phenomenology}(e,f) emerges only at high activity.}


The link between micro-- and macro-- scale non-reciprocity is therefore subtle: Figure~\ref{fig:Phenomenology} shows a structure which is highly non-reciprocal yet possesses decaying odd response, in contrast to the monotonic structure-function link exhibited by recent examples of living and synthetic active solids~\cite{tanOddDynamicsLiving2022, chaoSelectiveExcitationWorkgenerating2026,veenstraAdaptiveLocomotionActive2025}. To uncover the structural motifs that separate monotonic and non-monotonic materials,
\rev{we use our analytic coarse-graining procedure to numerically generate a family of honeycomb lattices containing a varying fraction $p$ of active plaquettes, and calculate their theoretical elastic moduli and internal strains under compression (Fig.~\ref{fig:DesignerDilution} and Supplemental movie S2~\cite{SupplementaryMaterial}).}
We vary $p$ either through periodic dilutions of a fully tiled lattice [Figs.~\ref{fig:DesignerDilution}(a-d)] or via random removals of active plaquettes [Figs.~\ref{fig:DesignerDilution}(e-h)]---in Methods \S\ref{sec:Computation} we detail our analytical results for periodic lattices, and numerical procedure for calculating the moduli of random structures.

For $p \approx 1$ [Figs.~\ref{fig:DesignerDilution}(a-b,e-f)], \rev{our numerics show} that active plaquettes strain in system-spanning clusters, with a corresponding nonzero odd response [Fig.~\ref{fig:DesignerDilution}(i,k)]. However, at $p=p_c=1/2$, we cross the site percolation threshold for the triangular Bravais lattice on which the hexagonal plaquettes sit. Below the percolation threshold active plaquettes decouple from one another [Figs.~\ref{fig:DesignerDilution}(c-d,g-h)]: Each plaquette independently rigidifies, and odd response is lost [Figs.~\ref{fig:DesignerDilution}(i,k)]. Just beneath $p_c$ [Figs.~\ref{fig:DesignerDilution}(b-c,f-g)], these suddenly rigid plaquettes occupy a substantial fraction of the sample, causing a spike in the passive shear modulus $\mu$ [Figs.~\ref{fig:DesignerDilution}(j, k)] which coincides with the jump in the active modulus $K^o$ [Figs.~\ref{fig:DesignerDilution}(i, k)]. This spike in $\mu$ then decays as $p$ tends to zero [Fig.~\ref{fig:DesignerDilution}(k)] and rigid units are further removed from the lattice. \rev{These twin behaviours---a jump in $K^o$ and a spike in $\mu$---are not restricted to on-lattice disorder, but also occur even in positionally disordered structures (Methods \S\ref{sec:PositionalDisorder})}. In summary, the lattice connectivity itself mediates a transition into an anomalous non-monotonic regime in $K^o$ as a function of $\kappa^a$, alongside an active stiffening of the passive modulus $\mu$.

\begin{figure}[t!]
\centering
\includegraphics[width=\linewidth]{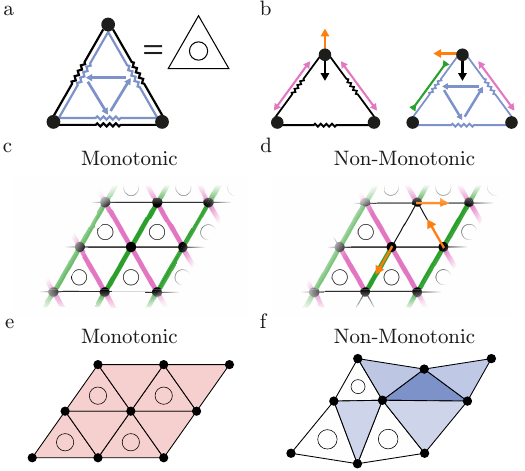}
  \caption{
  {{\bf The anatomy of vanishing odd response.}  
  (a) A minimal model for vanishing odd response and plaquette locking: A triangular plaquette made of extensional springs with stiffness $k$ (black), and non-reciprocal springs (blue) with tension law $T_i= k^a(\delta L_{i+1} - \delta L_{i-1})$. Blue directed arrows show the neighbors of a given bond. (b) Compressing passive springs generates a symmetric tension field (pink) and restoring force (orange) parallel to the applied load (black). Compressing non-reciprocal springs generates a tension field (green, pink arrows) and restoring force which are skewed. 
  (c,d) System-spanning active force chains are only possible in certain lattice geometries. Panel (c) shows a successful chain in a $p=1/2$ activated lattice. In the $p=3/8$ activated lattice (d)  we cannot construct a chain. When active plaquettes are sheared, we necessarily find unbalanced nodal forces. This lattice cannot support odd response at high activity. (e,f) Force-balanced displacement fields under vertical applied shear for $k^a \gg k$, colored by shear strain magnitude. In the under-percolated $p=3/8$ lattice, active plaquettes do not shear.
  }}
\label{fig:ActiveForceBalance}
\end{figure}

\begin{figure*}[t!]
\centering
\includegraphics[width=\linewidth]{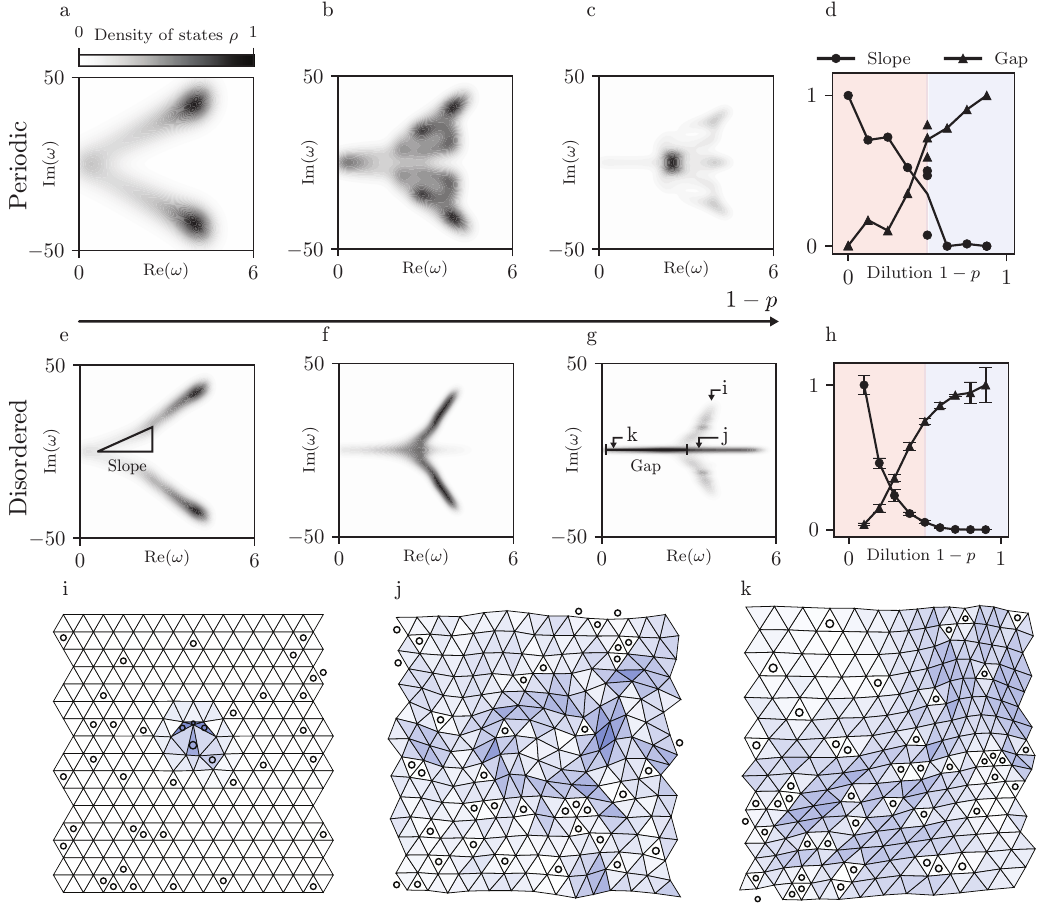}
      \caption{{\bf Under-percolated lattices exhibit an active phonon gap in their vibrational spectrum}.
      As active plaquettes are removed from \rev{ball-spring models of}  a fully tiled triangular lattice, the vibrational eigenspectrum $\omega$ splits into two discrete lobes---an effectively passive component with $\mathrm{Im}(\omega)=0$ which contains lattice phonons, and highly localized vibrations at complex $\omega$. \rev{The eigenvalues $\omega$ are expressed in units of the spring stiffness $k$}. (a-c) The density of states $\rho(\omega)$ for three periodic dilutions of a fully tiled active triangular lattice at $k^a/k=5$ [(a) $p=1$, (b) $p=5/8$, (c) $p=3/8$]. For $p\approx 1$ we find phonon bands continuously connected to the origin, whose slope in the complex plane quantifies $K^o$. Yet as $p$ is decreased, these bands collapse to the $\mathrm{Re}(\omega)$ axis, leaving lobes of active modes at complex $\omega$ gapped from the origin. Eigenvalues $\omega$ are given in units where $k=1$. (e-g) Random dilutions [(e), $p=0.9$, (f) $p=0.5$, (g) $p=0.2$] exhibit the same spectral collapse as their structured counterparts. (d,h) A bilinear fit of $\rho(\omega)$ quantifies both the slope at the origin and active phonon gap. We find that $K^o$ vanishes as the gap opens and $p$ crosses $p_c=1/2$. In panel (d), data points correspond to all isotropic dilutions of a $2 \times 4$ supercell of active plaquettes (Methods \S\ref{sec:Computation}). \rev{Three periodic lattices exist at $p=1/2$, with inter-lattice variations in these metrics---we show data for each lattice, with the line denoting an average. In panel (h), the error bars quantify fluctuations in fits between samples. For ease of comparison, we normalize both slope and gap data by their largest value to lie in $[0,1]$.} (i-k) The eigenstates of a dilute lattice [panel (g)] reveal that modes at complex $\omega$ are composed of highly localised oscillations of individual plaquettes (i), whereas modes living on the $\mathrm{Re}(\omega)$ axis exhibit the swirls (j) and quasi plane waves (k) of an effectively passive spring network, interspersed with unsheared active plaquettes that act as a source of disorder. Color shows the shear strain in each plaquette under the corresponding normalized eigenvector displacement field.} 
      \label{fig:DynamicalDesignerDilution}
\end{figure*}

Motivated by the crucial role of $p$ in the onset of anomolous odd response, we compare the force balance in two lattices that lie on either side of percolation (Fig.~\ref{fig:ActiveForceBalance}). We first boil the complexity of our honeycomb
\rev{lattices} down to a minimal model: 
a triangular lattice composed of longitudinal springs with finite stiffness $k$, where each triangular plaquette also hosts non-reciprocal longitudinal springs [Fig.~\ref{fig:ActiveForceBalance}(a)]. These springs exert tensions~$k^a(\delta L_{i+1}-\delta L_{i-1})$ in response to length changes $\delta L$ of neighboring bonds. Straining a single plaquette of these non-reciprocal springs generates an antisymmetric bond tension field which is skewed at an angle to the applied load [Fig.~\ref{fig:ActiveForceBalance}(b)]. We now attempt to tile this skewed tension field across a lattice, in order to build system-spanning tension chains that generate odd elastic response.

We first consider a half tiled lattice [Fig.~\ref{fig:ActiveForceBalance}(c)], for which we predict a monotonic odd response $K^o=3 k^a/4$. Examining the tension chains in this lattice under shear, we find that our plaquette-level tensions can be successfully tiled across the entire sample. We now attempt the same tiling process for an underpercolated lattice with $p=3/8$, for which our theory predicts \rev{$K^o \sim k^2/(k^a)$} at large $k^a$ (Methods \S\ref{sec:Computation}) [Fig.~\ref{fig:ActiveForceBalance}(d)]. We discover a crucial feature in the tension field within this dilute lattice: we inevitably encounter masses at which active forces cannot be balanced. At finite passive stiffness $k$, these unbalanced active forces are counteracted by passive elasticity. However, as $k^a/k \rightarrow \infty$ we have a separation of scales, and active forces must balance against one another.
Therefore, instead of balancing active forces across unit cells, the $p=3/8$ lattice balances them within individual plaquettes.

The consequence of this local active force balance is shown in Figs.~\ref{fig:ActiveForceBalance}(e,f), where we examine the equilibrium configurations of both triangular lattices under shear:
\rev{our model exhibits}
the same locked states, and non-affine checkerboards, as we observe in our 
\rev{experimentally constructed}
 metamaterials [Fig.~\ref{fig:Phenomenology}(d,e)] (Methods \S\ref{sec:Honeycomb}).
Intuitively, non-reciprocal tensions are screened out at high activity. The resulting inter-cell tension field is then found by minimizing the energy of an effectively passive lattice, which must have $K^o=0$. This reduced lattice also has fewer degrees-of-freedom than the original lattice, because of the additional constraints of active force balance. Adding these constraints rigidifies the lattice, increasing $\mu$.

If the odd modulus of a non-reciprocal material vanishes, what are the remaining signatures of activity? In Figure~\ref{fig:DynamicalDesignerDilution}, we show that the vibrational spectrum of an under-percolated active lattice differs dramatically from passive materials, even in the limit of vanishing $K^o$. We consider again lattices of non-reciprocal triangular plaquettes, such as those shown in Fig.~\ref{fig:ActiveForceBalance}, and focus on the high activity limit $k^a/k \rightarrow \infty$. By diagonalizing the lattice dynamical matrix, we obtain the vibrational eigenspectrum $\omega$ (Methods \S\ref{sec:Computation}). Because the dynamical matrix is non-Hermitian, $\omega$ is in general complex: under an overdamped dynamics, $\mathrm{Re}(\omega)$ would correspond to relaxation and $\mathrm{Im}(\omega)$ to oscillations.

For large $p$, we find two branches of complex eigenmodes [Figs.~\ref{fig:DynamicalDesignerDilution}(a,e)], inherited from the phonon structure of a fully tiled active lattice. These oscillatory phonons are the long-wavelength signature of microscopic non-reciprocity, with the slope of each branch in the complex plane quantifying the odd modulus,~$\mathrm{Im}(\omega)/\mathrm{Re}(\omega) \sim K^o/\mu$~\cite{scheibnerNonHermitianBandTopology2020}. As active plaquettes are removed from the lattice, these phononic bands collapse to the $\mathrm{Re}(\omega)$ axis as $K^o$ vanishes [Figs.~\ref{fig:DynamicalDesignerDilution}(b-c,f-g)].
\rev{Indeed in Fig.~\ref{fig:TriangularLatticeElasticModuli} and Table \ref{tab:TriangularLattice} of the Methods, we show that these triangular lattices show a decay of $K^o$ with decreasing $p$, and a spike in $\mu$ at intermediate $p$, that is analogous to that shown in Fig.~\ref{fig:DesignerDilution}(i,j).}
Yet even as this collapse occurs, we find that a second branch of modes remains complex: The spectrum splits into two pieces, and an active phonon gap in $\mathrm{Re}({\omega})$ opens. Within the gap, the vibrations of the material appear effectively passive. Beyond the gap, oscillations emerge at complex frequencies. In Figs.~\ref{fig:DynamicalDesignerDilution}(d,h) we quantify the appearance of this gap as a function of plaquette fraction $p$. We find that vanishing of the slope of the phononic bands and the emergence of the active phonon gap occur simultaneously, both coinciding with the $p=1/2$ site percolation threshold for removal of triangular plaquettes. 

What is the spatial structure of the lattice vibrations that underpin this active phonon gap? In Figs.~\ref{fig:DynamicalDesignerDilution}(i-k) we take a single underpercolated lattice and visualize the deformation fields which live in each branch of the eigenspectrum [Fig.~\ref{fig:DynamicalDesignerDilution}(g)]. We find that complex modes with nonzero $\mathrm{Im}(\omega)$ are composed of highly non-affine displacement fields localized around a single active plaquette: even in an overdamped scenario this plaquette would oscillate with substantial shear strain [Fig.~\ref{fig:DynamicalDesignerDilution}(i)]. In stark contrast, we find that eigenmodes lying on the real axis exhibit no shear in the active plaquettes at all. Instead, the shear of the passive units alone exhibits the mid-frequency swirls [Fig.~\ref{fig:DynamicalDesignerDilution}(j)] and low-frequency quasi-plane waves [Fig.~\ref{fig:DynamicalDesignerDilution}(k)] which are characteristic of a passive granular material or disordered spring network~\cite{heckeJammingSoftParticles2009}. When active plaquettes are too dilute to couple with one another, activity no longer enables the collective motion and wave propagation captured in a continuum description, but instead powers highly localized oscillations.

In this work, we have explored non-reciprocity across scales in active solids, and discovered a regime where more is less: increasing activity on the microscale can decrease it at the continuum level. The fundamental reason is multiscale order: as active degrees of freedom fail to percolate the system, microscopic non-reciprocity cannot upscale to the continuum, instead remaining trapped in optical modes and high wavenumbers. 
We anticipate our results to apply broadly to active solids upon sufficient dilution~\cite{armonModelingEpithelialTissues2021, tanOddDynamicsLiving2022, chaoSelectiveExcitationWorkgenerating2026,burlaStressManagementComposite2019, roncerayStressdependentAmplificationActive2019,zhangPulsatingActiveMatter2023,xuAutonomousWavesGlobal2023}, 
and speculate that active percolation transitions radically affect core functional properties of such materials such as soliton propagation~\cite{veenstraNonreciprocalTopologicalSolitons2024}, synchronisation~\cite{xuAutonomousWavesGlobal2023, baconnierSelectiveCollectiveActuation2022a}, locomotion, or learning~\cite{ yanArchitectureCoevolutionAllosteric2017,sternSupervisedLearningPhysical2021,mandalLearningDynamicalBehaviors2024,duMetamaterialsThatLearn2025}, as well as offering enhanced mechanical tunability of non-equilibrium moduli~\cite{goodrichPrincipleIndependentBondLevel2015, rocksDesigningAllosteryinspiredResponse2017,alvaradoMolecularMotorsRobustly2013,burlaStressManagementComposite2019,roncerayStressdependentAmplificationActive2019,
wysejacksonStructuralOriginsCartilage2022}.
\rev{Recent experiments directly probe the structure-function link in biologically active solids, measuring non-reciprocal moduli in epithelia~\cite{chenChiralityScalesTissue2025} alongside acoustic and optical modes in active spinners~\cite{chaoSelectiveExcitationWorkgenerating2026}.
Given such advances, our results suggest correlating the distribution of active elements with measurements of mode structure as a useful quantifier for active response in living systems, just as structural connectivity data predicts passive rigidity in cell networks~\cite{petridouRigidityPercolationUncovers2021}.}

\rev{The phenomena we uncover here are also reminiscent of recent work on vanishing non-reciprocity upon coarse-graining active fluids~\cite{dinelliNonreciprocityScalesActive2023}, albeit with a distinct physical mechanism. This commonality invites the question of a fluidic analogue to the percolation of active units~\cite{jorgeActiveHydraulicsLaws2024}, and highlights the subtleties of predicting the large-scale properties of active networks: our results form an amusing analogue to Braess' paradox~\cite{nicolaouMechanicalMetamaterialsNegative2012,ducarmeExoticMechanicalProperties2025}, in which adding more roads to a network increases travel times, or indeed adding more struts to a bridge makes it weaker. Recent analogues to active matter in quantum and optomechanical platforms observe many-body synchronised states which are powered by non-reciprocity~\cite{raskatlaContinuousSpaceTimeCrystal2024, hanaiNonreciprocalFrustrationTime2024}. An exciting future question is to ask whether an analogous active percolation transition exists in this quantum context. }

\acknowledgments{\textit{Acknowledgments}---We thank Edan Lerner and Moumita Das for helpful discussions. J.B.~acknowledges funding from the European Union’s Horizon research and innovation programme under the Marie Sklodowska-Curie Grant Agreement No. 101106500. This research was supported in part by grant NSF PHY-2309135 to the Kavli Institute for Theoretical Physics (KITP). A.S.~acknowledges funding from UKRI through award No.~EP/T000961/1. Codes and data supporting this study are available in a public Zenodo repository at https://doi.org/10.5281/zenodo.15241055.
}

%

\clearpage

\section{Methods}
In this Methods section, in \S\ref{sec:CoarseGraining} and \S\ref{sec:Moduli} we develop the general framework for constructing dynamical matrices of non-reciprocal lattices and coarse-graining to extract elastic moduli. In \S\ref{sec:Monotonic} we give an argument demonstrating that the link between microscale stiffness and macroscale elasticity must be monotonic for passive materials, and give conditions on when the odd modulus $K^o$ anomalously vanishes. \S\ref{sec:Honeycomb} shows how active force chains vanish in our honeycomb metamaterials. \S\ref{sec:Computation} describes the computational implementation of our analytical calculations and the numerical methods we use to analyze periodic and random ball-spring lattices. Finally,\S\ref{sec:Construction} and \S\ref{sec:ExperimentalProtocol} provide experimental details for the construction of our robotic metamaterials and experimental protocols.

\section{From microscopics to the elastic limit for non-reciprocal lattices}
\label{sec:CoarseGraining}
\begin{figure}[t]
\centering
\includegraphics[width=\linewidth]{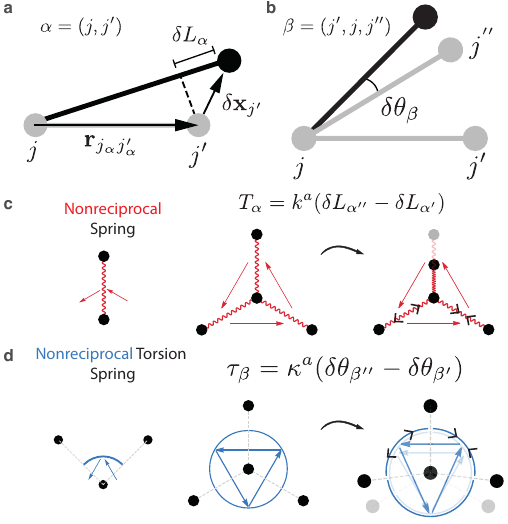}
    \caption{(a,b) Schematics of the variables associated with the $\alpha$\textsuperscript{th} longitudinal spring given by the ball-index pair $[j_\alpha,j_\alpha']$ and the $\beta$\textsuperscript{th} torsional spring given by the ball-index triplet $[j_\beta',j_\beta,j_\beta'']$. (c,d) Schematics of non-reciprocal tension-length springs and torsion-angle springs. The connectivity network that encodes neighboring lengths (angles) is shown via red (blue) arrows. In each panel we show an example deformation, with the corresponding longitudinal (angular) tensions shown as black arrows.}
\label{fig:Notation}
\end{figure}
In this section we describe how to construct the dynamical matrix for an arbitrary network containing non-reciprocal active bonds, and how to coarse-grain this matrix to extract elastic moduli. Our approach follows~\cite{scheibnerOddElasticity2020, lubenskyPhononsElasticityCritically2015} generalised to include the non-reciprocal force laws described in the main text.

\subsection{Non-reciprocal lattices}
A lattice is described by a set of balls (labeled $j=1,2,...$), and a set of springs connecting these balls. Here we consider two types of spring: longitudinal and torsional. Each longitudinal spring (labeled $\alpha=1,2,...$) identifies an ordered pair of ball indices $[j_\alpha,j_\alpha']$, and each torsional spring (labeled $\beta=1,2,...$) identifies an ordered triple of ball indices $[j_\beta',j_\beta,j_\beta'']$, as shown in Fig.~\ref{fig:Notation}(a,b). 

A longitudinal spring $\alpha$ has a length change $\delta L_\alpha$ and linear tension $ T_\alpha$. A torsional spring $\beta$ has an angle change $\delta \theta_\beta $ and angular tension (or torque) $\tau_\beta$. 
Here, the angle subtended by this torsional bond is defined in a right-handed (counter-clockwise) sense, in other words from $j'$ to $j''$ about $j$ in Fig.~\ref{fig:Notation}(b).
In Eq.~\eqref{eq:TensionDisplacement} below we will prescribe non-reciprocal relations between the length/angle changes and tensions.

Within this framework, the linearized map between ball displacements $\delta \bm X=(\delta \bm x_1^\intercal, \delta \bm x_2^\intercal, ...)^\intercal$ and ball forces $\bm F =(\bm f_1^\intercal, \bm f_2^\intercal, ... )^\intercal$ can be written as
 \begin{equation}
 M\,\delta\ddot{\bm X} =  {\bm F} =-D \delta \bm X =  -C^\intercal \Bbbk C \delta \bm X,
 \label{eq:EOM}
 \end{equation}
where $M=\mathrm{diag}(m_1,m_2,...)$ is the matrix of ball masses. The dynamical matrix $D$ splits into three separate matrices, 
\begin{equation}
D=C^\intercal \Bbbk C.
\end{equation}
First, the compatibility matrix $C$ maps ball displacements $\delta \bm X$ to bond elongations $\bm \delta \bm E = (\delta \bm L^\intercal, \delta \bm \theta^\intercal)^\intercal = (\delta L_1, \delta L_2, ..., \delta \theta_1, \delta \theta_2, ....)^\intercal$.
For a lattice composed of both longitudinal and torsional springs, $C$ splits into a longitudinal  part, $C_L$ and a torsional part $C_\theta$, with 
 \begin{equation}
 \delta\bm E = 
 \begin{pmatrix}
    \delta L \\
    \delta \theta
 \end{pmatrix}
 = 
 \begin{pmatrix}
    C_L \\
    C_\theta
 \end{pmatrix}
 \delta \bm X
 = C\,\delta\bm X.
 \end{equation}
We will give explicit constructions of $C_L$ and $C_\theta$ in \S\ref{subsec:CompMat}.
Next, the connectivity matrix $\Bbbk$ maps elongations $\delta \bm E$ to tensions $\bm T = (T_1, T_2,...,\tau_1,\tau_1,...)^\intercal$ 
 \begin{equation}\label{eq:elongations2tensions}
 \bm T = - \Bbbk\,  \delta\bm E,
 \end{equation}
such that (for positive diagonal entries in $\Bbbk$) an change in elongation leads to a tension that opposes that elongation. We will discuss more exotic constructions of $\Bbbk$ in \S\ref{subsec:StiffMat}.
Finally the equilibrium matrix $C^\intercal$ maps tensions to forces
\begin{equation}
    \bm F = C^\intercal\bm T.
\end{equation}

\subsection{The microscopic stiffness matrix $\Bbbk$ } \label{subsec:StiffMat}
In a standard ball-spring lattice made up of Hookean longitudinal and torsional springs, $\Bbbk$ would simply be a diagonal matrix of spring constants, $\mathbbm k=\mathrm{diag}(k_1,k_2,\dots,\kappa_1/a^2,\kappa_2/a^2,\dots)$, where the $k_\alpha$ and $\kappa_\beta$ are the longitudinal and torsional spring constants and $a$ is a typical length-scale of the lattice. More generally, whenever spring couplings are derived from a potential, $\Bbbk$ is constrained to be symmetric.

Our lattices break this reciprocal symmetry by explicitly including an antisymmetric component in $\Bbbk$, 
\begin{equation}
\Bbbk = 
\begin{pmatrix}
 & \vdots & \ddots  & \vdots & &\vdots  \\
\dots & k^a & \dots & k & \dots &  -k^a & \dots \\
 & \vdots &  & \vdots & \ddots &\vdots  \\
\end{pmatrix}.
\end{equation}
The microscopic spring stiffness matrix $\Bbbk$ encodes information sharing between different springs: If $\Bbbk$ contains non-diagonal elements, elongations in one spring can generate tensions in another spring. Interpreting $\Bbbk$ as a graph-theoretical adjacency matrix for this inter-spring communication network, the antisymmetric component corresponds to directed cycles within this network. Figure~\ref{fig:Notation}(c) shows a minimal example of this network: a length three cycle connecting extensional springs, with the connectivity network shown as directed red arrows. Figure~\ref{fig:Notation}(d) shows a similar example for torsional springs. For any cycle of extensional springs $( \dots \alpha', \alpha, \alpha'', \dots)$ or torsional springs $( \dots \beta', \beta, \beta'', \dots)$ we then have the force laws
\begin{align}
T_\alpha = k^a (\delta L_{\alpha''} - \delta L_{\alpha'}),\\
\tau_\beta = \kappa^a (\delta \theta_{\beta''} - \delta \theta_{\beta'}),
\label{eq:TensionDisplacement}
\end{align}
with the corresponding entries in $\Bbbk$ reading $\Bbbk_{\alpha \alpha'' } = -\Bbbk_{\alpha \alpha '} = k^a$, $\Bbbk_{\beta \beta'' } = -\Bbbk_{\beta \beta'} = \kappa^a$. 

\subsection{The compatibility matrix $C$}\label{subsec:CompMat}
Given an arrangement of springs, and relative ball positions $\bm r_{jj'}$ as shown in Fig.~\ref{fig:Notation}(a--b), the expressions for the elements of $C$ follow from direct geometric calculations of linearized length and angle changes for longitudinal and torsional springs in terms of ball displacements. The displacement of a single extensional spring $\alpha=[j_\alpha, j_\alpha']$ gives a linearised length change $\delta L_\alpha$ of 
\begin{equation}
\delta L_\alpha = 
\begin{pmatrix}
-\hat{\bm r}_{jj'} & \hat{\bm r}_{jj'}
\end{pmatrix}
\begin{pmatrix}
\delta \bm x_j \\
\delta \bm x_{j'}
\end{pmatrix},
\label{eq:LinearizedLengthChange}
\end{equation}
as shown in Fig.~\ref{fig:Notation}(a). In Eq.~\eqref{eq:LinearizedLengthChange} $\hat{\bm r}$ denote the normalized bond vector ${\bm r}$, and we omit the spring-index subscripts $\alpha$ when they are clear from context. Similarly, the displacement of a single torsional spring $\beta =[j_\beta',j_\beta,j_\beta'']$ [Fig.~\ref{fig:Notation}(b)] gives a linearised angle change $\delta\theta_\beta$ of
\begin{equation}
\delta \theta_\beta = 
\begin{pmatrix}
-\frac{\hat{\bm r}^\perp_{jj'}}{r_{jj'}}& \frac{\hat{\bm r}^\perp_{jj'}}{{r_{jj'}}} - \frac{\hat{\bm r}^\perp_{jj''}}{r_{jj''}}  &\frac{\hat{\bm r}^\perp_{jj''}}{{r_{jj''}}}
\end{pmatrix}
\begin{pmatrix}
\delta \bm x_{j'} \\
\delta \bm x_j \\
\delta \bm x_{j''}
\end{pmatrix},
\end{equation}
where ${\bm r}^\perp = {\bmh z} \times {\bm r}$ gives a counterclockwise (right-handed) rotation, and $r = |{\bm r}|$. From these relations we build the compatibility matrix row by row.
The $\alpha^{\mathrm{th}}$ row of $C_L$ is given by
\begin{equation}
(C_L)_\alpha
= \left(
        {\bm e}_{j_\alpha'} - 
        {\bm e}_{j_\alpha}
    \right)\otimes {\bm r}_{j_\alpha j_\alpha'}
    \label{eq:CL}
\end{equation}
where ${\bm e}_{j}$ is a row vector with a $1$ in position $j$, and zeros elsewhere [more formally ${\bm e}_j$ is a dual vector on the space of ball displacements]. 
Here, $\otimes$ is the Kronecker product between row-vectors, i.e.,~$(u_1,u_2,\dots)\otimes(v_1,v_2,\dots) = (u_1v_1, u_1v_2, \dots, u_2v_1, u_2v_2, 
\dots)$.
In a similar fashion, for a torsional bond $\beta$ we have
\begin{align}
\begin{split}
(C_\theta)_\beta
    =&\left( \bm e_{j_\beta''} - \bm e_{j_\beta} \right)\otimes
    \frac{\bm \hat {\bm r}^\perp_{j_\beta j_\beta''}}{r_{j_\beta j_\beta''}}
    \\
    &-\left( \bm e_{j_\beta'} - \bm e_{j_\beta}\right)\otimes
    \frac{\bm \hat {\bm r}^\perp_{j_\beta j_\beta' }}{r_{j_\beta j_\beta'}}.
    \label{eq:Ctheta}
\end{split}
\end{align}

\subsection{Periodic Lattices}
The expressions given above apply to any network of balls and springs, periodic or not. We now specialise to periodic lattices. In a two-dimensional periodic lattice with primitive vectors $\bm a_1$, $\bm a_2$ a single ball is no longer indexed by an integer $j$ but by the pair $(j, \bm \ell)$, where the lattice index $\bm \ell\in\mathbbm Z^2$ enumerates the unit cells, and the basis index $j=1 \hdots N$ enumerates the $N$ balls within the unit cell.
The equilibrium positions of balls are $\bm x_{j,\bm \ell} = \bm x_{j,0} + \bm R(\bm \ell)$, in terms of a lattice vector $\bm R(\bm \ell) = \ell_1\bm a_1+\ell_2\bm a_2$ and a position $\bm x_{j,\bm0}$ of the $j$\textsuperscript{th} ball in the central ($\bm \ell=\bm0$) unit cell. 
Similarly to ball indices, we now enumerate the longitudinal spring which connects balls $(j_\alpha,\bm \ell_\alpha+ \bm \ell)$ and $(j_\alpha',\bm \ell_\alpha'+\bm \ell)$ with the pair $(\alpha, \bm \ell)$. With the same notation, the torsional spring indexed by $(\beta,\bm \ell)$ subtends the counter-clockwise angle between balls $(j_\beta',\bm \ell_\beta'+\bm \ell)$ to $(j_\beta'',\bm \ell_\beta''+\bm \ell)$ about the ball $(j_\beta,\bm \ell_\beta+\bm \ell)$.

If the dynamics implemented by the connectivity matrix $\Bbbk$ satisfy the lattice periodicity, we can always assign the springs to the unit cell such that each active spring only communicates with active springs within the same unit cell, that is $\Bbbk_{(\alpha,\bm\ell),(\alpha',\bm\ell)} = \Bbbk_{\alpha\alpha'}\delta_{\bm\ell\bm\ell'}$.

With this preparation, the dynamical equation Eq.~(\ref{eq:EOM}) for the periodic lattice can be written as
\begin{equation}\label{eq:latticeEoM}
    {\bm F}_\ell
    = -\sum_{\bm\ell'}D_{\bm\ell'}\delta\bm X_{\bm\ell+\bm\ell'}
    ,\quad
    D_{\bm\ell'} 
    = \sum_{\bm\ell}C_{\bm\ell}^\intercal\mathbbm k C_{\bm\ell+\bm\ell'},
\end{equation}
where we have used periodicity to define $\bm\ell$-parametrized family of compatibility matrices
\begin{equation}
    C_{\bm\ell} 
    := \frac{\delta\bm E_{\bm0}}{\delta\bm X_{\bm\ell}}
    = \frac{\delta\bm E_{\bm\ell'}}{\delta\bm X_{\bm\ell+\bm\ell'}}
    \quad\forall\bm\ell'.
\end{equation}
We re-write these expressions in terms of Bloch coefficients $\delta\bm X(\bm q) = \sum_{\bm\ell} \delta\bm X_{\bm\ell}\,e^{-i\bm q\cdot\bm R(\bm\ell)}$ with wave-vector $\bm q$. Taking the Fourier transform of Eq.~(\ref{eq:latticeEoM}), we obtain 
\begin{equation}\label{eq:BlochEoM}
    \bm f(\bm q)
    = -D(\bm q)\delta\bm X(\bm q) \\
    = -C(\bm q)^\dagger\mathbbm k C(\bm q)\delta\bm X(\bm q),
\end{equation}
where the 
Bloch compatibility matrix is
\begin{equation}
    C(\bm q) 
    = \sum_{\bm\ell}\, \frac{\delta\bm E_{\bm0}}{\delta\bm X_{\bm\ell}}\, e^{i\bm q\cdot\bm R(\bm\ell)}
    = \frac{\delta}{\delta\bm X(\bm q)} \begin{pmatrix} \bm L_{\bm0} \\ \bm\theta_{\bm0} \end{pmatrix}.
\end{equation}
Explicit row-wise expressions for elements of the $\bm q$-space compatibility matrix $C(\bm q)$ mirror those given in Eqs. \eqref{eq:CL},\eqref{eq:Ctheta}, but contain additional Floquet factors when springs cross neighboring unit cells. For longitudinal springs,
\begin{align} \begin{split}
&(C_L)_\alpha= \frac{\delta L_{\alpha,\bm0}}{\delta\bm X(\bm q)}= \\ 
&\left(
        \bm e_{j_\alpha'}e^{i\bm q\cdot\bm R(\bm\ell_\alpha')} - 
        \bm e_{j_\alpha}e^{i\bm q\cdot\bm R(\bm\ell_\alpha)}
    \right)\otimes \bmh r_{j_\alpha\bm\ell_\alpha j_\alpha'\bm\ell_\alpha'},
    \label{eq:CLq}
\end{split} \end{align}
for the $\alpha$\textsuperscript{th} row of $C(\bm q)$.
Similarly, for torsion springs we find
\begin{align}\begin{split}
   &(C_\theta)_\beta= \frac{\delta\theta_{\beta,\bm0}}{\delta\bm X(\bm q)} = \\
    &\left( \bm e_{j_\beta''}\,e^{i\bm q\cdot\bm R(\bm\ell_\beta'')} - \bm e_{j_\beta}\,e^{i\bm q\cdot\bm R(\bm\ell_\beta)} \right)\otimes
   \frac{\bmh r^\perp_{j_\beta\bm\ell_\beta j_\beta''\bm\ell_\beta''}}{r_{j_\beta\bm\ell_\beta j_\beta''\bm\ell_\beta''}}
    \\
    &-\left( \bm e_{j_\beta'}\,e^{i\bm q\cdot\bm R(\bm\ell_\beta')} - \bm e_{j_\beta}\,e^{i\bm q\cdot\bm R(\bm\ell_\beta)} \right)\otimes
    \frac{\bmh r^\perp_{j_\beta\bm\ell_\beta j_\beta'\bm\ell_\beta'}}{r_{j_\beta\bm\ell_\beta j_\beta'\bm\ell_\beta'}}.
    \label{eq:Cthetaq}
\end{split}\end{align}

\subsection{Coarse graining procedure}
We now describe how to take the dynamical matrix $D(\bm q) = C(\bm q)^\dagger \Bbbk C(\bm q)$ from Eq.~\eqref{eq:BlochEoM}, and coarse grain to extract elastic moduli. The approach is to separate displacement fields into slow and fast variables. The slow variable is the centre of mass coordinate of each unit cell. The fast variables are then relative displacement vectors inside each unit cell. We will assume that these internal degrees of freedom relax on a timescale much shorter than the dynamics of the centre of mass coordinate. 

The key observation is that the kernel of $C(\bm 0)$ naturally defines the slow vibrational modes of the lattice~\cite{lubenskyPhononsElasticityCritically2015}. 
That is, for deformation-dependent lattices, $\mathrm{Ker}[C(\bm0)]$ always contains the uniform translations:
\begin{equation}
{C}(\bm0)\mathbbm 1 = 0
\quad\text{where}\quad
\mathbbm1 = (\bm b_1, \bm b_2) = \frac{1}{\sqrt N}\begin{pmatrix}
    1&0 \\ 0&1 \\ 1&0 \\ 0&1 \\ \vdots & \vdots
\end{pmatrix}.
\label{eq:Kernel}
\end{equation}
The vectors $\bm b_i$ span the space of rigid-body translations of masses along the $d=2$ spatial dimensions, which keep all lengths and angles invariant. In the following derivation we assume that there are no additional zero modes beyond these two translations.

We may use the right-singular eigenvectors to split the space of ball displacements:
\begin{equation}
C(\bm0)^\dagger C(\bm0) = W \Sigma^2 W^\dagger
,\quad
W=\begin{pmatrix} 
\mathbbm1 & V
\end{pmatrix},
\end{equation}
where $\Sigma$ is the diagonal matrix of right-singular values, and $W\in SO(2N)$.
The first two right-singular vectors in $W$ form the block $\mathbbm 1$ spanning the kernel of $C(\bm0)$. 
The remaining columns form a $2N\times(2N-2)$ matrix $V$ and span the orthogonal complement to the kernel.

The basis change $W$ decomposes the space of ball displacements $\delta {\bm X}(\bm q)$ and forces $\bm f(\bm q)$ as
\begin{align}
    \begin{pmatrix} \bm u(\bm q) \\ \bm v(\bm q) \end{pmatrix}
    =
    W^\dagger
    \delta{\bm X}(\bm q),
  \quad  
    m\begin{pmatrix}
    \ddot{\bm u}(\bm q) \\
    \ddot{\bm v}(\bm q)
    \end{pmatrix}
    =
    W^\dagger
    {\bm f}(\bm q),
\end{align}
into a centre-of-mass like coordinate
\begin{equation}
\bm u(\bm q) 
= \mathbbm1^\dagger\delta\bm X(\bm q)
= \frac{\delta {\bm x}_1(\bm q)+...+\delta {\bm x}_N(\bm q)}{\sqrt{N}}
\end{equation}
and a vector of $2(N-1)$ internal displacements $\bm v(\bm q)$, which we will proceed to eliminate.
The square root normalization ensures orthonormality of the basis defined by $W$.
Once we have a linear equation in $\bm u$ only, we can rescale it by a further $\sqrt N$ and treat $\bm u(\bm q)$ as the Fourier components of the average motion of the unit cells. 
In our new basis, the dynamical equation Eq.~(\ref{eq:BlochEoM}) reads
\begin{equation}
\label{eq:CoMForceDisplacement}
    m\begin{pmatrix}
    \ddot{\bm u} \\
    \ddot{\bm v}
    \end{pmatrix}
    = - \begin{pmatrix}
        C_u^\dagger \,\Bbbk\, C_u & C_u^\dagger \,\Bbbk\, C_v \\
        C_v^\dagger \,\Bbbk\, C_u & C_v^\dagger \,\Bbbk\, C_v
    \end{pmatrix}
    \begin{pmatrix} 
    \bm u \\ 
    \bm v
    \end{pmatrix},
\end{equation}
where we have denoted $C_u(\bm q) = C(\bm q)\mathbbm1$ and $C_v(\bm q) = C(\bm q)V$, and suppressed explicit $\bm q$ dependence for compactness.

To take the elastic limit, we consider the lowest order-expansion in the wavevector $\bm q$.
Note that, since the columns of $\mathbbm1$ span the kernel of $C(\bm 0)$, $C_u\sim O(q)$ and $C_v\sim O(1)$.
Hence the top-left block in Eq.~(\ref{eq:CoMForceDisplacement}) is $O(q^2)$, and tracks long-wavelength deformations of the centre-of-mass coordinate. 
By contrast, the bottom-right block is $O(1)$, and tracks internal modes within a unit cell.

We now eliminate the fast degree of freedom $\bm v( \bm q)$, by assuming the $O(1)$ lattice dynamics relax forces until internal forces balance, i.e.~$\ddot{\bm v}(\bm q)=0$. This assumption lets us solve for a quasi-static $\bm v( \bm q)$:
\begin{equation}
\bm v
= -\left[C_v^{\dagger} \,\Bbbk\, C_v\right]^{-1} 
C_v^{\dagger} \,\Bbbk\, C_u 
\bm u.
\label{eq:V(q)}
\end{equation}
Substituting $\bm v (\bm q)$ from Eq.~\eqref{eq:V(q)} back into the force-displacement relation Eq.~\eqref{eq:CoMForceDisplacement} we obtain the dynamical equation for the centre-of-mass coordinate
\begin{equation}\label{eq:fu}
    \rho\ddot{\bm u} = -\frac{N}{\mathcal A}C_u^\dagger\left(
    \Bbbk - \Bbbk\,C_v \left[C_v^{\dagger} \,\Bbbk\, C_v\right]^{-1} 
C_v^{\dagger} \,\Bbbk
    \right)C_u \bm u,
\end{equation}
where $\rho = Nm/\mathcal A$ is the mass density in terms of the unit cell area $\mathcal A=|\bm a_1\times\bm a_2|$. Note that to lowest order in $q$, the entire expression within the round brackets is $O(1)$, and in particular the expression in square brackets is invertible at $\bm q=\bm0$.

Equation~\eqref{eq:fu} represents a coarse-grained force-displacement relation that only tracks the slow mode $\bm u(\bm q)$: we have eliminated internal degrees-of-freedom and coarse-grained to an effectively affine lattice from which we may extract elastic moduli.
\section{Elastic moduli}
\label{sec:Moduli}
In this section we describe how to extract elastic moduli from the coarse-grained force-displacement relation, with particular attention to the linearly isotropic case.
We also show how to identify the even and odd shear moduli in the linearly anisotropic case, by extending the definition of principal axes into the active domain.

\subsection{Extracting the elastic moduli}
The coarse-grained dynamical equation Eq.~\eqref{eq:fu} is (to lowest order) quadratic in $\bm q$, thus its Fourier transform is a second-order PDE in $\bm u $. 
For linearly isotropic materials (in particular for random lattices or periodic lattices with three-fold rotational symmetry), Ref.~\cite{scheibnerOddElasticity2020} showed by symmetry arguments that this PDE must take the form of the active Navier-Cauchy equation
\begin{equation}
    \rho\ddot{\bm u} 
    = \begin{pmatrix} B & -A \\ A & B \end{pmatrix}
    \bm\nabla(\bm\nabla\cdot\bm u) 
    + \begin{pmatrix} \mu & -K^o \\ K^o & \mu \end{pmatrix}
    \nabla^2\bm u,
\end{equation}
in terms of two passive moduli, the bulk $B$ and shear $\mu$ moduli, alongside two active moduli, the compression-torque $A$ and odd $K^o$ moduli.

Accordingly, we obtain Hooke's law for active isotropic media, which relates the induced stresses $\rho\ddot{u}_i=\partial_j\sigma_{ij}$ to the imposed strain $\varepsilon_{ij} = \tfrac{1}{2}(\partial_iu_j + \partial_ju_i)$
\begin{equation}
    \underline\sigma = 2\begin{pmatrix} B & -A \\ A & B \end{pmatrix}
    \,\mathrm{vol}(\underline\varepsilon)
    + 2\begin{pmatrix} \mu & -K^o \\ K^o & \mu \end{pmatrix}
    \mathrm{dev}(\underline\varepsilon)
\end{equation}
in terms of the volumetric $\mathrm{vol}(\underline\varepsilon)=\tfrac{1}{2}tr(\underline\varepsilon) I_2$ and deviatoric $\mathrm{dev}(\underline\varepsilon) = \underline\varepsilon - \mathrm{vol}(\underline\varepsilon)$ strain tensors.

In more generality, the stress-strain relation can be written as $\sigma_{ij} = K_{ijkl}\varepsilon_{kl}$ in terms of a stiffness tensor $K$. Comparing with Eq.~\eqref{eq:fu}, we find that the stiffness tensor is
\begin{equation}
K_{i \alpha j \beta} = \frac{N}{\mathcal A}
    \left[
    \frac{\partial (C\bm b_i)^\dagger}{\partial q_\alpha}
    \,\widetilde{\Bbbk}\,
    \frac{\partial (C\bm b_j)}{\partial q_\beta}
    \right]_{\bm q=\bm0} + \epsilon_{\alpha\beta}\chi_{ij},
\label{eq:Stiffness}
\end{equation} 
where $\epsilon_{\alpha\beta}$ is the Levi-Civita symbol, the $\chi_{ij}$ are constants of integration set by imposing boundary conditions on the stress, and the effective microscopic stiffness matrix is
\begin{equation}
\widetilde{\Bbbk} 
= \left[
\Bbbk- \Bbbk\,CV \left[(CV)^\dagger \,\Bbbk\, CV\right]^{-1} (CV)^\dagger
\,\Bbbk 
\right]_{\bm q=\bm0}.
\label{eq:EffectiveStiffness}
\end{equation}
From Eqs.~(\ref{eq:Stiffness}--\ref{eq:EffectiveStiffness}) we can see the major symmetry of $K$ follows directly from the symmetry of $\Bbbk$.

Here, we use the symmetrized strain tensor $\varepsilon$ without loss of generality, since the microscopic dynamics are invariant under rigid rotations.
Further, since uniform scaling keeps angles invariant and induces no length-differences in pairs of springs that have the same rest-length, $B$ and $A$ are independent of activity for all lattices mentioned in the main text (in particular, this means that $A=0$).
Therefore, all activity is contained in the shear-response of the lattice.

\subsection{Anisotropy and principal axes}
\begin{figure}[t]
    \centering
    \begin{tikzpicture}[scale=0.7]
    \draw[white, fill=white] (-1.25,-1.25) rectangle (1.25,1.25);
    \draw[thick] (0,0) circle (1.25);
    \draw[gray, dashed] (0,0) circle (1);
    \draw[thick] (0,0) circle (1.25);
    \draw[line width=2, ->] (-1+.2,0) -- (-1-.2,0);
    \draw[line width=2, ->] (+1-.2,0) -- (+1+.2,0);
    \draw[line width=2, ->] (0,-1+.2) -- (0,-1-.2);
    \draw[line width=2, ->] (0,+1-.2) -- (0,+1+.2);
    \node[scale=1.5] at (0,0) {$\odot$};
    \end{tikzpicture}
    \quad
    \begin{tikzpicture}[scale=0.7]
    \draw[white, fill=white] (-1.25,-1.25) rectangle (1.25,1.25);
    \draw[gray, dashed] (0,0) circle (1);
    \draw[thick] (0,0) circle (1);
    \draw[line width=2, ->] (-1,+.2) -- (-1,-.2);
    \draw[line width=2, ->] (+1,-.2) -- (+1,+.2);
    \draw[line width=2, ->] (-.2,-1) -- (+.2,-1);
    \draw[line width=2, ->] (+.2,+1) -- (-.2,+1);
    \node[scale=1.5] at (0,0) {$\circlearrowleft$};
    \end{tikzpicture}
    \quad
    \begin{tikzpicture}[scale=0.7]
    \draw[white, fill=white] (-1.25,-1.25) rectangle (1.25,1.25);
    \draw[gray, dashed] (0,0) circle (1);
    \draw[rotate=0] (0,0) ellipse (1.25 and .8);
    \draw[line width=2, ->] (-1+.2,0) -- (-1-.2,0);
    \draw[line width=2, ->] (+1-.2,0) -- (+1+.2,0);
    \draw[line width=2, ->] (0,-1-.2) -- (0,-1+.2);
    \draw[line width=2, ->] (0,+1+.2) -- (0,+1-.2);
    \node[scale=1.5] at (0,0) {$+$};
    \end{tikzpicture}
    \quad
    \begin{tikzpicture}[scale=0.7]
    \draw[white, fill=white] (-1.25,-1.25) rectangle (1.25,1.25);
    \draw[gray, dashed] (0,0) circle (1);
    \draw[rotate=45] (0,0) ellipse (1.25 and .8);
    \draw[line width=2, ->] (-1,+.2) -- (-1,-.2);
    \draw[line width=2, ->] (+1,-.2) -- (+1,+.2);
    \draw[line width=2, ->] (+.2,-1) -- (-.2,-1);
    \draw[line width=2, ->] (-.2,+1) -- (+.2,+1);
    \node[scale=1.5] at (0,0) {$\times$};
    \end{tikzpicture}
    \caption{
    The basis of linear deformations corresponding to the matrices in Eq.~\eqref{eq:2x2decomposition}.
    The arrows represent the deformations $\bm u = \underline\varepsilon\bm x$ or traction forces $\bm f = \underline\sigma\bmh n$.
    }
    \label{fig6:infinitesimalstrains}
\end{figure}

By measuring a single arbitrary shear response, a shear-active isotropic material is indistinguishable from a passive anisotropic material.
To mark the difference, we briefly mention how to identify active shear in anisotropic materials.

The elasticity tensor is a linear operation on the vector space of $2\times2$ matrices.
We will parametrize this vector space by four matrices that represent dilation, rotation, and two shears whose axes differ by a $45^\circ$ angle:
\begin{equation}\label{eq:2x2decomposition}
    \underline\varepsilon 
    = \varepsilon_\odot \left(\begin{smallmatrix}1&0\\0&1\end{smallmatrix}\right)
    + \varepsilon_\circlearrowleft \left(\begin{smallmatrix}0&-1\\1&0\end{smallmatrix}\right)
    + \varepsilon_+ \left(\begin{smallmatrix}1&0\\0&-1\end{smallmatrix}\right)
    + \varepsilon_\times \left(\begin{smallmatrix}0&1\\1&0\end{smallmatrix}\right).
\end{equation}
Similarly decomposing the stress tensor, we can collect the coefficients as vectors
\begin{equation}
    \bm\varepsilon = (\varepsilon_\odot, \varepsilon_\circlearrowleft, \varepsilon_+, \varepsilon_\times)^T
    ,\quad
    \bm\sigma = (\sigma_\odot, \sigma_\circlearrowleft, \sigma_+, \sigma_\times)^T.
\end{equation}
In this basis, the stress-strain relation is $\bm\sigma = K\bm\varepsilon$ in terms of an elasticity matrix $K$. 
Since $\underline\varepsilon$ is symmetric by definition, $\varepsilon_\circlearrowleft$ is always zero.
This is reflected by fixing the constant of integration $\chi$ such that the second column of $K$ is zero.

For simplicity, let us assume shear and bulk modes are decoupled, so we can focus on the shear sector in isolation.
We may rotate space until the symmetric part of $K$ diagonalizes, and the shear stress-strain relation takes the canonical form
\begin{equation}\label{eq:anisotropic-shear-stress-strain}
    \begin{pmatrix} \sigma_+ \\ \sigma_\times \end{pmatrix}
    = \begin{pmatrix} \mu_1 & -K^o \\ K^o & \mu_2 \end{pmatrix}
    \begin{pmatrix} \varepsilon_+ \\ \varepsilon_\times \end{pmatrix},
\end{equation}
with two even (shear) moduli $\mu_{1,2}$ and one odd (shear) modulus $K^o$.

We will consider uniaxial compression along a varying angle $\varphi$, i.e.
\begin{equation}
    \underline\varepsilon =  \varepsilon(\bmh n\bmh n^\intercal - \bmh n_\perp\bmh n_\perp^\intercal)
\end{equation}
where $\bmh n = (\cos\varphi,\sin\varphi)^\intercal$ and $\bmh n_\perp = (-\sin\varphi, \cos\varphi)^\intercal$.
As we change the angle $\varphi$ of the shear axis, we will keep track of the normal $f_\parallel$ and tangential $f_\perp$ components of the induced force at the contact plane
\begin{equation}
    \bm f = \underline\sigma\cdot\bmh n 
    = \epsilon \left(f_\parallel \bmh n + f_\perp \bmh n_\perp\right).
\end{equation}
Using Eq.~\eqref{eq:anisotropic-shear-stress-strain}, we find
\begin{align}
    f_\parallel &= \tfrac{1}{2}(\mu_1+\mu_2) +\tfrac{1}{2}(\mu_1-\mu_2)\cos(4\varphi),
    \\
    f_\perp &= \hspace{23pt}K^o\hspace{12pt} -\tfrac{1}{2}(\mu_1-\mu_2)\sin(4\varphi).
\end{align}
We note that $\mu_{1,2}$ and $K^o$ are respectively the extrema of $f_\parallel$ and the average of $f_\perp$. 

We thus find that one can measure the active shear modulus statically by measuring the tangential force resulting from compression along a principal axis, which for active elastic solids we may define as an axis under which the normal force is extremal.
For passive materials, we recover that principal axes are characterized by a vanishing tangential force, as $f_\perp$ averages to zero if $K^o=0$.
Finally, in linearly isotropic (active) materials $\mu_1=\mu_2$, and every axis is principal.
\section{Broken and preserved monotonicity}
In this section, we provide a short proof for the monotonic relationship between microscopic stiffness and macroscopic rigidity in passive materials, and give mathematical intuition behind the non-monotonicity that can arise in active materials.

\label{sec:Monotonic}

\subsection{Passive monotonicity}
For passive solids, the stiffness matrix $\Bbbk$ is symmetric, allowing the force-displacement relation
\begin{equation}
    \bm F = -C^\intercal\,\Bbbk\,\delta\bm L= - C^\intercal\,\Bbbk\, C\delta\bm X + O^2(\delta\bm X),
\end{equation}
to be derived as the gradient of a potential energy
 \begin{equation}
 \label{eq:MicroPotentialEnergy}
     \mathcal E(\delta\bm X) 
     = \tfrac{1}{2}(C\delta\bm X)^\intercal\Bbbk(C\delta\bm X) + O^3(\delta\bm X).
 \end{equation}

Because we have a potential energy, we can follow a more direct coarse-graining procedure.
In equilibrium, the coarse-grained free energy of an affine deformation $\bm\varepsilon$
 \begin{equation}
     \mathcal F(\bm\varepsilon) 
     = \min\nolimits_{\Omega(\bm\varepsilon)} \mathcal E(\delta\bm X)
     = \tfrac{1}{2}\bm\varepsilon^\intercal K\bm\varepsilon + O^3(\bm\varepsilon),
\end{equation}
is the minimum of the energy over the affine space of ball displacements that can express this deformation.
For periodic lattices, we can characterize this space as 
\begin{equation}
    \Omega(\bm\varepsilon) 
    = \{ \delta\bm X: \bm u_{\bm\ell} = \underline\varepsilon\,\bm R(\bm\ell) \},
\end{equation}
possibly augmented by other restrictions (such as the bond incompressibility in our hexagonal lattice).
Here, $\bm u_{\bm\ell} = \frac{1}{N}\sum_j\delta\bm X_{j,\bm\ell}$ is the average motion of the $\bm\ell$-th unit cell, and $\bm R(\bm \ell) = \ell_1\bm a_1+\ell_2\bm a_2$ the $\bm\ell$-th lattice vector.
Note that $\Omega(\bm\varepsilon)$ is an affine rather than linear space (in particular $\bm0\not\in\Omega(\bm\varepsilon)$ for non-zero $\varepsilon$).

Now let us consider two scenarios (distinguished by subscripts 1 and 2) between which we increase the local stiffnesses.
We use notation $K_1\leq K_2$ to mean that $K_2-K_2$ is positive non-semidefinite, or equivalently that the eigenvalues of $K_2-K_1$ are all non-negative.
\newtheorem{theorem}{Theorem}[section]
\begin{theorem}
    $\ \Bbbk_1\leq \Bbbk_2 \Rightarrow K_1\leq K_2$
\end{theorem}
\emph{Proof.}
Assume that $\Bbbk_2-\Bbbk_1$ is positive semi-definite. This is equivalent to the statement that $\mathcal E_1(\delta\bm X)\leq \mathcal E_2(\delta\bm X)$ for all $\delta\bm X$.
It suffices to show that $\mathcal F_1(\bm\varepsilon)\leq\mathcal F_2(\bm\varepsilon)$ for all $\bm\varepsilon$.
Let $\delta\bm X_i$ be the minimizer of $\mathcal E_i$ over $\Omega(\bm\varepsilon)$, then indeed
\begin{equation}
    \mathcal E_1(\delta\bm X_1)\leq \mathcal E_1(\delta\bm X_2)\leq \mathcal E_2(\delta\bm X_2),
\end{equation}
by minimizer and microscopic monotonicity respectively.
Thus $K_2-K_1$ is positive-semi definite. $\square$

In particular for diagonal $\Bbbk$, increasing any of its diagonal elements (microscopic stiffnesses) cannot decrease the eigenvalues of $K$ (the elastic moduli).

\subsection{High-activity non-monoticity}
At non-zero activity, we assume our lattice still finds an equilibrium state, given by the force-balance condition
\begin{equation}\label{eq:ForceBalance}
    \bm 0 
    = \bm F_p + \bm F_a
    = C^\intercal(\Bbbk_p+\Bbbk_a)C\delta\bm X,
\end{equation}
where $\Bbbk_p$ and $\Bbbk_a$ are the passive (symmetric) and active (antisymmetric) components of the microscopic stiffness matrix $\Bbbk$.

To explain the vanishing of $K^o$ and stiffening of $\mu$ in under-percolated lattices, we focus at force-balance in the high-activity regime.
For simplicity, we restrict our attention to systems without stiff scaffolding (e.g.~$\kappa\ll\kappa_a\ll k$) where the analysis would be a bit more involved, but the conclusions similar.
The force-balance Eq.~(\ref{eq:ForceBalance}) is then dominated by the active force, and is to first approximation satisfied by affine displacements belonging to the set
\begin{equation}
    \Omega_a
    = \{\delta\bm X\in\Omega: \bm F_a=\bm0\}
    = \Omega \cap \mathrm{ker}(C^\intercal\Bbbk_a C),
\end{equation}
where we have suppressed $\bm\varepsilon$-dependence for compactness.
The equilibrium displacement is given by the minimizer of the passive potential (Eq.~(\ref{eq:MicroPotentialEnergy}) with $\Bbbk=\Bbbk_p$) over this reduced space of active-force-balancing affine displacements $\Omega_a$.
An increase in $\mu$ indicates that the energy minimizer over $\Omega(\bm\varepsilon)$, where $\bm\varepsilon$ is a shear, is no longer contained in the restricted space $\Omega_a(\bm\varepsilon)$.

Net active forces $\bm F_a$ can vanish either because generated non-zero active forces balance out, or because no active forces are generated at all.
In the latter case, any stress must be generated by passive elements, and therefore odd elastic moduli such as $K^o$ must vanish.
We call displacements that do not generate active forces \emph{active floppy modes}, and analyze them as members of
\begin{equation}
    \Omega_a^0
    = \{\delta\bm X\in\Omega: \bm T_a=\bm0\}
    = \Omega \cap \mathrm{ker}(\Bbbk_a C),
\end{equation}
since absence of active tensions implies absence of active forces.
Thus, we find that $K^o$ must vanish if the minimum over $\Omega_a$ is contained in the subset $\Omega_a^0$.

As we dilute the active elements in our lattice (that is, decrease the activation fraction $p$), the rank of $\Bbbk_a$ decreases.
As a result, the space of active floppy modes (the kernel of $\Bbbk_a C$) broadens, and commensurately the space of generatable active force configurations (the image of $\Bbbk_a C$) shrinks.
When, for a given $\varepsilon$, the image $\Bbbk_a C(\Omega(\bm\varepsilon))$ has shrunk to the point that it no longer intersects the kernel of $C^\intercal$, all active-force-balancing microscopic displacements that realize the macroscopic affine deformation $\bm\varepsilon$ must be floppy.
Thus the minimum over $\Omega_a$ must be contained in $\Omega_a^0$ (as $\Omega_a=\Omega_a^0$), and the odd moduli associated with $\varepsilon$ must vanish.

\section{Active Force Balance in Honeycomb Metamaterials}
\label{sec:Honeycomb}
In the main text, we rationalize the emergence of plaquette locking and a non-monotonic $K^o(\kappa^a)$ in our honeycomb metamaterials by first simplifying to a triangular lattice (Fig.~\ref{fig:ActiveForceBalance}). This minimal model captures the essence of the phenomena, but here we describe the equivalent breakdown of active tension chains in our honeycomb metamaterials.

\begin{figure}[t]
\centering
\includegraphics[width=\linewidth]{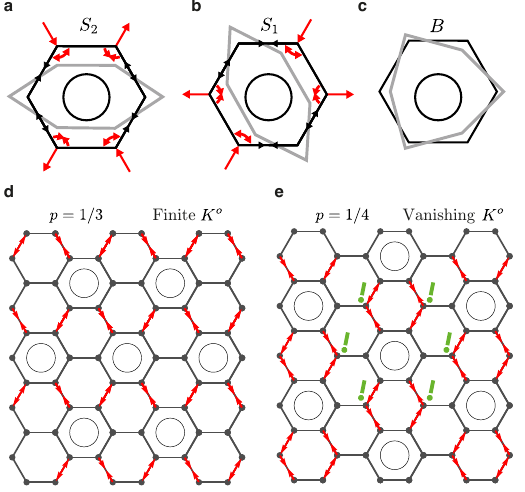}
  \caption{
  {{\bf Vanishing odd respose in honeycomb metamaterials}  
  (a--c) The active forces (external thick red arrows) generated by a basis for the space of infinitesimal motions of the equilateral hexagon (shown exaggerated in gray outline), up to rotation and translation. The change in angle generates active torques (red internal arrows), which lead to forces on the vertices (thin light-red arrows), whose components along the bars (black arrows) are resisted by the incompressibility of the rigid linkages. (d) A macroscopic force chain built from these antisymmetric plaquette-level nodal forces in a $p=1/3$ lattice. This lattice supports finite odd response at large activity. (e) In a $p=1/4$ tiled lattice we cannot succesfully construct active force chains without unbalanced nodal forces (exclamation points). This lattice cannot support $K^o$ at high activity.}
  }
\label{fig:HoneycombActiveForceBalance}
\end{figure}

Figure~\ref{fig:HoneycombActiveForceBalance}(a-c) shows a basis for the deformations of an incompressible hexagon, composed of two shear modes and one breathing mode. In a non-reciprocal plaquette, these shear modes generate angular tensions [Figure~\ref{fig:HoneycombActiveForceBalance}(a,b), red internal arrows] and hence nodal forces [Figure~\ref{fig:HoneycombActiveForceBalance}(a,b), red external arrows] which are skewed with respect to the applied shear. The remaining breathing mode does not couple to non-reciprocity. In Figs.~\ref{fig:HoneycombActiveForceBalance}(c,d) we attempt to tile these plaquette level active forces across the two lattices shown in Fig.~\ref{fig:DesignerDilution}(b,c), which exhibit the transition into non-monotonic response and plaquette locking. In Fig.~\ref{fig:HoneycombActiveForceBalance}(d) every third plaquette is active ($p=1/3$) and our theory predicts that odd response is monotonic, $K^o =2\kappa^a/a^2$. Note that $p=1/3$ is actually below the $p=1/2$ transition shown in Fig.~\ref{fig:DesignerDilution}(k). In this case, the specific periodic structure of the tiling nevertheless allows  a finite odd response not seen in a typical random tiling.
By contrast, in the more dilute lattice shown in Fig.~\ref{fig:HoneycombActiveForceBalance}(e), which is experimentally realized in Fig.~\ref{fig:Phenomenology}, we have $p=1/4$ and we predict that odd response vanishes at large activity, Eq.~\eqref{eq:TheoryKek20}. Analogously to the triangular lattices shown in Fig.~\ref{fig:ActiveForceBalance}, we discover that tension chains can be successfully constructed in Fig.~\ref{fig:HoneycombActiveForceBalance}(d), but they cannot in  Fig.~\ref{fig:HoneycombActiveForceBalance}(e): we necessarily encounter unbalanced nodal forces [Fig.~\ref{fig:HoneycombActiveForceBalance}(e), exclamation points]. The actual force-balanced displacement fields adopted by the lattice are shown in Fig.~\ref{fig:DesignerDilution}(b,c).

\section{Data Analysis and Computational Methodology}
\label{sec:Computation}
In this section we describe our computational methodology for building the dynamical matrices of lattices and extracting their elastic moduli which is used throughout the main text. We also provide details of the data analysis used to generate the visualisations shown in Fig.~\ref{fig:Phenomenology}, alongside the theoretical results shown in Fig.~\ref{fig:DesignerDilution}--Fig.\ref{fig:DynamicalDesignerDilution}.

 \subsection{Analysis of Honeycomb Lattice Data}
 \subsubsection{Theoretical Honeycomb Lattice Data}
Throughout Fig.~\ref{fig:Phenomenology} we compare experimental data to theoretical predictions for honeycomb lattices. These theoretical data [curves in Fig.~\ref{fig:Phenomenology}(c,f), image in Fig.~\ref{fig:Phenomenology}(e)] are generated using the coarse-graining procedure leading to Eq.~\eqref{eq:TheoryKek20}, described in \S\ref{sec:CoarseGraining}. We implement the symbolic expressions for $\bf q$-space compatibility matrices Eqs.~\eqref{eq:CLq},~\eqref{eq:Cthetaq} to generate dynamical matrices as in Eq.~\eqref{eq:BlochEoM} using Mathematica 12. Elastic moduli are computed through the symbolic expression Eq.~\eqref{eq:Stiffness}. In practice the symbolic inversion of $C^\dagger \Bbbk C$ in Eq.~\eqref{eq:Stiffness} limits a naive analytic calculation to around $8$ balls per unit cell. For larger lattices we replace analytic inversions with numerical matrix solves. 

For the lattice shown in Fig.~\ref{fig:Phenomenology} we find the odd and shear moduli in Eq.~\eqref{eq:TheoryKek20} of the main text, with $c_1=27/64$, $c_2=45/64$.
 In Fig.~\ref{fig:DesignerDilution} we show several progressively diluted honeycomb lattices, of which panel (c) corresponds to the lattice shown in Fig.~\ref{fig:Phenomenology} (a). Analytical results for the other panels are:
 \begin{align}
  K^o=  6 \kappa^a, \ \mu = 2\sqrt{3} \kappa \ \mathrm{panel\ a } \\
  K^o= 2 \kappa^a, \  \mu = 2\sqrt{3} \kappa \ \mathrm{panel\ b }, 
 \end{align}
 in units of the honeycomb side spacing $a=1$. The unit cell of Fig.~\ref{fig:DesignerDilution}(d) is too large for an analytical inversion, and we proceed numerically as for random lattices.

To generate the strain fields shown in Fig.~\ref{fig:Phenomenology}(e,f), we apply the simple shear $\epsilon$ to a unit cell, let the internal degrees of freedom relax [i.e. use $\bm v$ in Eq.~\eqref{eq:V(q)} for the internal ball displacements], and then compute hexagonal shear strains using Eq.~\eqref{eq:Sprojection}. In Fig.~\ref{fig:DesignerDilution}(a-h), we instead apply a vertical pure shear for visual clarity, otherwise the procedure is identical.

 \subsection{Analysis of Triangular Lattice Data}
\begin{figure*}[t!]
\centering
\includegraphics[width=\linewidth]{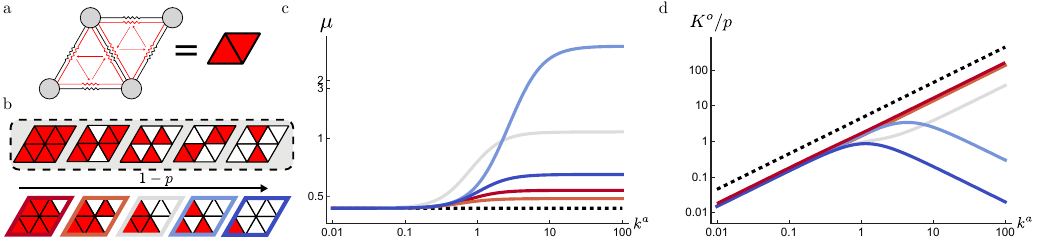}
    \caption{ {\bf \rev{ Odd and shear moduli of periodic triangular active lattices}}
    (a) Sketch of periodic tile of extension-active triangles, with black passive springs with tension laws $T_{p,i} = -k\delta L_i$, and red active springs with tension laws $T_{a,i} = -k^a(\delta L_{i-1}-\delta L_{i+1})$.
    (b) The ten isotropic dilutions of a $4\times4$ triangular unit cell (table ~\ref{tab:TriangularLattice}). Colors map to the curves shown in panels (c,d). (c,d) Shear $\mu$ and odd $K^o$ moduli of the ten dilutions shown in (b), as a function of $k^a$. We rescale $K^o$ by $p$ in panel (b) to collapse the monotonic lattice responses onto one another. Just as in our honeycomb lattices, the overall trend is for $K^o$ to vanish for small $p$, and $\mu$ to exhibit a peak at intermediate values of $p$. However certain lattices defy this trend. For example, our $p=2/8$ lattice remains monotonic despite being unpercolated, because active bonds form system-spanning chains.}
\label{fig:TriangularLatticeElasticModuli}
\end{figure*}

\begin{table*}
    \centering
      \begin{tabular}{l|l|l|l|l}
        Lattice & $p$ & {Odd Modulus $K^o$}  &  {Shear Modulus $\mu$} & {Classification}  \\\hline
        
        \includegraphics[valign=c]{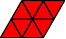}
        & $8/8$ & 
        $\frac{3k^a}{2}$
        & $\frac{\sqrt3}{4}$
        & Monotonic 
        \\[10pt]
        
        \includegraphics[valign=c]{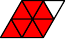}
        & $7/8$ 
        & $\frac{3 \left(14 k^4 {k^a}+79 k^2 {k^a}^3+80 {k^a}^5\right)}{8 \left(4 k^4+23 k^2
   {k^a}^2+25 {k^a}^4\right)} \rightarrow k^a$
        &$ \frac{\sqrt{3} \left(8 k^5+49 k^3 {k^a}^2+62 k {k^a}^4\right)}{8 \left(4 k^4+23 k^2 {k^a}^2+25 {k^a}^4\right)}$ 
        & Monotonic 
         \\[10pt]
            
        \includegraphics[valign=c]{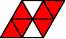}
        & $6/8$
        & $\frac{9k^a}{8}$ 
        & $\frac{\sqrt3}{4}$
        & Monotonic 
        \\[10pt]
        
        \includegraphics[valign=c]{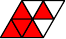}
        & $5/8$
        &$ \frac{3 \left(20 k^4 {k^a}+79 k^2 {k^a}^3+56 {k^a}^5\right)}{4 \left(16 k^4+68 k^2 {k^a}^2+49 {k^a}^4\right)} \rightarrow k^a $ 
        &$ \frac{\sqrt{3} \left(16 k^5+74 k^3 {k^a}^2+55 k {k^a}^4\right)}{4 \left(16 k^4+68 k^2
   {k^a}^2+49 {k^a}^4\right)} $
        & Monotonic 
   \\[10pt]
            
        \includegraphics[valign=c]{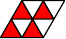}
        & $4/8$ 
        & $\frac{3k^a}{4}$ 
        & $\frac{\sqrt3}{4}$
        & Monotonic 
        \\[10pt]
        
        \includegraphics[valign=c]{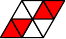}
        & $4/8$
        & $\frac{3k^a}{4}$ 
        & $\frac{\sqrt3}{4}$ 
        & Monotonic 
        \\[10pt]
        
        \includegraphics[valign=c]{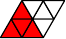}
        & $4/8$ 
        & $ \frac{3 \left(4 k^2 {k^a}+{k^a}^3\right)}{16 \left(k^2+{k^a}^2\right)}  \rightarrow k^a$
        & $\frac{\sqrt{3} \left(2 k^3+5 k {k^a}^2\right)}{8 \left(k^2+{k^a}^2\right)}$
        & Monotonic 
        \\[10pt]
            
        \includegraphics[valign=c]{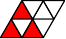}
        & $3/8$ 
        & $   \frac{9 \left(4 k^4 {k^a}+5 k^2 {k^a}^3\right)}{4 \left(16 k^4+20 k^2 {k^a}^2+{k^a}^4\right)}  \rightarrow \frac{1}{k^a}$ 
        & $ \frac{\sqrt{3} \left(16 k^5+26 k^3 {k^a}^2+7 k {k^a}^4\right)}{4 \left(16 k^4+20 k^2
   {k^a}^2+{k^a}^4\right)}$
        & Non-Monotonic 
   \\[10pt]
            
        \includegraphics[valign=c]{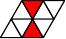}
        & $2/8$  
        & $\frac{3k^a}{8}$ 
        & $\frac{\sqrt3}{4}$ 
        & Monotonic 
        \\[10pt]
        
        \includegraphics[valign=c]{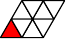}
        & $1/8$ 
        & $\frac{3 k^2 {k^a}}{16 k^2+12 {k^a}^2} \rightarrow \frac{1}{k^a}$ 
        & $\frac{\sqrt{3} k \left(8 k^2+9 {k^a}^2\right)}{8 \left(4 k^2+3 {k^a}^2\right)}$  
        & Non-Monotonic 
    \end{tabular}
    \caption{Odd $K^o$ and shear $\mu$ moduli for the ten linearly-isotropic dilutions of the tile consisting of eight extension-active triangles, classified according to the asymptotic behaviour of $K^o$ at large $k^a$ which is also shown.}
    \label{tab:TriangularLattice}
\end{table*}
In Figures~\ref{fig:ActiveForceBalance} and~\ref{fig:DynamicalDesignerDilution} we consider a minimal model of a non-reciprocal lattice, composed of a triangular lattice of longitudinal springs [$T_i= -k \delta L_i$] and non-reciprocal longitudinal springs [$T_i=k^a(\delta L_{i+1} - \delta L_{i-1})$]. Our general notation for such springs is shown in Fig.~\ref{fig:Notation}(c), with a specific triangular building block shown in Fig.~\ref{fig:ActiveForceBalance}. 
 
\rev{\subsubsection{Eigenspectra and Elastic Moduli of Periodic Triangular Lattices} }
 Figure~\ref{fig:DynamicalDesignerDilution}(a-d) analyzes the eigenspectra for the ten isotropic dilutions of a $2\times 4$ supercell of such triangular plaquettes. The unit cells of all such dilutions, alongside their dilution factor $p$ and elastic moduli, are given in Table~\ref{tab:TriangularLattice} and shown in Fig.~\ref{fig:TriangularLatticeElasticModuli}. The mapping between Figure~\ref{fig:DynamicalDesignerDilution} and Table~\ref{tab:TriangularLattice} is: Panel (a): $p=8/8$ (Table~\ref{tab:TriangularLattice} , first row). Panel (b): $p=5/8$ (fourth row). Panel (c): $p=3/8$ (eighth row). For all eigenspectra shown in Fig.~\ref{fig:DynamicalDesignerDilution}, $k^a/k=5$, and the triangular lattice side length is set to $a=1$. For the slope and gap metrics we set $k^a/k=100$.
 
 Note that there are several periodic dilutions all with $p=4/8=1/2$, shown as gray dots in Fig.~\ref{fig:DynamicalDesignerDilution}(d), which all have different eigenspectra and elastic moduli. We take their average to draw the line in Fig.~\ref{fig:DynamicalDesignerDilution}(d). This diversity is a reflection of the specific spatial structure of each lattice: periodic lattices sitting on the edge of percolation thresholds (isostatic lattices) often show distinct elastic properties depending on the details of their geometry~\cite{lubenskyPhononsElasticityCritically2015}. We note also that the $p=2/8$ lattice of Table~\ref{tab:TriangularLattice} presents an interesting edge case in which the eigenspectrum near the origin appears to scale as a power law; such edge cases are washed out in the averaging of random lattices. 

 \rev{In Figure~\ref{fig:TriangularLatticeElasticModuli}, we show the elastic moduli of these triangular lattices as a function of $k^a$---analogous curves for honeycomb lattices are shown in Fig.~\ref{fig:DesignerDilution}(i,j). Just as in our honeycomb lattices, the trend is for $K^o$ to vanish as $p$ decreases, with a spike in $\mu$ at intermediate values of $p$. Note however that certain lattices defy this trend, for example the $p=2/8$ lattice discussed above is underpercolated yet shows a diverging odd modulus, because plaquettes are still able to form system-spanning clusters.}

\subsubsection{Curve Fits} 
 To fit the slope and gap data shown in Figs.~\ref{fig:DynamicalDesignerDilution} (d,h), we first extract a one-dimensional representation of the eigenspectrum in the complex plane by taking marginals over the $\mathrm{Re}(\omega)$ axis. For the random data, we apply a bilinear fit to the eigenspectrum lying in the upper half plane, with the $\mathrm{Re}(\omega)$ intersection of the second line a free parameter. This intersection defines the gap. For the structured data, it suffices to fit two lines separately: once to data about the origin, defining the slope, and once to the entire dataset, whose intersection with the $\mathrm{Re}(\omega)$ axis defines the gap. \rev{The error bars shown in Figs.~\ref{fig:DynamicalDesignerDilution}(h) are obtain by fitting the gap and slope for the eigenspectrum of each sample, and then taking the mean and standard deviation of these fluctuating quantities. }
 
\subsubsection{Eigenspectra and Eigenvectors of Random Lattices} 
We generate random lattices as described below. The eigenspectra plots of Fig.~\ref{fig:DynamicalDesignerDilution} are then density plots of all the eigenvalues $\omega$ taken from $50$ realisations. The realizations of displacement fields shown in Figs.~\ref{fig:DynamicalDesignerDilution}(i-k) are a random subset of this data taken from each lobe of the spectrum for an under-percolated lattice. The shear strain in a triangular plaquette is defined as 
\begin{align}
    S_1 =& \frac1{\sqrt 2} ( \delta \theta_1 - \delta \theta_2 ) \quad \mathrm{Triangular}, \\ 
    S_2=&  \frac1{\sqrt 6} (\delta \theta_1 + \delta \theta_2 - 2\delta \theta_3) \quad \mathrm{Triangular} 
    \label{eq:TriangularSprojection},
\end{align}
and is visualized identically to the methodology used for honeycomb lattices.

 \subsection{Generation of Random Lattices}
 To generate the random lattices shown in Figs.~\ref{fig:DesignerDilution}, ~\ref{fig:DynamicalDesignerDilution} we generate the $\bf q$-space dynamical matrix of a periodic lattice which has the (large) unit cell that coincides with the random lattice we wish to study. We then extract elastic moduli as above. We generate these dynamical matrices by first generating the dynamical matrix of a fully tiled lattice, and then removing active plaquettes from $D(\bf q)$ with a probability $p$ to construct our randomized $D(\bf q)$. In Fig.~\ref{fig:DesignerDilution} we generate lattices of $25\times25$ honeycomb plaquettes, and in Fig.~\ref{fig:DynamicalDesignerDilution} we generate lattices of $40\times40$ triangular plaquettes. As above, the analytical inversions in Eq.~\ref{eq:Stiffness} are infeasible here and we use numerical matrix solves instead. In each case we generate $50$ random realisations to compute statistics on---we check that this is a sufficient number of realizations for our statistics to converge. The data shown in  Fig.~\ref{fig:DesignerDilution} has the complication that we wish to work in the limit $k\rightarrow \infty$, however we generate $D(\bf q)$ for finite $k$. In practice, we generate several datasets for $k = 10^3, 10^4, 10^5$. We check that the data converges, that the condition number of $D(\bm q)$ is acceptable for the number of digits of accuracy we require on $\mu$ and $K^o$, and that Mathematica issues no warnings on the condition number.

\rev{\section{Odd percolation occurs in positionally disordered lattices} \label{sec:PositionalDisorder}}
\begin{figure}[t!]
\centering
\includegraphics{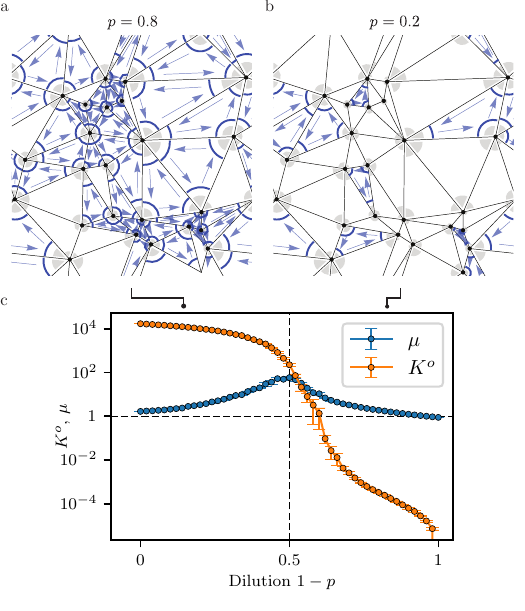}
      \caption{{\bf \rev{Odd Percolation in a disordered Delaunay triangulations.}} Just as in dilutions of honeycomb lattices, random triangulations of non-reciprocal torsional bonds also show a jump in the odd modulus $K^o$, and a spike in shear modulus $\mu$, in the vicinity of the site percolation threshold $p=1/2$. (a,b) Two example Delaunay triangulations of randomly generated point clouds, with each triangle randomly assigned non-reciprocal torsional actuation of strength $\kappa^a$ with probability $p$. Each bond has a longitudinal strength $k~1/a$. (c) As the dilution crosses the site percolation threshold, the shear modulus $\mu$ spikes and the odd modulus $K^o$ jumps, in analogy to Fig. 2 of the main text. The moduli are measured in units of $k \langle a\rangle^2$, where $\langle a \rangle $ is the average bond length. We fix $\kappa^a/k \langle a \rangle ^2 = 100$ for all $p$.}
      \label{fig:Delaunay}
\end{figure}
In the main text we have focused on the effects of on-lattice disorder, but the percolation scenario we describe also occurs in positionally disordered systems in which there is no lattice. To illustrate this point, we examine the odd moduli of random materials built from point clouds. We take $N=100$ Poisson distributed points in 2D and construct their Delaunay triangulation [Fig.~\ref{fig:Delaunay}(a,b)]. We assign to each edge a longitudinal bond stiffness $k\sim 1/a$, where $a$ is the bond length---in contrast to our honeycomb and triangular lattices $a$ now varies, and this protocol mimics a stiffness that arises from a material modulus and hence scales with overall spring length. Each plaquette is now activated with non-reciprocal torsion bonds of strength $\kappa^a$ with a probability $p$. At each $p$ we build $100$ samples, generating a fresh triangulation and assignment of active plaquettes each time.

Examining the moduli of these random triangulations as a function of $p$ [Fig.~\ref{fig:Delaunay}(c)] we again find a jump in $K^o$ and a spike in $\mu$ as $p$ crosses the site percolation threshold, in analogy to our results on periodic and random lattices (Fig.~\ref{fig:DesignerDilution}). We conclude that the basic phenomenology of the transition described in the main text is robust to the class of disorder considered---the key ingredients are a characteristic activity scale, and a sufficiently dilute material. 

\rev{\section{ Quantifying Finite Size Effects in our Metamaterials } \label{sec:Deviation}}

\begin{figure*}[t!]
\centering
\includegraphics{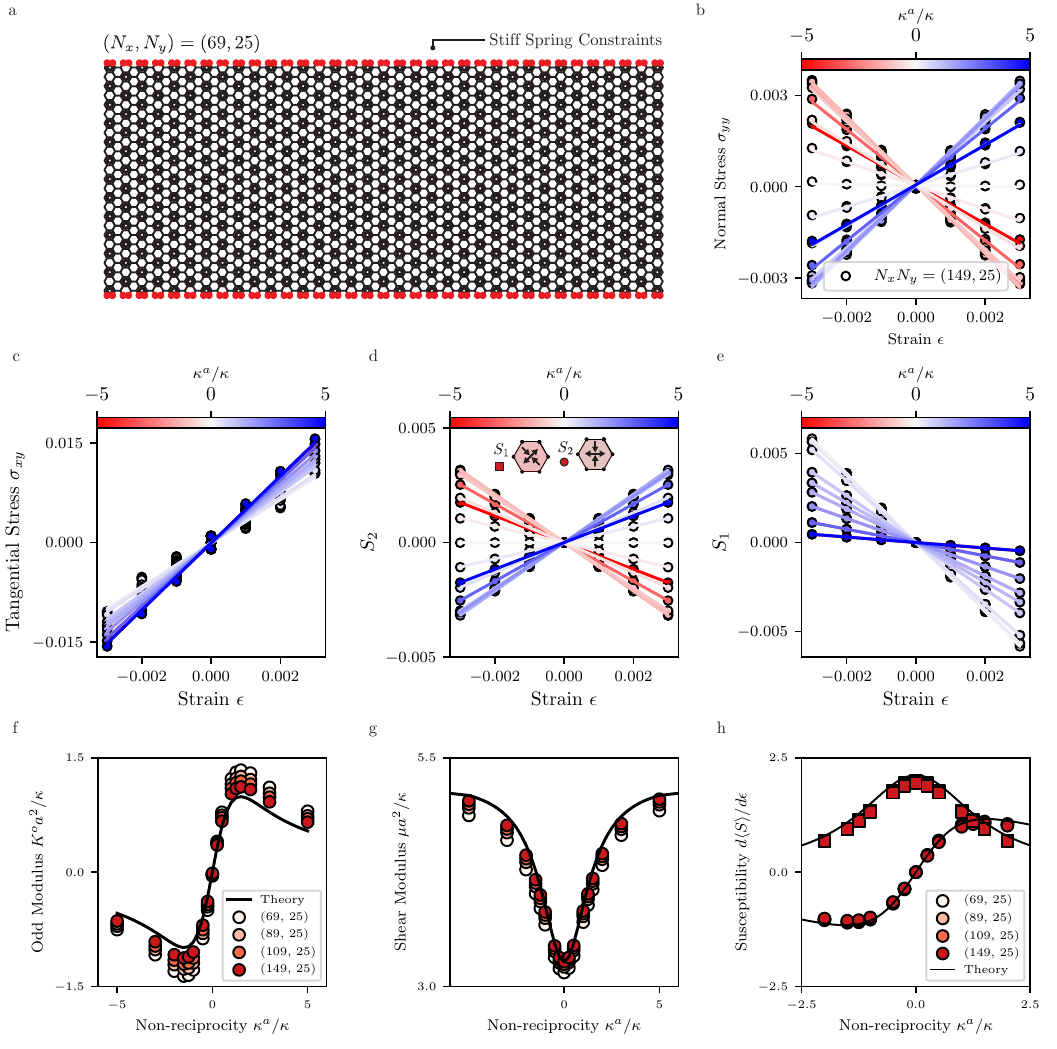}
      \caption{{\bf \rev{Recovering our periodic theory in a large but finite lattice.}} (a) An example ball-spring lattice, composed of an $(N_x, N_y)=(69,25)$ grid of honeycombs, with active plaquettes shown as hollow circles, and constraint points indicated in red. (b-e) We mimic our experimental protocol of horizontally shearing the lattice and measuring normal and tangential forces on constraint nodes, alongside averaged internal shear modes. For small applied strain $\epsilon$, these relations are all linear. (f-h) We recover our theoretical predictions for $K^o$, $\mu$ and $d\langle S\rangle/d\epsilon$ as a function of activity $\kappa^a$ for large lattices.}
      \label{fig:FiniteSizeScaling1}
\end{figure*}
\begin{figure*}[t!]
\centering
\includegraphics{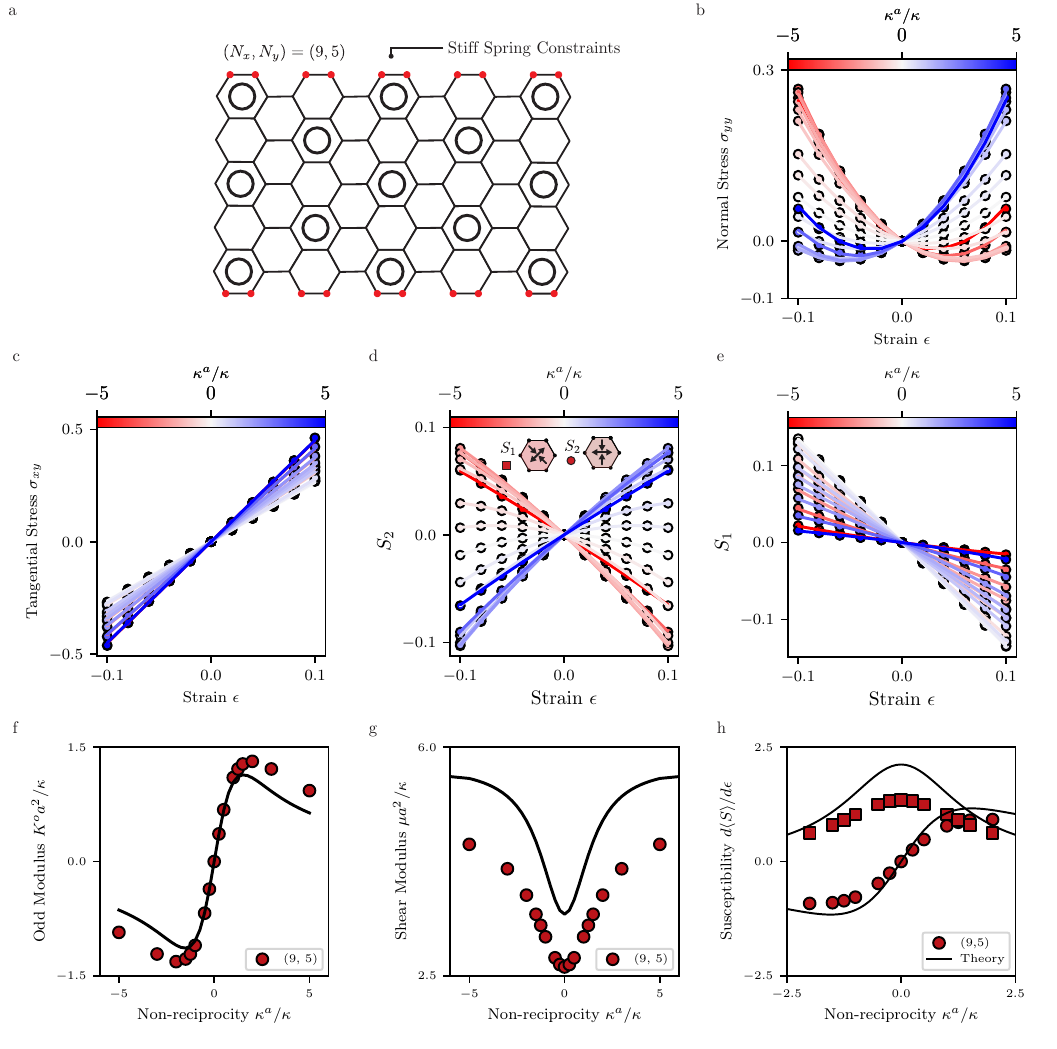}
      \caption{{\bf \rev{Finite size effects in a numerical mimic of our experimental lattice.}} (a) A simulated $(N_x,N_y)=(9,5)$ lattice mimicking our experiments. (b-e) We apply large strains $\epsilon\sim 0.1$ to mirror our experimental protocol, and track boundary forces and internal shear modes. At larger strains these relations become nonlinear---we extract tangent moduli and susceptibilities by fitting a quadratic and taking the linear coefficient. (f-h) For the $(9,5)$ lattice we see finite size corrections which slightly increase $K^o$ at high activity and shift $\mu$ downwards for all activities. Correspondingly, $S_1 \sim \epsilon$ is downwards-shifted for all $\kappa^a$, and $S_2$ is slightly off at high activities.}
      \label{fig:FiniteSizeScaling2}
\end{figure*}

Figure~\ref{fig:Phenomenology} shows that our periodic theory captures the key features of our metamaterials experiments: a decaying odd response, active stiffening, and plaquette locking, at high activity. These features are signatures of the percolation transition described in the main text. However there are some quantitative discrepancies: our theory overestimates $K^o$, $\mu$ and the plaquette shears $S_1$ and $S_2$ relative to experiment.  
One potential source of systematic bias is finite size effects: The experiments shown in Fig.~\ref{fig:Phenomenology} are performed on a $9\times5$ lattice of honeycombs, comprising $O(10)$ unit cells of the shown lattice (Methods \S \ref{sec:ExperimentalProtocol}), whereas our theory assumes a periodic lattice. To quantify finite size effects we develop ball spring simulations of our lattices, in which we explicitly simulate our experimental protocol of applying a horizontal shear, and measuring the forces on the boundary nodes (Figs.~\ref{fig:FiniteSizeScaling1},~\ref{fig:FiniteSizeScaling2}). 

We simulate the overdamped dynamics of a ball-spring mesh, where the forces on each node mimic our experiments. Each linkage exerts passive torques on the nodes with a torsional stiffness $\kappa$, and within an active plaquette non-reciprocal torques are additionally exerted as in Fig.~\ref{fig:Notation}. Longitudinal bonds are implemented with a finite but large spring stiffness $k$. We implement our boundary conditions with strain control: each node on the top and bottom of the lattice is constrained by a stiff spring, representing the rigid bar in our experiments. These constraining springs are horizontally displaced to mimic bar displacement [Fig.~\ref{fig:FiniteSizeScaling1}(a)]. Our simulation protocol mimics our experiments: we ramp through applied strains $\epsilon$, and track boundary nodal forces [Fig.~\ref{fig:FiniteSizeScaling1}(b,c)] and internal shear modes [Fig.~\ref{fig:FiniteSizeScaling1}(d,e)]. At a given strain $\epsilon$ we test for internal convergence of net forces before moving to the next strain. Our simulations are performed using $\kappa=1$, $k=100$, and $k=500$ for the constraining boundary springs. We use a forward Euler dynamics, where the frictional coefficient and timestep are interchangeable; we use timestep $dt=10^{-3}$.

We first verify that we recover the odd modulus $K^o$, shear modulus $\mu$ and internal deformation fields $S_1$ and $S_2$ that are predicted by our periodic theory in a finite sample, provided that the system size is large [Fig.~\ref{fig:FiniteSizeScaling1}(f-h)]. However, these simulations indeed indicate that there are finite size effects which can cause our simulated measurements to quantitatively deviate from the predictions of the periodic theory, whilst retaining the qualitative trends the theory predicts.

With this verification in hand, we now simulate lattices of the same dimensions as our experimental setup (Fig.~\ref{fig:FiniteSizeScaling2}), using the same protocol for generating and analyzing data that we use in our experiments (Methods \S\ref{sec:ExperimentalProtocol}, Figs.~\ref{fig:DataProcessing}). As in our experiments, we apply large strains [$\epsilon \sim 0.1$], at which the stress-strain relations  [Fig.~\ref{fig:FiniteSizeScaling2}(b,c)] and the relations between applied strain and internal shear modes [Fig.~\ref{fig:FiniteSizeScaling2}(d,e)] both become nonlinear. Linear elastic moduli and susceptibilities are extracted using the tangent modulus, by fitting a quadratic and taking the linear coefficient. 

We find that our finite-size simulations push $\mu$, $S_1$ and $S_2$ towards our experimental results, whilst leaving $K^o$ within experimental error bars (Fig.~\ref{fig:Phenomenology}). In particular, the systematic bias in $\mu$ [Fig.~\ref{fig:FiniteSizeScaling2}](g)] and $S_1$ [Fig.~\ref{fig:FiniteSizeScaling2}](h)], which exist even at $\kappa^a=0$, are well-accounted for by our finite size simulations. We conclude that the quantitative details of the relations $S_1, S_2$, $K^o$ and $\mu$ against $\kappa^a/\kappa$ are indeed sensitive to finite size scaling and the imperfections of our experimental realization, yet the qualitative form of these relations is robust to these details.

\section{Construction of the robotic materials} \label{sec:Construction}
The metamaterial shown in Fig.~\ref{fig:Phenomenology}(a) is composed of motorised vertices connected by plastic arms. 
Each vertex consists of a DC coreless motor (Motraxx CL1628) embedded in a cylindrical heatsink, an angular encoder (CUI AMT113S), and a microcontroller (ESP32) connected to a custom electronic board. The electronic board enables power conversion, interfacing between the sensor and motor, and communication between vertices.
Each vertex has a diameter 50 mm, height 90 mm, and mass $0.2$ kg. 
The power necessary to drive the motor is provided 
by an external 48V DC power source.

Rigid 3D-Printed arms connect each motor's drive shaft to the heat sink of the adjacent unit. 
The angle formed between the two arms at vertex $i$ is denoted by $\theta_i$. The on-board sensor measures $\theta_i$ at a 
sample rate of $100$ Hz and communicates the measurement to nearest neighbours. 
In response to the incoming signal, vertex $i$ exerts an active torsional force $\tau_i^a$
\begin{equation}
\tau_i^a = \kappa^a \left(\delta \theta_{i+1}-\delta \theta_{i-1} \right), \label{eq:fb} 
\end{equation}
where $\delta \theta_i = \theta_i - \theta^0$ and $\theta^0$ is the rest angle. The active stiffness $\kappa^a$ was programmed by the microcontroller and calibrated by measuring torque-displacement slopes for different values of the electronic feedback. 
The coreless motor saturates at a maximum torque of $\tau_\mathrm{max} = 12$ mN\,m. Each link has a length of $a=7.5\,\mathrm{cm}$. Adjacent vertices are also connected by a rubber band [blue in Fig.~\ref{fig:Phenomenology}(a)] with a thickness 4 mm which provides a passive elastic torsional stiffness of $\kappa=48$ mN m/rad. 
For further details see Ref.~\cite{veenstraAdaptiveLocomotionActive2025}. 

\section{Experimental protocol}
\label{sec:ExperimentalProtocol}
\begin{figure}[t]
\centering
\includegraphics{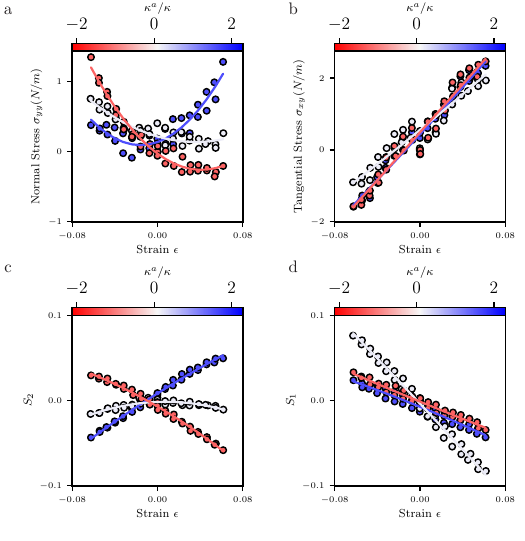}
      \caption{{\bf \rev{Experimental data processing pipeline}}. (a,b) Example data showing normal and tangential stresses $\sigma_{yy}$, $\sigma_{xy}$ vs. applied strain $\epsilon$, corresponding to Fig.~\ref{fig:Phenomenology}. We show traces at $\kappa^a/\kappa = 0, \pm 0.8$. The fits shown are quadratic. The slopes at the origin of these fits, $d \sigma_{yy}/d\epsilon$ and  $d \sigma_{xy}/d\epsilon$, define $K^o$ and $\mu$. (c,d) Average shear strains $S_1$ and $S_2$ across active plaquettes as a function of applied strain $\epsilon$, with the slope of quadratic fits defining the susceptibility $d\langle S \rangle /d\epsilon$ of each shear mode. } 
      
      \label{fig:DataProcessing}
\end{figure}
\rev{All experiments take place on top of a custom-made low friction air table as the motors we use do not have enough torque to lift the weight of the unit cells. The table consists of two $1.5 \, \si{m}\times 1.5\,  \si{m}$  plexiglas plates that sandwich 5 mm-wide air channels. The top plate is pierced by an array of  holes (pitch 10~mm, diameter 1~mm) conveying air pressurised at 10~bars. Each motorised vertex floats on a thin layer of pressurised air without making contact with the table. The experiments are filmed via a Nikon D5600 camera equipped with a 50 mm lens, recording 2 Mpx images at 30 frames per second. A marker is placed on the top of each vertex, and we use the python package OpenCV to detect and track the position of each vertex with a spatial resolution of 0.8 mm.}

\rev{In Fig.~\ref{fig:Phenomenology} of the main text, the top and bottom rows of hexagons in the lattice are constrained by 3D-printed fixtures attached to their top (bottom) vertices. These fixtures are in turn attached to rigid horizontal bars attached to a translation stage. Embedded in the horizontal bars are force sensors recording the normal force at a sample rate of 80Hz and a precision of $\pm 0.2N$, allowing us to simultaneously apply a simple transverse shear and measure net forces on the bar. The metamaterial is then sheared to a maximum strain of 0.1 and allowed to equilibrate during 2 seconds at intervals of 0.01 strain, after which shear and stress data are recorded.}

\rev{To generate the plots shown in Fig.~\ref{fig:Phenomenology} we start with our recorded timeseries of vertex positions within the honeycomb lattice, combined with a timeseries of force and displacement data for the horizontal bar during a compression cycle. These data allow us to calculate the imposed macroscopic strain $\epsilon$, the measured normal stress $\sigma_{yy}$ and tangential stress $\sigma_{xy}$ and the internal shear state $S_1$-$S_2$ of every hexagon. Examples of these raw data are shown in Fig.~\ref{fig:DataProcessing}.}

\rev{ To extract the odd and shear moduli shown in Fig.~\ref{fig:Phenomenology}(c-d), we take the stress/strain curves given by the bar data and extract the slope at the origin. These curves are nonlinear at large strain, and to extract tangent moduli we fit a second order polynomial and take the slope at $\epsilon=0$. To extract $S_1$ and $S_2$, we calculate}
\begin{align}
    S_1 =& \frac1{\sqrt 2} ( \delta \theta_1 - \delta \theta_3 ), \\ 
    S_2=&  \frac1{\sqrt 6} (\delta \theta_1 + \delta \theta_3 - 2 \delta \theta_5),
    \label{eq:Sprojection}
\end{align}
\rev{given in terms of the six angular deviations $\delta \theta_i$, $i = 1 \hdots 6$ within each hexagon from our tracking data. Schematics of each of these modes of deformation are shown in Fig.~\ref{fig:Phenomenology}(b) of the main text. We take the amplitudes of these two shear modes, averaged across every active hexagonal plaquette in the lattice, to generate a series of $S_1$-$S_2$ data against applied strain $\epsilon$ for a given compression experiment. This data is shown in Fig.~\ref{fig:DataProcessing}(c,d). Now, exactly as in the measurement of the odd modulus, we extract the slope of this strain curve against the imposed macroscopic strain $\epsilon$. This slope $d \langle S\rangle/d\epsilon$ defines the susceptibility, which we plot as a function of activity in Fig.~\ref{fig:Phenomenology}(g,h) of the main text. This procedure is also used on our simulation data Figs.~\ref{fig:FiniteSizeScaling1},~\ref{fig:FiniteSizeScaling2}. }

\rev{There are two differences between our experiments and simulations. First, the force sensors on each side of the lattice record different force signals due to substrate and solid friction. The data shown in Fig.~\ref{fig:Phenomenology}(c-d) takes their average and the error bars show the spread between sensors. Second, our experimentally constructed metamaterial has a bias towards a nonzero odd modulus even at zero activity [Fig.~\ref{fig:DataProcessing}(a)]. We do not attribute this bias purely to finite size effects, because it is not present in our simulations~\ref{fig:FiniteSizeScaling2}. Instead, it is likely caused by an intrinsic chirality in the purely passive construction of our metamaterial. We measure odd moduli relative to this slight baseline bias. The experimental hinge length $a$, passive stiffness $\kappa$, and bar length are all taken from independently calibrated data as described in \S \ref{sec:Construction}. }


\begin{thebibliography}{62}%
\makeatletter
\providecommand \@ifxundefined [1]{%
 \@ifx{#1\undefined}
}%
\providecommand \@ifnum [1]{%
 \ifnum #1\expandafter \@firstoftwo
 \else \expandafter \@secondoftwo
 \fi
}%
\providecommand \@ifx [1]{%
 \ifx #1\expandafter \@firstoftwo
 \else \expandafter \@secondoftwo
 \fi
}%
\providecommand \natexlab [1]{#1}%
\providecommand \enquote  [1]{``#1''}%
\providecommand \bibnamefont  [1]{#1}%
\providecommand \bibfnamefont [1]{#1}%
\providecommand \citenamefont [1]{#1}%
\providecommand \href@noop [0]{\@secondoftwo}%
\providecommand \href [0]{\begingroup \@sanitize@url \@href}%
\providecommand \@href[1]{\@@startlink{#1}\@@href}%
\providecommand \@@href[1]{\endgroup#1\@@endlink}%
\providecommand \@sanitize@url [0]{\catcode `\\12\catcode `\$12\catcode `\&12\catcode `\#12\catcode `\^12\catcode `\_12\catcode `\%12\relax}%
\providecommand \@@startlink[1]{}%
\providecommand \@@endlink[0]{}%
\providecommand \url  [0]{\begingroup\@sanitize@url \@url }%
\providecommand \@url [1]{\endgroup\@href {#1}{\urlprefix }}%
\providecommand \urlprefix  [0]{URL }%
\providecommand \Eprint [0]{\href }%
\providecommand \doibase [0]{https://doi.org/}%
\providecommand \selectlanguage [0]{\@gobble}%
\providecommand \bibinfo  [0]{\@secondoftwo}%
\providecommand \bibfield  [0]{\@secondoftwo}%
\providecommand \translation [1]{[#1]}%
\providecommand \BibitemOpen [0]{}%
\providecommand \bibitemStop [0]{}%
\providecommand \bibitemNoStop [0]{.\EOS\space}%
\providecommand \EOS [0]{\spacefactor3000\relax}%
\providecommand \BibitemShut  [1]{\csname bibitem#1\endcsname}%
\let\auto@bib@innerbib\@empty
\bibitem [{\citenamefont {Marchetti}\ \emph {et~al.}(2013)\citenamefont {Marchetti}, \citenamefont {Joanny}, \citenamefont {Ramaswamy}, \citenamefont {Liverpool}, \citenamefont {Prost}, \citenamefont {Rao},\ and\ \citenamefont {Simha}}]{marchettiHydrodynamicsSoftActive2013}%
  \BibitemOpen
  \bibfield  {author} {\bibinfo {author} {\bibfnamefont {M.~C.}\ \bibnamefont {Marchetti}}, \bibinfo {author} {\bibfnamefont {J.~F.}\ \bibnamefont {Joanny}}, \bibinfo {author} {\bibfnamefont {S.}~\bibnamefont {Ramaswamy}}, \bibinfo {author} {\bibfnamefont {T.~B.}\ \bibnamefont {Liverpool}}, \bibinfo {author} {\bibfnamefont {J.}~\bibnamefont {Prost}}, \bibinfo {author} {\bibfnamefont {M.}~\bibnamefont {Rao}},\ and\ \bibinfo {author} {\bibfnamefont {R.~A.}\ \bibnamefont {Simha}},\ }\bibfield  {title} {\bibinfo {title} {Hydrodynamics of soft active matter},\ }\href {https://doi.org/10.1103/RevModPhys.85.1143} {\bibfield  {journal} {\bibinfo  {journal} {Reviews of Modern Physics}\ }\textbf {\bibinfo {volume} {85}},\ \bibinfo {pages} {1143} (\bibinfo {year} {2013})}\BibitemShut {NoStop}%
\bibitem [{\citenamefont {Gompper}\ \emph {et~al.}(2025)\citenamefont {Gompper}, \citenamefont {Stone}, \citenamefont {Kurzthaler}, \citenamefont {Saintillan}, \citenamefont {Peruani}, \citenamefont {Fedosov}, \citenamefont {Auth}, \citenamefont {{Cottin-Bizonne}}, \citenamefont {Ybert}, \citenamefont {Cl{\'e}ment}, \citenamefont {Darnige}, \citenamefont {Lindner}, \citenamefont {Goldstein}, \citenamefont {Liebchen}, \citenamefont {Binysh}, \citenamefont {Souslov}, \citenamefont {Isa}, \citenamefont {{di Leonardo}}, \citenamefont {Frangipane}, \citenamefont {Gu}, \citenamefont {Nelson}, \citenamefont {Brauns}, \citenamefont {Marchetti}, \citenamefont {Cichos}, \citenamefont {Heuthe}, \citenamefont {Bechinger}, \citenamefont {Korman}, \citenamefont {Feinerman}, \citenamefont {Cavagna}, \citenamefont {Giardina}, \citenamefont {Jeckel},\ and\ \citenamefont {Drescher}}]{gompper2025MotileActive2025}%
  \BibitemOpen
  \bibfield  {author} {\bibinfo {author} {\bibfnamefont {G.}~\bibnamefont {Gompper}}, \bibinfo {author} {\bibfnamefont {H.~A.}\ \bibnamefont {Stone}}, \bibinfo {author} {\bibfnamefont {C.}~\bibnamefont {Kurzthaler}}, \bibinfo {author} {\bibfnamefont {D.}~\bibnamefont {Saintillan}}, \bibinfo {author} {\bibfnamefont {F.}~\bibnamefont {Peruani}}, \bibinfo {author} {\bibfnamefont {D.~A.}\ \bibnamefont {Fedosov}}, \bibinfo {author} {\bibfnamefont {T.}~\bibnamefont {Auth}}, \bibinfo {author} {\bibfnamefont {C.}~\bibnamefont {{Cottin-Bizonne}}}, \bibinfo {author} {\bibfnamefont {C.}~\bibnamefont {Ybert}}, \bibinfo {author} {\bibfnamefont {E.}~\bibnamefont {Cl{\'e}ment}}, \bibinfo {author} {\bibfnamefont {T.}~\bibnamefont {Darnige}}, \bibinfo {author} {\bibfnamefont {A.}~\bibnamefont {Lindner}}, \bibinfo {author} {\bibfnamefont {R.~E.}\ \bibnamefont {Goldstein}}, \bibinfo {author} {\bibfnamefont {B.}~\bibnamefont {Liebchen}}, \bibinfo {author} {\bibfnamefont {J.}~\bibnamefont {Binysh}}, \bibinfo {author}
  {\bibfnamefont {A.}~\bibnamefont {Souslov}}, \bibinfo {author} {\bibfnamefont {L.}~\bibnamefont {Isa}}, \bibinfo {author} {\bibfnamefont {R.}~\bibnamefont {{di Leonardo}}}, \bibinfo {author} {\bibfnamefont {G.}~\bibnamefont {Frangipane}}, \bibinfo {author} {\bibfnamefont {H.}~\bibnamefont {Gu}}, \bibinfo {author} {\bibfnamefont {B.~J.}\ \bibnamefont {Nelson}}, \bibinfo {author} {\bibfnamefont {F.}~\bibnamefont {Brauns}}, \bibinfo {author} {\bibfnamefont {M.~C.}\ \bibnamefont {Marchetti}}, \bibinfo {author} {\bibfnamefont {F.}~\bibnamefont {Cichos}}, \bibinfo {author} {\bibfnamefont {V.-L.}\ \bibnamefont {Heuthe}}, \bibinfo {author} {\bibfnamefont {C.}~\bibnamefont {Bechinger}}, \bibinfo {author} {\bibfnamefont {A.}~\bibnamefont {Korman}}, \bibinfo {author} {\bibfnamefont {O.}~\bibnamefont {Feinerman}}, \bibinfo {author} {\bibfnamefont {A.}~\bibnamefont {Cavagna}}, \bibinfo {author} {\bibfnamefont {I.}~\bibnamefont {Giardina}}, \bibinfo {author} {\bibfnamefont {H.}~\bibnamefont {Jeckel}},\ and\ \bibinfo
  {author} {\bibfnamefont {K.}~\bibnamefont {Drescher}},\ }\bibfield  {title} {\bibinfo {title} {The 2025 motile active matter roadmap},\ }\href {https://doi.org/10.1088/1361-648X/adac98} {\bibfield  {journal} {\bibinfo  {journal} {Journal of Physics: Condensed Matter}\ }\textbf {\bibinfo {volume} {37}},\ \bibinfo {pages} {143501} (\bibinfo {year} {2025})}\BibitemShut {NoStop}%
\bibitem [{\citenamefont {Cates}\ and\ \citenamefont {Tailleur}(2015)}]{catesMotilityInducedPhaseSeparation2015}%
  \BibitemOpen
  \bibfield  {author} {\bibinfo {author} {\bibfnamefont {M.~E.}\ \bibnamefont {Cates}}\ and\ \bibinfo {author} {\bibfnamefont {J.}~\bibnamefont {Tailleur}},\ }\bibfield  {title} {\bibinfo {title} {Motility-{{Induced Phase Separation}}},\ }\href {https://doi.org/10.1146/annurev-conmatphys-031214-014710} {\bibfield  {journal} {\bibinfo  {journal} {Annual Review of Condensed Matter Physics}\ }\textbf {\bibinfo {volume} {6}},\ \bibinfo {pages} {219} (\bibinfo {year} {2015})}\BibitemShut {NoStop}%
\bibitem [{\citenamefont {Bruno}(2013)}]{brunoImpossibilitySpontaneouslyRotating2013}%
  \BibitemOpen
  \bibfield  {author} {\bibinfo {author} {\bibfnamefont {P.}~\bibnamefont {Bruno}},\ }\bibfield  {title} {\bibinfo {title} {Impossibility of {{Spontaneously Rotating Time Crystals}}: {{A No-Go Theorem}}},\ }\href {https://doi.org/10.1103/PhysRevLett.111.070402} {\bibfield  {journal} {\bibinfo  {journal} {Physical Review Letters}\ }\textbf {\bibinfo {volume} {111}},\ \bibinfo {pages} {070402} (\bibinfo {year} {2013})}\BibitemShut {NoStop}%
\bibitem [{\citenamefont {Watanabe}\ and\ \citenamefont {Oshikawa}(2015)}]{watanabeAbsenceQuantumTime2015}%
  \BibitemOpen
  \bibfield  {author} {\bibinfo {author} {\bibfnamefont {H.}~\bibnamefont {Watanabe}}\ and\ \bibinfo {author} {\bibfnamefont {M.}~\bibnamefont {Oshikawa}},\ }\bibfield  {title} {\bibinfo {title} {Absence of {{Quantum Time Crystals}}},\ }\href {https://doi.org/10.1103/PhysRevLett.114.251603} {\bibfield  {journal} {\bibinfo  {journal} {Physical Review Letters}\ }\textbf {\bibinfo {volume} {114}},\ \bibinfo {pages} {251603} (\bibinfo {year} {2015})}\BibitemShut {NoStop}%
\bibitem [{\citenamefont {Fruchart}\ \emph {et~al.}(2021)\citenamefont {Fruchart}, \citenamefont {Hanai}, \citenamefont {Littlewood},\ and\ \citenamefont {Vitelli}}]{fruchartNonreciprocalPhaseTransitions2021}%
  \BibitemOpen
  \bibfield  {author} {\bibinfo {author} {\bibfnamefont {M.}~\bibnamefont {Fruchart}}, \bibinfo {author} {\bibfnamefont {R.}~\bibnamefont {Hanai}}, \bibinfo {author} {\bibfnamefont {P.~B.}\ \bibnamefont {Littlewood}},\ and\ \bibinfo {author} {\bibfnamefont {V.}~\bibnamefont {Vitelli}},\ }\bibfield  {title} {\bibinfo {title} {Non-reciprocal phase transitions},\ }\href {https://doi.org/10.1038/s41586-021-03375-9} {\bibfield  {journal} {\bibinfo  {journal} {Nature}\ }\textbf {\bibinfo {volume} {592}},\ \bibinfo {pages} {363} (\bibinfo {year} {2021})}\BibitemShut {NoStop}%
\bibitem [{\citenamefont {Liu}\ \emph {et~al.}(2023)\citenamefont {Liu}, \citenamefont {Ou}, \citenamefont {MacDonald},\ and\ \citenamefont {Zheludev}}]{liuPhotonicMetamaterialAnalogue2023}%
  \BibitemOpen
  \bibfield  {author} {\bibinfo {author} {\bibfnamefont {T.}~\bibnamefont {Liu}}, \bibinfo {author} {\bibfnamefont {J.-Y.}\ \bibnamefont {Ou}}, \bibinfo {author} {\bibfnamefont {K.~F.}\ \bibnamefont {MacDonald}},\ and\ \bibinfo {author} {\bibfnamefont {N.~I.}\ \bibnamefont {Zheludev}},\ }\bibfield  {title} {\bibinfo {title} {Photonic metamaterial analogue of a continuous time crystal},\ }\href {https://doi.org/10.1038/s41567-023-02023-5} {\bibfield  {journal} {\bibinfo  {journal} {Nature Physics}\ }\textbf {\bibinfo {volume} {19}},\ \bibinfo {pages} {986} (\bibinfo {year} {2023})}\BibitemShut {NoStop}%
\bibitem [{\citenamefont {Volpe}\ \emph {et~al.}(2025)\citenamefont {Volpe}, \citenamefont {Ara{\'u}jo}, \citenamefont {Guix}, \citenamefont {Miodownik}, \citenamefont {Martin}, \citenamefont {Alvarez}, \citenamefont {Simmchen}, \citenamefont {Leonardo}, \citenamefont {Pellicciotta}, \citenamefont {Martinet}, \citenamefont {Palacci}, \citenamefont {Ng}, \citenamefont {Saxena}, \citenamefont {Sapienza}, \citenamefont {Nadine}, \citenamefont {Mano}, \citenamefont {Mahdavi}, \citenamefont {Beck~Adiels}, \citenamefont {Forth}, \citenamefont {Santangelo}, \citenamefont {Palagi}, \citenamefont {Seok}, \citenamefont {{Webster-Wood}}, \citenamefont {Wang}, \citenamefont {Yao}, \citenamefont {Aghakhani}, \citenamefont {Barois}, \citenamefont {Kellay}, \citenamefont {Coulais}, \citenamefont {{van Hecke}}, \citenamefont {Pierce}, \citenamefont {Wang}, \citenamefont {Chong}, \citenamefont {Goldman}, \citenamefont {Reina}, \citenamefont {Trianni}, \citenamefont {Volpe}, \citenamefont {Beckett}, \citenamefont {Nair},\ and\
  \citenamefont {Armstrong}}]{volpeRoadmapAnimateMatter2025}%
  \BibitemOpen
  \bibfield  {author} {\bibinfo {author} {\bibfnamefont {G.}~\bibnamefont {Volpe}}, \bibinfo {author} {\bibfnamefont {N.~A.~M.}\ \bibnamefont {Ara{\'u}jo}}, \bibinfo {author} {\bibfnamefont {M.}~\bibnamefont {Guix}}, \bibinfo {author} {\bibfnamefont {M.}~\bibnamefont {Miodownik}}, \bibinfo {author} {\bibfnamefont {N.}~\bibnamefont {Martin}}, \bibinfo {author} {\bibfnamefont {L.}~\bibnamefont {Alvarez}}, \bibinfo {author} {\bibfnamefont {J.}~\bibnamefont {Simmchen}}, \bibinfo {author} {\bibfnamefont {R.~D.}\ \bibnamefont {Leonardo}}, \bibinfo {author} {\bibfnamefont {N.}~\bibnamefont {Pellicciotta}}, \bibinfo {author} {\bibfnamefont {Q.}~\bibnamefont {Martinet}}, \bibinfo {author} {\bibfnamefont {J.}~\bibnamefont {Palacci}}, \bibinfo {author} {\bibfnamefont {W.~K.}\ \bibnamefont {Ng}}, \bibinfo {author} {\bibfnamefont {D.}~\bibnamefont {Saxena}}, \bibinfo {author} {\bibfnamefont {R.}~\bibnamefont {Sapienza}}, \bibinfo {author} {\bibfnamefont {S.}~\bibnamefont {Nadine}}, \bibinfo {author} {\bibfnamefont
  {J.~F.}\ \bibnamefont {Mano}}, \bibinfo {author} {\bibfnamefont {R.}~\bibnamefont {Mahdavi}}, \bibinfo {author} {\bibfnamefont {C.}~\bibnamefont {Beck~Adiels}}, \bibinfo {author} {\bibfnamefont {J.}~\bibnamefont {Forth}}, \bibinfo {author} {\bibfnamefont {C.}~\bibnamefont {Santangelo}}, \bibinfo {author} {\bibfnamefont {S.}~\bibnamefont {Palagi}}, \bibinfo {author} {\bibfnamefont {J.~M.}\ \bibnamefont {Seok}}, \bibinfo {author} {\bibfnamefont {V.~A.}\ \bibnamefont {{Webster-Wood}}}, \bibinfo {author} {\bibfnamefont {S.}~\bibnamefont {Wang}}, \bibinfo {author} {\bibfnamefont {L.}~\bibnamefont {Yao}}, \bibinfo {author} {\bibfnamefont {A.}~\bibnamefont {Aghakhani}}, \bibinfo {author} {\bibfnamefont {T.}~\bibnamefont {Barois}}, \bibinfo {author} {\bibfnamefont {H.}~\bibnamefont {Kellay}}, \bibinfo {author} {\bibfnamefont {C.}~\bibnamefont {Coulais}}, \bibinfo {author} {\bibfnamefont {M.}~\bibnamefont {{van Hecke}}}, \bibinfo {author} {\bibfnamefont {C.~J.}\ \bibnamefont {Pierce}}, \bibinfo {author}
  {\bibfnamefont {T.}~\bibnamefont {Wang}}, \bibinfo {author} {\bibfnamefont {B.}~\bibnamefont {Chong}}, \bibinfo {author} {\bibfnamefont {D.~I.}\ \bibnamefont {Goldman}}, \bibinfo {author} {\bibfnamefont {A.}~\bibnamefont {Reina}}, \bibinfo {author} {\bibfnamefont {V.}~\bibnamefont {Trianni}}, \bibinfo {author} {\bibfnamefont {G.}~\bibnamefont {Volpe}}, \bibinfo {author} {\bibfnamefont {R.}~\bibnamefont {Beckett}}, \bibinfo {author} {\bibfnamefont {S.~P.}\ \bibnamefont {Nair}},\ and\ \bibinfo {author} {\bibfnamefont {R.}~\bibnamefont {Armstrong}},\ }\bibfield  {title} {\bibinfo {title} {Roadmap for animate matter},\ }\href {https://doi.org/10.1088/1361-648X/adebd3} {\bibfield  {journal} {\bibinfo  {journal} {Journal of Physics: Condensed Matter}\ }\textbf {\bibinfo {volume} {37}},\ \bibinfo {pages} {333501} (\bibinfo {year} {2025})}\BibitemShut {NoStop}%
\bibitem [{\citenamefont {Tan}\ \emph {et~al.}(2022)\citenamefont {Tan}, \citenamefont {Mietke}, \citenamefont {Li}, \citenamefont {Chen}, \citenamefont {Higinbotham}, \citenamefont {Foster}, \citenamefont {Gokhale}, \citenamefont {Dunkel},\ and\ \citenamefont {Fakhri}}]{tanOddDynamicsLiving2022}%
  \BibitemOpen
  \bibfield  {author} {\bibinfo {author} {\bibfnamefont {T.~H.}\ \bibnamefont {Tan}}, \bibinfo {author} {\bibfnamefont {A.}~\bibnamefont {Mietke}}, \bibinfo {author} {\bibfnamefont {J.}~\bibnamefont {Li}}, \bibinfo {author} {\bibfnamefont {Y.}~\bibnamefont {Chen}}, \bibinfo {author} {\bibfnamefont {H.}~\bibnamefont {Higinbotham}}, \bibinfo {author} {\bibfnamefont {P.~J.}\ \bibnamefont {Foster}}, \bibinfo {author} {\bibfnamefont {S.}~\bibnamefont {Gokhale}}, \bibinfo {author} {\bibfnamefont {J.}~\bibnamefont {Dunkel}},\ and\ \bibinfo {author} {\bibfnamefont {N.}~\bibnamefont {Fakhri}},\ }\bibfield  {title} {\bibinfo {title} {Odd dynamics of living chiral crystals},\ }\href {https://doi.org/10.1038/s41586-022-04889-6} {\bibfield  {journal} {\bibinfo  {journal} {Nature}\ }\textbf {\bibinfo {volume} {607}},\ \bibinfo {pages} {287} (\bibinfo {year} {2022})}\BibitemShut {NoStop}%
\bibitem [{\citenamefont {Baconnier}\ \emph {et~al.}(2022)\citenamefont {Baconnier}, \citenamefont {Shohat}, \citenamefont {L{\'o}pez}, \citenamefont {Coulais}, \citenamefont {D{\'e}mery}, \citenamefont {D{\"u}ring},\ and\ \citenamefont {Dauchot}}]{baconnierSelectiveCollectiveActuation2022a}%
  \BibitemOpen
  \bibfield  {author} {\bibinfo {author} {\bibfnamefont {P.}~\bibnamefont {Baconnier}}, \bibinfo {author} {\bibfnamefont {D.}~\bibnamefont {Shohat}}, \bibinfo {author} {\bibfnamefont {C.~H.}\ \bibnamefont {L{\'o}pez}}, \bibinfo {author} {\bibfnamefont {C.}~\bibnamefont {Coulais}}, \bibinfo {author} {\bibfnamefont {V.}~\bibnamefont {D{\'e}mery}}, \bibinfo {author} {\bibfnamefont {G.}~\bibnamefont {D{\"u}ring}},\ and\ \bibinfo {author} {\bibfnamefont {O.}~\bibnamefont {Dauchot}},\ }\bibfield  {title} {\bibinfo {title} {Selective and collective actuation in active solids},\ }\href {https://doi.org/10.1038/s41567-022-01704-x} {\bibfield  {journal} {\bibinfo  {journal} {Nature Physics}\ }\textbf {\bibinfo {volume} {18}},\ \bibinfo {pages} {1234} (\bibinfo {year} {2022})}\BibitemShut {NoStop}%
\bibitem [{\citenamefont {Zhang}\ and\ \citenamefont {Fodor}(2023)}]{zhangPulsatingActiveMatter2023}%
  \BibitemOpen
  \bibfield  {author} {\bibinfo {author} {\bibfnamefont {Y.}~\bibnamefont {Zhang}}\ and\ \bibinfo {author} {\bibfnamefont {{\'E}.}~\bibnamefont {Fodor}},\ }\bibfield  {title} {\bibinfo {title} {Pulsating {{Active Matter}}},\ }\href {https://doi.org/10.1103/PhysRevLett.131.238302} {\bibfield  {journal} {\bibinfo  {journal} {Physical Review Letters}\ }\textbf {\bibinfo {volume} {131}},\ \bibinfo {pages} {238302} (\bibinfo {year} {2023})}\BibitemShut {NoStop}%
\bibitem [{\citenamefont {Xu}\ \emph {et~al.}(2023)\citenamefont {Xu}, \citenamefont {Huang}, \citenamefont {Zhang},\ and\ \citenamefont {Wu}}]{xuAutonomousWavesGlobal2023}%
  \BibitemOpen
  \bibfield  {author} {\bibinfo {author} {\bibfnamefont {H.}~\bibnamefont {Xu}}, \bibinfo {author} {\bibfnamefont {Y.}~\bibnamefont {Huang}}, \bibinfo {author} {\bibfnamefont {R.}~\bibnamefont {Zhang}},\ and\ \bibinfo {author} {\bibfnamefont {Y.}~\bibnamefont {Wu}},\ }\bibfield  {title} {\bibinfo {title} {Autonomous waves and global motion modes in living active solids},\ }\href {https://doi.org/10.1038/s41567-022-01836-0} {\bibfield  {journal} {\bibinfo  {journal} {Nature Physics}\ }\textbf {\bibinfo {volume} {19}},\ \bibinfo {pages} {46} (\bibinfo {year} {2023})}\BibitemShut {NoStop}%
\bibitem [{\citenamefont {Armon}\ \emph {et~al.}(2021)\citenamefont {Armon}, \citenamefont {Bull}, \citenamefont {Moriel}, \citenamefont {Aharoni},\ and\ \citenamefont {Prakash}}]{armonModelingEpithelialTissues2021}%
  \BibitemOpen
  \bibfield  {author} {\bibinfo {author} {\bibfnamefont {S.}~\bibnamefont {Armon}}, \bibinfo {author} {\bibfnamefont {M.~S.}\ \bibnamefont {Bull}}, \bibinfo {author} {\bibfnamefont {A.}~\bibnamefont {Moriel}}, \bibinfo {author} {\bibfnamefont {H.}~\bibnamefont {Aharoni}},\ and\ \bibinfo {author} {\bibfnamefont {M.}~\bibnamefont {Prakash}},\ }\bibfield  {title} {\bibinfo {title} {Modeling epithelial tissues as active-elastic sheets reproduce contraction pulses and predict rip resistance},\ }\href {https://doi.org/10.1038/s42005-021-00712-2} {\bibfield  {journal} {\bibinfo  {journal} {Communications Physics}\ }\textbf {\bibinfo {volume} {4}},\ \bibinfo {pages} {1} (\bibinfo {year} {2021})}\BibitemShut {NoStop}%
\bibitem [{\citenamefont {{P{\'e}rez-Verdugo}}\ \emph {et~al.}(2024)\citenamefont {{P{\'e}rez-Verdugo}}, \citenamefont {Banks},\ and\ \citenamefont {Banerjee}}]{perez-verdugoExcitableDynamicsDriven2024}%
  \BibitemOpen
  \bibfield  {author} {\bibinfo {author} {\bibfnamefont {F.}~\bibnamefont {{P{\'e}rez-Verdugo}}}, \bibinfo {author} {\bibfnamefont {S.}~\bibnamefont {Banks}},\ and\ \bibinfo {author} {\bibfnamefont {S.}~\bibnamefont {Banerjee}},\ }\bibfield  {title} {\bibinfo {title} {Excitable dynamics driven by mechanical feedback in biological tissues},\ }\href {https://doi.org/10.1038/s42005-024-01661-2} {\bibfield  {journal} {\bibinfo  {journal} {Communications Physics}\ }\textbf {\bibinfo {volume} {7}},\ \bibinfo {pages} {1} (\bibinfo {year} {2024})}\BibitemShut {NoStop}%
\bibitem [{\citenamefont {Alert}\ \emph {et~al.}(2022)\citenamefont {Alert}, \citenamefont {Casademunt},\ and\ \citenamefont {Joanny}}]{alertActiveTurbulence2022}%
  \BibitemOpen
  \bibfield  {author} {\bibinfo {author} {\bibfnamefont {R.}~\bibnamefont {Alert}}, \bibinfo {author} {\bibfnamefont {J.}~\bibnamefont {Casademunt}},\ and\ \bibinfo {author} {\bibfnamefont {J.-F.}\ \bibnamefont {Joanny}},\ }\bibfield  {title} {\bibinfo {title} {Active {{Turbulence}}},\ }\href {https://doi.org/10.1146/annurev-conmatphys-082321-035957} {\bibfield  {journal} {\bibinfo  {journal} {Annual Review of Condensed Matter Physics}\ }\textbf {\bibinfo {volume} {13}},\ \bibinfo {pages} {143} (\bibinfo {year} {2022})}\BibitemShut {NoStop}%
\bibitem [{\citenamefont {Veenstra}\ \emph {et~al.}(2025)\citenamefont {Veenstra}, \citenamefont {Scheibner}, \citenamefont {Brandenbourger}, \citenamefont {Binysh}, \citenamefont {Souslov}, \citenamefont {Vitelli},\ and\ \citenamefont {Coulais}}]{veenstraAdaptiveLocomotionActive2025}%
  \BibitemOpen
  \bibfield  {author} {\bibinfo {author} {\bibfnamefont {J.}~\bibnamefont {Veenstra}}, \bibinfo {author} {\bibfnamefont {C.}~\bibnamefont {Scheibner}}, \bibinfo {author} {\bibfnamefont {M.}~\bibnamefont {Brandenbourger}}, \bibinfo {author} {\bibfnamefont {J.}~\bibnamefont {Binysh}}, \bibinfo {author} {\bibfnamefont {A.}~\bibnamefont {Souslov}}, \bibinfo {author} {\bibfnamefont {V.}~\bibnamefont {Vitelli}},\ and\ \bibinfo {author} {\bibfnamefont {C.}~\bibnamefont {Coulais}},\ }\bibfield  {title} {\bibinfo {title} {Adaptive locomotion of active solids},\ }\href {https://doi.org/10.1038/s41586-025-08646-3} {\bibfield  {journal} {\bibinfo  {journal} {Nature}\ }\textbf {\bibinfo {volume} {639}},\ \bibinfo {pages} {935} (\bibinfo {year} {2025})}\BibitemShut {NoStop}%
\bibitem [{\citenamefont {Saintyves}\ \emph {et~al.}(2024)\citenamefont {Saintyves}, \citenamefont {Spenko},\ and\ \citenamefont {Jaeger}}]{saintyvesSelforganizingRoboticAggregate2024}%
  \BibitemOpen
  \bibfield  {author} {\bibinfo {author} {\bibfnamefont {B.}~\bibnamefont {Saintyves}}, \bibinfo {author} {\bibfnamefont {M.}~\bibnamefont {Spenko}},\ and\ \bibinfo {author} {\bibfnamefont {H.~M.}\ \bibnamefont {Jaeger}},\ }\bibfield  {title} {\bibinfo {title} {A self-organizing robotic aggregate using solid and liquid-like collective states},\ }\bibfield  {journal} {\bibinfo  {journal} {Science Robotics}\ }\href {https://doi.org/10.1126/scirobotics.adh4130} {10.1126/scirobotics.adh4130} (\bibinfo {year} {2024})\BibitemShut {NoStop}%
\bibitem [{\citenamefont {Devlin}\ \emph {et~al.}(2025)\citenamefont {Devlin}, \citenamefont {Kim}, \citenamefont {Camp{\`a}s},\ and\ \citenamefont {Hawkes}}]{devlinMateriallikeRoboticCollectives2025}%
  \BibitemOpen
  \bibfield  {author} {\bibinfo {author} {\bibfnamefont {M.~R.}\ \bibnamefont {Devlin}}, \bibinfo {author} {\bibfnamefont {S.}~\bibnamefont {Kim}}, \bibinfo {author} {\bibfnamefont {O.}~\bibnamefont {Camp{\`a}s}},\ and\ \bibinfo {author} {\bibfnamefont {E.~W.}\ \bibnamefont {Hawkes}},\ }\bibfield  {title} {\bibinfo {title} {Material-like robotic collectives with spatiotemporal control of strength and shape},\ }\href {https://doi.org/10.1126/science.ads7942} {\bibfield  {journal} {\bibinfo  {journal} {Science}\ }\textbf {\bibinfo {volume} {387}},\ \bibinfo {pages} {880} (\bibinfo {year} {2025})}\BibitemShut {NoStop}%
\bibitem [{\citenamefont {Hanai}(2024)}]{hanaiNonreciprocalFrustrationTime2024}%
  \BibitemOpen
  \bibfield  {author} {\bibinfo {author} {\bibfnamefont {R.}~\bibnamefont {Hanai}},\ }\bibfield  {title} {\bibinfo {title} {Nonreciprocal {{Frustration}}: {{Time Crystalline Order-by-Disorder Phenomenon}} and a {{Spin-Glass-like State}}},\ }\href {https://doi.org/10.1103/PhysRevX.14.011029} {\bibfield  {journal} {\bibinfo  {journal} {Physical Review X}\ }\textbf {\bibinfo {volume} {14}},\ \bibinfo {pages} {011029} (\bibinfo {year} {2024})}\BibitemShut {NoStop}%
\bibitem [{\citenamefont {Avni}\ \emph {et~al.}(2025)\citenamefont {Avni}, \citenamefont {Fruchart}, \citenamefont {Martin}, \citenamefont {Seara},\ and\ \citenamefont {Vitelli}}]{avniDynamicalPhaseTransitions2025}%
  \BibitemOpen
  \bibfield  {author} {\bibinfo {author} {\bibfnamefont {Y.}~\bibnamefont {Avni}}, \bibinfo {author} {\bibfnamefont {M.}~\bibnamefont {Fruchart}}, \bibinfo {author} {\bibfnamefont {D.}~\bibnamefont {Martin}}, \bibinfo {author} {\bibfnamefont {D.}~\bibnamefont {Seara}},\ and\ \bibinfo {author} {\bibfnamefont {V.}~\bibnamefont {Vitelli}},\ }\bibfield  {title} {\bibinfo {title} {Dynamical phase transitions in the nonreciprocal {{Ising}} model},\ }\href {https://doi.org/10.1103/PhysRevE.111.034124} {\bibfield  {journal} {\bibinfo  {journal} {Physical Review E}\ }\textbf {\bibinfo {volume} {111}},\ \bibinfo {pages} {034124} (\bibinfo {year} {2025})}\BibitemShut {NoStop}%
\bibitem [{\citenamefont {Loos}\ \emph {et~al.}(2023)\citenamefont {Loos}, \citenamefont {Klapp},\ and\ \citenamefont {Martynec}}]{loosLongRangeOrderDirectional2023}%
  \BibitemOpen
  \bibfield  {author} {\bibinfo {author} {\bibfnamefont {S.~A.~M.}\ \bibnamefont {Loos}}, \bibinfo {author} {\bibfnamefont {S.~H.~L.}\ \bibnamefont {Klapp}},\ and\ \bibinfo {author} {\bibfnamefont {T.}~\bibnamefont {Martynec}},\ }\bibfield  {title} {\bibinfo {title} {Long-{{Range Order}} and {{Directional Defect Propagation}} in the {{Nonreciprocal XY Model}} with {{Vision Cone Interactions}}},\ }\href {https://doi.org/10.1103/PhysRevLett.130.198301} {\bibfield  {journal} {\bibinfo  {journal} {Physical Review Letters}\ }\textbf {\bibinfo {volume} {130}},\ \bibinfo {pages} {198301} (\bibinfo {year} {2023})}\BibitemShut {NoStop}%
\bibitem [{\citenamefont {Raskatla}\ \emph {et~al.}(2024)\citenamefont {Raskatla}, \citenamefont {Liu}, \citenamefont {Li}, \citenamefont {MacDonald},\ and\ \citenamefont {Zheludev}}]{raskatlaContinuousSpaceTimeCrystal2024}%
  \BibitemOpen
  \bibfield  {author} {\bibinfo {author} {\bibfnamefont {V.}~\bibnamefont {Raskatla}}, \bibinfo {author} {\bibfnamefont {T.}~\bibnamefont {Liu}}, \bibinfo {author} {\bibfnamefont {J.}~\bibnamefont {Li}}, \bibinfo {author} {\bibfnamefont {K.~F.}\ \bibnamefont {MacDonald}},\ and\ \bibinfo {author} {\bibfnamefont {N.~I.}\ \bibnamefont {Zheludev}},\ }\bibfield  {title} {\bibinfo {title} {Continuous {{Space-Time Crystal State Driven}} by {{Nonreciprocal Optical Forces}}},\ }\href {https://doi.org/10.1103/PhysRevLett.133.136202} {\bibfield  {journal} {\bibinfo  {journal} {Physical Review Letters}\ }\textbf {\bibinfo {volume} {133}},\ \bibinfo {pages} {136202} (\bibinfo {year} {2024})}\BibitemShut {NoStop}%
\bibitem [{\citenamefont {Li{\v s}ka}\ \emph {et~al.}(2024)\citenamefont {Li{\v s}ka}, \citenamefont {Zem{\'a}nkov{\'a}}, \citenamefont {J{\'a}kl}, \citenamefont {{\v S}iler}, \citenamefont {Simpson}, \citenamefont {Zem{\'a}nek},\ and\ \citenamefont {Brzobohat{\'y}}}]{liskaPTlikePhaseTransition2024}%
  \BibitemOpen
  \bibfield  {author} {\bibinfo {author} {\bibfnamefont {V.}~\bibnamefont {Li{\v s}ka}}, \bibinfo {author} {\bibfnamefont {T.}~\bibnamefont {Zem{\'a}nkov{\'a}}}, \bibinfo {author} {\bibfnamefont {P.}~\bibnamefont {J{\'a}kl}}, \bibinfo {author} {\bibfnamefont {M.}~\bibnamefont {{\v S}iler}}, \bibinfo {author} {\bibfnamefont {S.~H.}\ \bibnamefont {Simpson}}, \bibinfo {author} {\bibfnamefont {P.}~\bibnamefont {Zem{\'a}nek}},\ and\ \bibinfo {author} {\bibfnamefont {O.}~\bibnamefont {Brzobohat{\'y}}},\ }\bibfield  {title} {\bibinfo {title} {{{PT-like}} phase transition and limit cycle oscillations in non-reciprocally coupled optomechanical oscillators levitated in vacuum},\ }\href {https://doi.org/10.1038/s41567-024-02590-1} {\bibfield  {journal} {\bibinfo  {journal} {Nature Physics}\ }\textbf {\bibinfo {volume} {20}},\ \bibinfo {pages} {1622} (\bibinfo {year} {2024})}\BibitemShut {NoStop}%
\bibitem [{\citenamefont {Reisenbauer}\ \emph {et~al.}(2024)\citenamefont {Reisenbauer}, \citenamefont {Rudolph}, \citenamefont {Egyed}, \citenamefont {Hornberger}, \citenamefont {Zasedatelev}, \citenamefont {Abuzarli}, \citenamefont {Stickler},\ and\ \citenamefont {Deli{\'c}}}]{reisenbauerNonHermitianDynamicsNonreciprocity2024}%
  \BibitemOpen
  \bibfield  {author} {\bibinfo {author} {\bibfnamefont {M.}~\bibnamefont {Reisenbauer}}, \bibinfo {author} {\bibfnamefont {H.}~\bibnamefont {Rudolph}}, \bibinfo {author} {\bibfnamefont {L.}~\bibnamefont {Egyed}}, \bibinfo {author} {\bibfnamefont {K.}~\bibnamefont {Hornberger}}, \bibinfo {author} {\bibfnamefont {A.~V.}\ \bibnamefont {Zasedatelev}}, \bibinfo {author} {\bibfnamefont {M.}~\bibnamefont {Abuzarli}}, \bibinfo {author} {\bibfnamefont {B.~A.}\ \bibnamefont {Stickler}},\ and\ \bibinfo {author} {\bibfnamefont {U.}~\bibnamefont {Deli{\'c}}},\ }\bibfield  {title} {\bibinfo {title} {Non-{{Hermitian}} dynamics and non-reciprocity of optically coupled nanoparticles},\ }\href {https://doi.org/10.1038/s41567-024-02589-8} {\bibfield  {journal} {\bibinfo  {journal} {Nature Physics}\ }\textbf {\bibinfo {volume} {20}},\ \bibinfo {pages} {1629} (\bibinfo {year} {2024})}\BibitemShut {NoStop}%
\bibitem [{\citenamefont {Ruesink}\ \emph {et~al.}(2016)\citenamefont {Ruesink}, \citenamefont {Miri}, \citenamefont {Al{\`u}},\ and\ \citenamefont {Verhagen}}]{ruesinkNonreciprocityMagneticfreeIsolation2016}%
  \BibitemOpen
  \bibfield  {author} {\bibinfo {author} {\bibfnamefont {F.}~\bibnamefont {Ruesink}}, \bibinfo {author} {\bibfnamefont {M.-A.}\ \bibnamefont {Miri}}, \bibinfo {author} {\bibfnamefont {A.}~\bibnamefont {Al{\`u}}},\ and\ \bibinfo {author} {\bibfnamefont {E.}~\bibnamefont {Verhagen}},\ }\bibfield  {title} {\bibinfo {title} {Nonreciprocity and magnetic-free isolation based on optomechanical interactions},\ }\href {https://doi.org/10.1038/ncomms13662} {\bibfield  {journal} {\bibinfo  {journal} {Nature Communications}\ }\textbf {\bibinfo {volume} {7}},\ \bibinfo {pages} {13662} (\bibinfo {year} {2016})}\BibitemShut {NoStop}%
\bibitem [{\citenamefont {Weidemann}\ \emph {et~al.}(2020)\citenamefont {Weidemann}, \citenamefont {Kremer}, \citenamefont {Helbig}, \citenamefont {Hofmann}, \citenamefont {Stegmaier}, \citenamefont {Greiter}, \citenamefont {Thomale},\ and\ \citenamefont {Szameit}}]{weidemannTopologicalFunnelingLight2020}%
  \BibitemOpen
  \bibfield  {author} {\bibinfo {author} {\bibfnamefont {S.}~\bibnamefont {Weidemann}}, \bibinfo {author} {\bibfnamefont {M.}~\bibnamefont {Kremer}}, \bibinfo {author} {\bibfnamefont {T.}~\bibnamefont {Helbig}}, \bibinfo {author} {\bibfnamefont {T.}~\bibnamefont {Hofmann}}, \bibinfo {author} {\bibfnamefont {A.}~\bibnamefont {Stegmaier}}, \bibinfo {author} {\bibfnamefont {M.}~\bibnamefont {Greiter}}, \bibinfo {author} {\bibfnamefont {R.}~\bibnamefont {Thomale}},\ and\ \bibinfo {author} {\bibfnamefont {A.}~\bibnamefont {Szameit}},\ }\bibfield  {title} {\bibinfo {title} {Topological funneling of light},\ }\href {https://doi.org/10.1126/science.aaz8727} {\bibfield  {journal} {\bibinfo  {journal} {Science}\ }\textbf {\bibinfo {volume} {368}},\ \bibinfo {pages} {311} (\bibinfo {year} {2020})}\BibitemShut {NoStop}%
\bibitem [{\citenamefont {Fleury}\ \emph {et~al.}(2014)\citenamefont {Fleury}, \citenamefont {Sounas}, \citenamefont {Sieck}, \citenamefont {Haberman},\ and\ \citenamefont {Al{\`u}}}]{fleurySoundIsolationGiant2014}%
  \BibitemOpen
  \bibfield  {author} {\bibinfo {author} {\bibfnamefont {R.}~\bibnamefont {Fleury}}, \bibinfo {author} {\bibfnamefont {D.~L.}\ \bibnamefont {Sounas}}, \bibinfo {author} {\bibfnamefont {C.~F.}\ \bibnamefont {Sieck}}, \bibinfo {author} {\bibfnamefont {M.~R.}\ \bibnamefont {Haberman}},\ and\ \bibinfo {author} {\bibfnamefont {A.}~\bibnamefont {Al{\`u}}},\ }\bibfield  {title} {\bibinfo {title} {Sound {{Isolation}} and {{Giant Linear Nonreciprocity}} in a {{Compact Acoustic Circulator}}},\ }\href {https://doi.org/10.1126/science.1246957} {\bibfield  {journal} {\bibinfo  {journal} {Science}\ }\textbf {\bibinfo {volume} {343}},\ \bibinfo {pages} {516} (\bibinfo {year} {2014})}\BibitemShut {NoStop}%
\bibitem [{\citenamefont {Lim}\ \emph {et~al.}(2024)\citenamefont {Lim}, \citenamefont {VanSaders},\ and\ \citenamefont {Jaeger}}]{limAcousticManipulationMultibody2024}%
  \BibitemOpen
  \bibfield  {author} {\bibinfo {author} {\bibfnamefont {M.~X.}\ \bibnamefont {Lim}}, \bibinfo {author} {\bibfnamefont {B.}~\bibnamefont {VanSaders}},\ and\ \bibinfo {author} {\bibfnamefont {H.~M.}\ \bibnamefont {Jaeger}},\ }\bibfield  {title} {\bibinfo {title} {Acoustic manipulation of multi-body structures and dynamics},\ }\href {https://doi.org/10.1088/1361-6633/ad43f9} {\bibfield  {journal} {\bibinfo  {journal} {Reports on Progress in Physics}\ }\textbf {\bibinfo {volume} {87}},\ \bibinfo {pages} {064601} (\bibinfo {year} {2024})}\BibitemShut {NoStop}%
\bibitem [{\citenamefont {Beatus}\ \emph {et~al.}(2006)\citenamefont {Beatus}, \citenamefont {Tlusty},\ and\ \citenamefont {{Bar-Ziv}}}]{beatusPhononsOnedimensionalMicrofluidic2006}%
  \BibitemOpen
  \bibfield  {author} {\bibinfo {author} {\bibfnamefont {T.}~\bibnamefont {Beatus}}, \bibinfo {author} {\bibfnamefont {T.}~\bibnamefont {Tlusty}},\ and\ \bibinfo {author} {\bibfnamefont {R.}~\bibnamefont {{Bar-Ziv}}},\ }\bibfield  {title} {\bibinfo {title} {Phonons in a one-dimensional microfluidic crystal},\ }\href {https://doi.org/10.1038/nphys432} {\bibfield  {journal} {\bibinfo  {journal} {Nature Physics}\ }\textbf {\bibinfo {volume} {2}},\ \bibinfo {pages} {743} (\bibinfo {year} {2006})}\BibitemShut {NoStop}%
\bibitem [{\citenamefont {Bililign}\ \emph {et~al.}(2022)\citenamefont {Bililign}, \citenamefont {Balboa~Usabiaga}, \citenamefont {Ganan}, \citenamefont {Poncet}, \citenamefont {Soni}, \citenamefont {Magkiriadou}, \citenamefont {Shelley}, \citenamefont {Bartolo},\ and\ \citenamefont {Irvine}}]{bililignMotileDislocationsKnead2022}%
  \BibitemOpen
  \bibfield  {author} {\bibinfo {author} {\bibfnamefont {E.~S.}\ \bibnamefont {Bililign}}, \bibinfo {author} {\bibfnamefont {F.}~\bibnamefont {Balboa~Usabiaga}}, \bibinfo {author} {\bibfnamefont {Y.~A.}\ \bibnamefont {Ganan}}, \bibinfo {author} {\bibfnamefont {A.}~\bibnamefont {Poncet}}, \bibinfo {author} {\bibfnamefont {V.}~\bibnamefont {Soni}}, \bibinfo {author} {\bibfnamefont {S.}~\bibnamefont {Magkiriadou}}, \bibinfo {author} {\bibfnamefont {M.~J.}\ \bibnamefont {Shelley}}, \bibinfo {author} {\bibfnamefont {D.}~\bibnamefont {Bartolo}},\ and\ \bibinfo {author} {\bibfnamefont {W.~T.~M.}\ \bibnamefont {Irvine}},\ }\bibfield  {title} {\bibinfo {title} {Motile dislocations knead odd crystals into whorls},\ }\href {https://doi.org/10.1038/s41567-021-01429-3} {\bibfield  {journal} {\bibinfo  {journal} {Nature Physics}\ }\textbf {\bibinfo {volume} {18}},\ \bibinfo {pages} {212} (\bibinfo {year} {2022})}\BibitemShut {NoStop}%
\bibitem [{\citenamefont {Gu}\ \emph {et~al.}(2025)\citenamefont {Gu}, \citenamefont {Guiselin}, \citenamefont {Bain}, \citenamefont {Zuriguel},\ and\ \citenamefont {Bartolo}}]{guEmergenceCollectiveOscillations2025}%
  \BibitemOpen
  \bibfield  {author} {\bibinfo {author} {\bibfnamefont {F.}~\bibnamefont {Gu}}, \bibinfo {author} {\bibfnamefont {B.}~\bibnamefont {Guiselin}}, \bibinfo {author} {\bibfnamefont {N.}~\bibnamefont {Bain}}, \bibinfo {author} {\bibfnamefont {I.}~\bibnamefont {Zuriguel}},\ and\ \bibinfo {author} {\bibfnamefont {D.}~\bibnamefont {Bartolo}},\ }\bibfield  {title} {\bibinfo {title} {Emergence of collective oscillations in massive human crowds},\ }\href {https://doi.org/10.1038/s41586-024-08514-6} {\bibfield  {journal} {\bibinfo  {journal} {Nature}\ }\textbf {\bibinfo {volume} {638}},\ \bibinfo {pages} {112} (\bibinfo {year} {2025})}\BibitemShut {NoStop}%
\bibitem [{\citenamefont {You}\ \emph {et~al.}(2020)\citenamefont {You}, \citenamefont {Baskaran},\ and\ \citenamefont {Marchetti}}]{youNonreciprocityGenericRoute2020}%
  \BibitemOpen
  \bibfield  {author} {\bibinfo {author} {\bibfnamefont {Z.}~\bibnamefont {You}}, \bibinfo {author} {\bibfnamefont {A.}~\bibnamefont {Baskaran}},\ and\ \bibinfo {author} {\bibfnamefont {M.~C.}\ \bibnamefont {Marchetti}},\ }\bibfield  {title} {\bibinfo {title} {Nonreciprocity as a generic route to traveling states},\ }\href {https://doi.org/10.1073/pnas.2010318117} {\bibfield  {journal} {\bibinfo  {journal} {Proceedings of the National Academy of Sciences}\ }\textbf {\bibinfo {volume} {117}},\ \bibinfo {pages} {19767} (\bibinfo {year} {2020})}\BibitemShut {NoStop}%
\bibitem [{\citenamefont {Brauns}\ and\ \citenamefont {Marchetti}(2024)}]{braunsNonreciprocalPatternFormation2024}%
  \BibitemOpen
  \bibfield  {author} {\bibinfo {author} {\bibfnamefont {F.}~\bibnamefont {Brauns}}\ and\ \bibinfo {author} {\bibfnamefont {M.~C.}\ \bibnamefont {Marchetti}},\ }\bibfield  {title} {\bibinfo {title} {Nonreciprocal {{Pattern Formation}} of {{Conserved Fields}}},\ }\href {https://doi.org/10.1103/PhysRevX.14.021014} {\bibfield  {journal} {\bibinfo  {journal} {Physical Review X}\ }\textbf {\bibinfo {volume} {14}},\ \bibinfo {pages} {021014} (\bibinfo {year} {2024})}\BibitemShut {NoStop}%
\bibitem [{\citenamefont {Saha}\ \emph {et~al.}(2020)\citenamefont {Saha}, \citenamefont {{Agudo-Canalejo}},\ and\ \citenamefont {Golestanian}}]{sahaScalarActiveMixtures2020}%
  \BibitemOpen
  \bibfield  {author} {\bibinfo {author} {\bibfnamefont {S.}~\bibnamefont {Saha}}, \bibinfo {author} {\bibfnamefont {J.}~\bibnamefont {{Agudo-Canalejo}}},\ and\ \bibinfo {author} {\bibfnamefont {R.}~\bibnamefont {Golestanian}},\ }\bibfield  {title} {\bibinfo {title} {Scalar {{Active Mixtures}}: {{The Nonreciprocal Cahn-Hilliard Model}}},\ }\href {https://doi.org/10.1103/PhysRevX.10.041009} {\bibfield  {journal} {\bibinfo  {journal} {Physical Review X}\ }\textbf {\bibinfo {volume} {10}},\ \bibinfo {pages} {041009} (\bibinfo {year} {2020})}\BibitemShut {NoStop}%
\bibitem [{\citenamefont {Dinelli}\ \emph {et~al.}(2023)\citenamefont {Dinelli}, \citenamefont {O'Byrne}, \citenamefont {Curatolo}, \citenamefont {Zhao}, \citenamefont {Sollich},\ and\ \citenamefont {Tailleur}}]{dinelliNonreciprocityScalesActive2023}%
  \BibitemOpen
  \bibfield  {author} {\bibinfo {author} {\bibfnamefont {A.}~\bibnamefont {Dinelli}}, \bibinfo {author} {\bibfnamefont {J.}~\bibnamefont {O'Byrne}}, \bibinfo {author} {\bibfnamefont {A.}~\bibnamefont {Curatolo}}, \bibinfo {author} {\bibfnamefont {Y.}~\bibnamefont {Zhao}}, \bibinfo {author} {\bibfnamefont {P.}~\bibnamefont {Sollich}},\ and\ \bibinfo {author} {\bibfnamefont {J.}~\bibnamefont {Tailleur}},\ }\bibfield  {title} {\bibinfo {title} {Non-reciprocity across scales in active mixtures},\ }\href {https://doi.org/10.1038/s41467-023-42713-5} {\bibfield  {journal} {\bibinfo  {journal} {Nature Communications}\ }\textbf {\bibinfo {volume} {14}},\ \bibinfo {pages} {7035} (\bibinfo {year} {2023})}\BibitemShut {NoStop}%
\bibitem [{\citenamefont {Osat}\ and\ \citenamefont {Golestanian}(2023)}]{osatNonreciprocalMultifariousSelforganization2023}%
  \BibitemOpen
  \bibfield  {author} {\bibinfo {author} {\bibfnamefont {S.}~\bibnamefont {Osat}}\ and\ \bibinfo {author} {\bibfnamefont {R.}~\bibnamefont {Golestanian}},\ }\bibfield  {title} {\bibinfo {title} {Non-reciprocal multifarious self-organization},\ }\href {https://doi.org/10.1038/s41565-022-01258-2} {\bibfield  {journal} {\bibinfo  {journal} {Nature Nanotechnology}\ }\textbf {\bibinfo {volume} {18}},\ \bibinfo {pages} {79} (\bibinfo {year} {2023})}\BibitemShut {NoStop}%
\bibitem [{\citenamefont {Meredith}\ \emph {et~al.}(2020)\citenamefont {Meredith}, \citenamefont {Moerman}, \citenamefont {Groenewold}, \citenamefont {Chiu}, \citenamefont {Kegel}, \citenamefont {{van Blaaderen}},\ and\ \citenamefont {Zarzar}}]{meredithPredatorPreyInteractions2020}%
  \BibitemOpen
  \bibfield  {author} {\bibinfo {author} {\bibfnamefont {C.~H.}\ \bibnamefont {Meredith}}, \bibinfo {author} {\bibfnamefont {P.~G.}\ \bibnamefont {Moerman}}, \bibinfo {author} {\bibfnamefont {J.}~\bibnamefont {Groenewold}}, \bibinfo {author} {\bibfnamefont {Y.-J.}\ \bibnamefont {Chiu}}, \bibinfo {author} {\bibfnamefont {W.~K.}\ \bibnamefont {Kegel}}, \bibinfo {author} {\bibfnamefont {A.}~\bibnamefont {{van Blaaderen}}},\ and\ \bibinfo {author} {\bibfnamefont {L.~D.}\ \bibnamefont {Zarzar}},\ }\bibfield  {title} {\bibinfo {title} {Predator--prey interactions between droplets driven by non-reciprocal oil exchange},\ }\href {https://doi.org/10.1038/s41557-020-00575-0} {\bibfield  {journal} {\bibinfo  {journal} {Nature Chemistry}\ }\textbf {\bibinfo {volume} {12}},\ \bibinfo {pages} {1136} (\bibinfo {year} {2020})}\BibitemShut {NoStop}%
\bibitem [{\citenamefont {Scheibner}\ \emph {et~al.}(2020{\natexlab{a}})\citenamefont {Scheibner}, \citenamefont {Souslov}, \citenamefont {Banerjee}, \citenamefont {Sur{\'o}wka}, \citenamefont {Irvine},\ and\ \citenamefont {Vitelli}}]{scheibnerOddElasticity2020}%
  \BibitemOpen
  \bibfield  {author} {\bibinfo {author} {\bibfnamefont {C.}~\bibnamefont {Scheibner}}, \bibinfo {author} {\bibfnamefont {A.}~\bibnamefont {Souslov}}, \bibinfo {author} {\bibfnamefont {D.}~\bibnamefont {Banerjee}}, \bibinfo {author} {\bibfnamefont {P.}~\bibnamefont {Sur{\'o}wka}}, \bibinfo {author} {\bibfnamefont {W.~T.~M.}\ \bibnamefont {Irvine}},\ and\ \bibinfo {author} {\bibfnamefont {V.}~\bibnamefont {Vitelli}},\ }\bibfield  {title} {\bibinfo {title} {Odd elasticity},\ }\href {https://doi.org/10.1038/s41567-020-0795-y} {\bibfield  {journal} {\bibinfo  {journal} {Nature Physics}\ }\textbf {\bibinfo {volume} {16}},\ \bibinfo {pages} {475} (\bibinfo {year} {2020}{\natexlab{a}})}\BibitemShut {NoStop}%
\bibitem [{\citenamefont {Fruchart}\ \emph {et~al.}(2023)\citenamefont {Fruchart}, \citenamefont {Scheibner},\ and\ \citenamefont {Vitelli}}]{fruchartOddViscosityOdd2023}%
  \BibitemOpen
  \bibfield  {author} {\bibinfo {author} {\bibfnamefont {M.}~\bibnamefont {Fruchart}}, \bibinfo {author} {\bibfnamefont {C.}~\bibnamefont {Scheibner}},\ and\ \bibinfo {author} {\bibfnamefont {V.}~\bibnamefont {Vitelli}},\ }\bibfield  {title} {\bibinfo {title} {Odd {{Viscosity}} and {{Odd Elasticity}}},\ }\href {https://doi.org/10.1146/annurev-conmatphys-040821-125506} {\bibfield  {journal} {\bibinfo  {journal} {Annual Review of Condensed Matter Physics}\ }\textbf {\bibinfo {volume} {14}},\ \bibinfo {pages} {471} (\bibinfo {year} {2023})}\BibitemShut {NoStop}%
\bibitem [{\citenamefont {Goodrich}\ \emph {et~al.}(2015)\citenamefont {Goodrich}, \citenamefont {Liu},\ and\ \citenamefont {Nagel}}]{goodrichPrincipleIndependentBondLevel2015}%
  \BibitemOpen
  \bibfield  {author} {\bibinfo {author} {\bibfnamefont {C.~P.}\ \bibnamefont {Goodrich}}, \bibinfo {author} {\bibfnamefont {A.~J.}\ \bibnamefont {Liu}},\ and\ \bibinfo {author} {\bibfnamefont {S.~R.}\ \bibnamefont {Nagel}},\ }\bibfield  {title} {\bibinfo {title} {The {{Principle}} of {{Independent Bond-Level Response}}: {{Tuning}} by {{Pruning}} to {{Exploit Disorder}} for {{Global Behavior}}},\ }\href {https://doi.org/10.1103/PhysRevLett.114.225501} {\bibfield  {journal} {\bibinfo  {journal} {Physical Review Letters}\ }\textbf {\bibinfo {volume} {114}},\ \bibinfo {pages} {225501} (\bibinfo {year} {2015})}\BibitemShut {NoStop}%
\bibitem [{\citenamefont {Rocks}\ \emph {et~al.}(2017)\citenamefont {Rocks}, \citenamefont {Pashine}, \citenamefont {Bischofberger}, \citenamefont {Goodrich}, \citenamefont {Liu},\ and\ \citenamefont {Nagel}}]{rocksDesigningAllosteryinspiredResponse2017}%
  \BibitemOpen
  \bibfield  {author} {\bibinfo {author} {\bibfnamefont {J.~W.}\ \bibnamefont {Rocks}}, \bibinfo {author} {\bibfnamefont {N.}~\bibnamefont {Pashine}}, \bibinfo {author} {\bibfnamefont {I.}~\bibnamefont {Bischofberger}}, \bibinfo {author} {\bibfnamefont {C.~P.}\ \bibnamefont {Goodrich}}, \bibinfo {author} {\bibfnamefont {A.~J.}\ \bibnamefont {Liu}},\ and\ \bibinfo {author} {\bibfnamefont {S.~R.}\ \bibnamefont {Nagel}},\ }\bibfield  {title} {\bibinfo {title} {Designing allostery-inspired response in mechanical networks},\ }\href {https://doi.org/10.1073/pnas.1612139114} {\bibfield  {journal} {\bibinfo  {journal} {Proceedings of the National Academy of Sciences}\ }\textbf {\bibinfo {volume} {114}},\ \bibinfo {pages} {2520} (\bibinfo {year} {2017})}\BibitemShut {NoStop}%
\bibitem [{\citenamefont {Yan}\ \emph {et~al.}(2017)\citenamefont {Yan}, \citenamefont {Ravasio}, \citenamefont {Brito},\ and\ \citenamefont {Wyart}}]{yanArchitectureCoevolutionAllosteric2017}%
  \BibitemOpen
  \bibfield  {author} {\bibinfo {author} {\bibfnamefont {L.}~\bibnamefont {Yan}}, \bibinfo {author} {\bibfnamefont {R.}~\bibnamefont {Ravasio}}, \bibinfo {author} {\bibfnamefont {C.}~\bibnamefont {Brito}},\ and\ \bibinfo {author} {\bibfnamefont {M.}~\bibnamefont {Wyart}},\ }\bibfield  {title} {\bibinfo {title} {Architecture and coevolution of allosteric materials},\ }\href {https://doi.org/10.1073/pnas.1615536114} {\bibfield  {journal} {\bibinfo  {journal} {Proceedings of the National Academy of Sciences}\ }\textbf {\bibinfo {volume} {114}},\ \bibinfo {pages} {2526} (\bibinfo {year} {2017})}\BibitemShut {NoStop}%
\bibitem [{\citenamefont {Stern}\ \emph {et~al.}(2021)\citenamefont {Stern}, \citenamefont {Hexner}, \citenamefont {Rocks},\ and\ \citenamefont {Liu}}]{sternSupervisedLearningPhysical2021}%
  \BibitemOpen
  \bibfield  {author} {\bibinfo {author} {\bibfnamefont {M.}~\bibnamefont {Stern}}, \bibinfo {author} {\bibfnamefont {D.}~\bibnamefont {Hexner}}, \bibinfo {author} {\bibfnamefont {J.~W.}\ \bibnamefont {Rocks}},\ and\ \bibinfo {author} {\bibfnamefont {A.~J.}\ \bibnamefont {Liu}},\ }\bibfield  {title} {\bibinfo {title} {Supervised {{Learning}} in {{Physical Networks}}: {{From Machine Learning}} to {{Learning Machines}}},\ }\href {https://doi.org/10.1103/PhysRevX.11.021045} {\bibfield  {journal} {\bibinfo  {journal} {Physical Review X}\ }\textbf {\bibinfo {volume} {11}},\ \bibinfo {pages} {021045} (\bibinfo {year} {2021})}\BibitemShut {NoStop}%
\bibitem [{\citenamefont {Du}\ \emph {et~al.}(2025)\citenamefont {Du}, \citenamefont {Veenstra}, \citenamefont {van Mastrigt},\ and\ \citenamefont {Coulais}}]{duMetamaterialsThatLearn2025}%
  \BibitemOpen
  \bibfield  {author} {\bibinfo {author} {\bibfnamefont {Y.}~\bibnamefont {Du}}, \bibinfo {author} {\bibfnamefont {J.}~\bibnamefont {Veenstra}}, \bibinfo {author} {\bibfnamefont {R.}~\bibnamefont {van Mastrigt}},\ and\ \bibinfo {author} {\bibfnamefont {C.}~\bibnamefont {Coulais}},\ }\href {https://doi.org/10.48550/arXiv.2501.11958} {\bibinfo {title} {Metamaterials that learn to change shape}} (\bibinfo {year} {2025}),\ \Eprint {https://arxiv.org/abs/2501.11958} {arXiv:2501.11958} \BibitemShut {NoStop}%
\bibitem [{\citenamefont {Ronceray}\ \emph {et~al.}(2019)\citenamefont {Ronceray}, \citenamefont {Broedersz},\ and\ \citenamefont {Lenz}}]{roncerayStressdependentAmplificationActive2019}%
  \BibitemOpen
  \bibfield  {author} {\bibinfo {author} {\bibfnamefont {P.}~\bibnamefont {Ronceray}}, \bibinfo {author} {\bibfnamefont {C.~P.}\ \bibnamefont {Broedersz}},\ and\ \bibinfo {author} {\bibfnamefont {M.}~\bibnamefont {Lenz}},\ }\bibfield  {title} {\bibinfo {title} {Stress-dependent amplification of active forces in nonlinear elastic media},\ }\href {https://doi.org/10.1039/C8SM00949J} {\bibfield  {journal} {\bibinfo  {journal} {Soft Matter}\ }\textbf {\bibinfo {volume} {15}},\ \bibinfo {pages} {331} (\bibinfo {year} {2019})}\BibitemShut {NoStop}%
\bibitem [{\citenamefont {Duque}\ \emph {et~al.}(2024)\citenamefont {Duque}, \citenamefont {Bonfanti}, \citenamefont {Fouchard}, \citenamefont {Baldauf}, \citenamefont {Azenha}, \citenamefont {Ferber}, \citenamefont {Harris}, \citenamefont {Barriga}, \citenamefont {Kabla},\ and\ \citenamefont {Charras}}]{duqueRuptureStrengthLiving2024}%
  \BibitemOpen
  \bibfield  {author} {\bibinfo {author} {\bibfnamefont {J.}~\bibnamefont {Duque}}, \bibinfo {author} {\bibfnamefont {A.}~\bibnamefont {Bonfanti}}, \bibinfo {author} {\bibfnamefont {J.}~\bibnamefont {Fouchard}}, \bibinfo {author} {\bibfnamefont {L.}~\bibnamefont {Baldauf}}, \bibinfo {author} {\bibfnamefont {S.~R.}\ \bibnamefont {Azenha}}, \bibinfo {author} {\bibfnamefont {E.}~\bibnamefont {Ferber}}, \bibinfo {author} {\bibfnamefont {A.}~\bibnamefont {Harris}}, \bibinfo {author} {\bibfnamefont {E.~H.}\ \bibnamefont {Barriga}}, \bibinfo {author} {\bibfnamefont {A.~J.}\ \bibnamefont {Kabla}},\ and\ \bibinfo {author} {\bibfnamefont {G.}~\bibnamefont {Charras}},\ }\bibfield  {title} {\bibinfo {title} {Rupture strength of living cell monolayers},\ }\href {https://doi.org/10.1038/s41563-024-02027-3} {\bibfield  {journal} {\bibinfo  {journal} {Nature Materials}\ }\textbf {\bibinfo {volume} {23}},\ \bibinfo {pages} {1563} (\bibinfo {year} {2024})}\BibitemShut {NoStop}%
\bibitem [{\citenamefont {Chao}\ \emph {et~al.}(2026)\citenamefont {Chao}, \citenamefont {Gokhale}, \citenamefont {Lin}, \citenamefont {Hastewell}, \citenamefont {Bacanu}, \citenamefont {Chen}, \citenamefont {Li}, \citenamefont {Liu}, \citenamefont {Lee}, \citenamefont {Dunkel},\ and\ \citenamefont {Fakhri}}]{chaoSelectiveExcitationWorkgenerating2026}%
  \BibitemOpen
  \bibfield  {author} {\bibinfo {author} {\bibfnamefont {Y.-C.}\ \bibnamefont {Chao}}, \bibinfo {author} {\bibfnamefont {S.}~\bibnamefont {Gokhale}}, \bibinfo {author} {\bibfnamefont {L.}~\bibnamefont {Lin}}, \bibinfo {author} {\bibfnamefont {A.}~\bibnamefont {Hastewell}}, \bibinfo {author} {\bibfnamefont {A.}~\bibnamefont {Bacanu}}, \bibinfo {author} {\bibfnamefont {Y.}~\bibnamefont {Chen}}, \bibinfo {author} {\bibfnamefont {J.}~\bibnamefont {Li}}, \bibinfo {author} {\bibfnamefont {J.}~\bibnamefont {Liu}}, \bibinfo {author} {\bibfnamefont {H.}~\bibnamefont {Lee}}, \bibinfo {author} {\bibfnamefont {J.}~\bibnamefont {Dunkel}},\ and\ \bibinfo {author} {\bibfnamefont {N.}~\bibnamefont {Fakhri}},\ }\bibfield  {title} {\bibinfo {title} {Selective excitation of work-generating cycles in non-reciprocal living solids},\ }\href {https://doi.org/10.1038/s41567-026-03178-7} {\bibfield  {journal} {\bibinfo  {journal} {Nature Physics}\ ,\ \bibinfo {pages} {1}} (\bibinfo {year} {2026})}\BibitemShut {NoStop}%
\bibitem [{Sup()}]{SupplementaryMaterial}%
  \BibitemOpen
  \href@noop {} {}\bibinfo {note} {See Supplemental Material at [URL] for Supplementary movies S1 and S2, which show experimental data and analysis corresponding to Fig. 1, and deformations of the ball-spring models shown in Fig. 2}\BibitemShut {NoStop}%
\bibitem [{\citenamefont {Choi}\ \emph {et~al.}(2024)\citenamefont {Choi}, \citenamefont {Huang},\ and\ \citenamefont {Goldenfeld}}]{choiNoisedrivenOddElastic2024}%
  \BibitemOpen
  \bibfield  {author} {\bibinfo {author} {\bibfnamefont {S.~H.}\ \bibnamefont {Choi}}, \bibinfo {author} {\bibfnamefont {Z.-F.}\ \bibnamefont {Huang}},\ and\ \bibinfo {author} {\bibfnamefont {N.}~\bibnamefont {Goldenfeld}},\ }\href {https://doi.org/10.48550/arXiv.2411.09615} {\bibinfo {title} {Noise-driven odd elastic waves in living chiral active matter}} (\bibinfo {year} {2024}),\ \Eprint {https://arxiv.org/abs/2411.09615} {arXiv:2411.09615} \BibitemShut {NoStop}%
\bibitem [{\citenamefont {Scheibner}\ \emph {et~al.}(2020{\natexlab{b}})\citenamefont {Scheibner}, \citenamefont {Irvine},\ and\ \citenamefont {Vitelli}}]{scheibnerNonHermitianBandTopology2020}%
  \BibitemOpen
  \bibfield  {author} {\bibinfo {author} {\bibfnamefont {C.}~\bibnamefont {Scheibner}}, \bibinfo {author} {\bibfnamefont {W.~T.~M.}\ \bibnamefont {Irvine}},\ and\ \bibinfo {author} {\bibfnamefont {V.}~\bibnamefont {Vitelli}},\ }\bibfield  {title} {\bibinfo {title} {Non-{{Hermitian Band Topology}} and {{Skin Modes}} in {{Active Elastic Media}}},\ }\href {https://doi.org/10.1103/PhysRevLett.125.118001} {\bibfield  {journal} {\bibinfo  {journal} {Physical Review Letters}\ }\textbf {\bibinfo {volume} {125}},\ \bibinfo {pages} {118001} (\bibinfo {year} {2020}{\natexlab{b}})}\BibitemShut {NoStop}%
\bibitem [{\citenamefont {van Hecke}(2009)}]{heckeJammingSoftParticles2009}%
  \BibitemOpen
  \bibfield  {author} {\bibinfo {author} {\bibfnamefont {M.}~\bibnamefont {van Hecke}},\ }\bibfield  {title} {\bibinfo {title} {Jamming of soft particles: Geometry, mechanics, scaling and isostaticity},\ }\href {https://doi.org/10.1088/0953-8984/22/3/033101} {\bibfield  {journal} {\bibinfo  {journal} {Journal of Physics: Condensed Matter}\ }\textbf {\bibinfo {volume} {22}},\ \bibinfo {pages} {033101} (\bibinfo {year} {2009})}\BibitemShut {NoStop}%
\bibitem [{\citenamefont {Burla}\ \emph {et~al.}(2019)\citenamefont {Burla}, \citenamefont {Tauber}, \citenamefont {Dussi}, \citenamefont {{van der Gucht}},\ and\ \citenamefont {Koenderink}}]{burlaStressManagementComposite2019}%
  \BibitemOpen
  \bibfield  {author} {\bibinfo {author} {\bibfnamefont {F.}~\bibnamefont {Burla}}, \bibinfo {author} {\bibfnamefont {J.}~\bibnamefont {Tauber}}, \bibinfo {author} {\bibfnamefont {S.}~\bibnamefont {Dussi}}, \bibinfo {author} {\bibfnamefont {J.}~\bibnamefont {{van der Gucht}}},\ and\ \bibinfo {author} {\bibfnamefont {G.~H.}\ \bibnamefont {Koenderink}},\ }\bibfield  {title} {\bibinfo {title} {Stress management in composite biopolymer networks},\ }\href {https://doi.org/10.1038/s41567-019-0443-6} {\bibfield  {journal} {\bibinfo  {journal} {Nature Physics}\ }\textbf {\bibinfo {volume} {15}},\ \bibinfo {pages} {549} (\bibinfo {year} {2019})}\BibitemShut {NoStop}%
\bibitem [{\citenamefont {Veenstra}\ \emph {et~al.}(2024)\citenamefont {Veenstra}, \citenamefont {Gamayun}, \citenamefont {Guo}, \citenamefont {Sarvi}, \citenamefont {Meinersen},\ and\ \citenamefont {Coulais}}]{veenstraNonreciprocalTopologicalSolitons2024}%
  \BibitemOpen
  \bibfield  {author} {\bibinfo {author} {\bibfnamefont {J.}~\bibnamefont {Veenstra}}, \bibinfo {author} {\bibfnamefont {O.}~\bibnamefont {Gamayun}}, \bibinfo {author} {\bibfnamefont {X.}~\bibnamefont {Guo}}, \bibinfo {author} {\bibfnamefont {A.}~\bibnamefont {Sarvi}}, \bibinfo {author} {\bibfnamefont {C.~V.}\ \bibnamefont {Meinersen}},\ and\ \bibinfo {author} {\bibfnamefont {C.}~\bibnamefont {Coulais}},\ }\bibfield  {title} {\bibinfo {title} {Non-reciprocal topological solitons in active metamaterials},\ }\href {https://doi.org/10.1038/s41586-024-07097-6} {\bibfield  {journal} {\bibinfo  {journal} {Nature}\ }\textbf {\bibinfo {volume} {627}},\ \bibinfo {pages} {528} (\bibinfo {year} {2024})}\BibitemShut {NoStop}%
\bibitem [{\citenamefont {Mandal}\ \emph {et~al.}(2024)\citenamefont {Mandal}, \citenamefont {Huang}, \citenamefont {Fruchart}, \citenamefont {Moerman}, \citenamefont {Vaikuntanathan}, \citenamefont {Murugan},\ and\ \citenamefont {Vitelli}}]{mandalLearningDynamicalBehaviors2024}%
  \BibitemOpen
  \bibfield  {author} {\bibinfo {author} {\bibfnamefont {R.}~\bibnamefont {Mandal}}, \bibinfo {author} {\bibfnamefont {R.}~\bibnamefont {Huang}}, \bibinfo {author} {\bibfnamefont {M.}~\bibnamefont {Fruchart}}, \bibinfo {author} {\bibfnamefont {P.~G.}\ \bibnamefont {Moerman}}, \bibinfo {author} {\bibfnamefont {S.}~\bibnamefont {Vaikuntanathan}}, \bibinfo {author} {\bibfnamefont {A.}~\bibnamefont {Murugan}},\ and\ \bibinfo {author} {\bibfnamefont {V.}~\bibnamefont {Vitelli}},\ }\href {https://doi.org/10.48550/arXiv.2406.07856} {\bibinfo {title} {Learning dynamical behaviors in physical systems}} (\bibinfo {year} {2024}),\ \Eprint {https://arxiv.org/abs/2406.07856} {arXiv:2406.07856} \BibitemShut {NoStop}%
\bibitem [{\citenamefont {Alvarado}\ \emph {et~al.}(2013)\citenamefont {Alvarado}, \citenamefont {Sheinman}, \citenamefont {Sharma}, \citenamefont {MacKintosh},\ and\ \citenamefont {Koenderink}}]{alvaradoMolecularMotorsRobustly2013}%
  \BibitemOpen
  \bibfield  {author} {\bibinfo {author} {\bibfnamefont {J.}~\bibnamefont {Alvarado}}, \bibinfo {author} {\bibfnamefont {M.}~\bibnamefont {Sheinman}}, \bibinfo {author} {\bibfnamefont {A.}~\bibnamefont {Sharma}}, \bibinfo {author} {\bibfnamefont {F.~C.}\ \bibnamefont {MacKintosh}},\ and\ \bibinfo {author} {\bibfnamefont {G.~H.}\ \bibnamefont {Koenderink}},\ }\bibfield  {title} {\bibinfo {title} {Molecular motors robustly drive active gels to a critically connected state},\ }\href {https://doi.org/10.1038/nphys2715} {\bibfield  {journal} {\bibinfo  {journal} {Nature Physics}\ }\textbf {\bibinfo {volume} {9}},\ \bibinfo {pages} {591} (\bibinfo {year} {2013})}\BibitemShut {NoStop}%
\bibitem [{\citenamefont {Wyse~Jackson}\ \emph {et~al.}(2022)\citenamefont {Wyse~Jackson}, \citenamefont {Michel}, \citenamefont {Lwin}, \citenamefont {Fortier}, \citenamefont {Das}, \citenamefont {Bonassar},\ and\ \citenamefont {Cohen}}]{wysejacksonStructuralOriginsCartilage2022}%
  \BibitemOpen
  \bibfield  {author} {\bibinfo {author} {\bibfnamefont {T.}~\bibnamefont {Wyse~Jackson}}, \bibinfo {author} {\bibfnamefont {J.}~\bibnamefont {Michel}}, \bibinfo {author} {\bibfnamefont {P.}~\bibnamefont {Lwin}}, \bibinfo {author} {\bibfnamefont {L.~A.}\ \bibnamefont {Fortier}}, \bibinfo {author} {\bibfnamefont {M.}~\bibnamefont {Das}}, \bibinfo {author} {\bibfnamefont {L.~J.}\ \bibnamefont {Bonassar}},\ and\ \bibinfo {author} {\bibfnamefont {I.}~\bibnamefont {Cohen}},\ }\bibfield  {title} {\bibinfo {title} {Structural origins of cartilage shear mechanics},\ }\href {https://doi.org/10.1126/sciadv.abk2805} {\bibfield  {journal} {\bibinfo  {journal} {Science Advances}\ }\textbf {\bibinfo {volume} {8}},\ \bibinfo {pages} {eabk2805} (\bibinfo {year} {2022})}\BibitemShut {NoStop}%
\bibitem [{\citenamefont {Chen}\ \emph {et~al.}(2025)\citenamefont {Chen}, \citenamefont {Gökmen}, \citenamefont {Fruchart}, \citenamefont {Krumbein}, \citenamefont {Silberzan}, \citenamefont {Yashunsky},\ and\ \citenamefont {Vitelli}}]{chenChiralityScalesTissue2025}%
  \BibitemOpen
  \bibfield  {author} {\bibinfo {author} {\bibfnamefont {S.}~\bibnamefont {Chen}}, \bibinfo {author} {\bibfnamefont {D.~E.}\ \bibnamefont {Gökmen}}, \bibinfo {author} {\bibfnamefont {M.}~\bibnamefont {Fruchart}}, \bibinfo {author} {\bibfnamefont {M.}~\bibnamefont {Krumbein}}, \bibinfo {author} {\bibfnamefont {P.}~\bibnamefont {Silberzan}}, \bibinfo {author} {\bibfnamefont {V.}~\bibnamefont {Yashunsky}},\ and\ \bibinfo {author} {\bibfnamefont {V.}~\bibnamefont {Vitelli}},\ }\href {https://doi.org/10.48550/arXiv.2506.12276} {\bibinfo {title} {Chirality across scales in tissue dynamics}} (\bibinfo {year} {2025}),\ \Eprint {https://arxiv.org/abs/2506.12276} {arXiv:2506.12276} \BibitemShut {NoStop}%
\bibitem [{\citenamefont {Petridou}\ \emph {et~al.}(2021)\citenamefont {Petridou}, \citenamefont {{Corominas-Murtra}}, \citenamefont {Heisenberg},\ and\ \citenamefont {Hannezo}}]{petridouRigidityPercolationUncovers2021}%
  \BibitemOpen
  \bibfield  {author} {\bibinfo {author} {\bibfnamefont {N.~I.}\ \bibnamefont {Petridou}}, \bibinfo {author} {\bibfnamefont {B.}~\bibnamefont {{Corominas-Murtra}}}, \bibinfo {author} {\bibfnamefont {C.-P.}\ \bibnamefont {Heisenberg}},\ and\ \bibinfo {author} {\bibfnamefont {E.}~\bibnamefont {Hannezo}},\ }\bibfield  {title} {\bibinfo {title} {Rigidity percolation uncovers a structural basis for embryonic tissue phase transitions},\ }\href {https://doi.org/10.1016/j.cell.2021.02.017} {\bibfield  {journal} {\bibinfo  {journal} {Cell}\ }\textbf {\bibinfo {volume} {184}},\ \bibinfo {pages} {1914} (\bibinfo {year} {2021})}\BibitemShut {NoStop}%
\bibitem [{\citenamefont {Jorge}\ \emph {et~al.}(2024)\citenamefont {Jorge}, \citenamefont {Chardac}, \citenamefont {Poncet},\ and\ \citenamefont {Bartolo}}]{jorgeActiveHydraulicsLaws2024}%
  \BibitemOpen
  \bibfield  {author} {\bibinfo {author} {\bibfnamefont {C.}~\bibnamefont {Jorge}}, \bibinfo {author} {\bibfnamefont {A.}~\bibnamefont {Chardac}}, \bibinfo {author} {\bibfnamefont {A.}~\bibnamefont {Poncet}},\ and\ \bibinfo {author} {\bibfnamefont {D.}~\bibnamefont {Bartolo}},\ }\bibfield  {title} {\bibinfo {title} {Active hydraulics laws from frustration principles},\ }\href {https://doi.org/10.1038/s41567-023-02301-2} {\bibfield  {journal} {\bibinfo  {journal} {Nature Physics}\ }\textbf {\bibinfo {volume} {20}},\ \bibinfo {pages} {303} (\bibinfo {year} {2024})}\BibitemShut {NoStop}%
\bibitem [{\citenamefont {Nicolaou}\ and\ \citenamefont {Motter}(2012)}]{nicolaouMechanicalMetamaterialsNegative2012}%
  \BibitemOpen
  \bibfield  {author} {\bibinfo {author} {\bibfnamefont {Z.~G.}\ \bibnamefont {Nicolaou}}\ and\ \bibinfo {author} {\bibfnamefont {A.~E.}\ \bibnamefont {Motter}},\ }\bibfield  {title} {\bibinfo {title} {Mechanical metamaterials with negative compressibility transitions},\ }\href {https://doi.org/10.1038/nmat3331} {\bibfield  {journal} {\bibinfo  {journal} {Nature Materials}\ }\textbf {\bibinfo {volume} {11}},\ \bibinfo {pages} {608} (\bibinfo {year} {2012})}\BibitemShut {NoStop}%
\bibitem [{\citenamefont {Ducarme}\ \emph {et~al.}(2025)\citenamefont {Ducarme}, \citenamefont {Weber}, \citenamefont {{van Hecke}},\ and\ \citenamefont {Overvelde}}]{ducarmeExoticMechanicalProperties2025}%
  \BibitemOpen
  \bibfield  {author} {\bibinfo {author} {\bibfnamefont {P.}~\bibnamefont {Ducarme}}, \bibinfo {author} {\bibfnamefont {B.}~\bibnamefont {Weber}}, \bibinfo {author} {\bibfnamefont {M.}~\bibnamefont {{van Hecke}}},\ and\ \bibinfo {author} {\bibfnamefont {J.~T.~B.}\ \bibnamefont {Overvelde}},\ }\bibfield  {title} {\bibinfo {title} {Exotic mechanical properties enabled by countersnapping instabilities},\ }\href {https://doi.org/10.1073/pnas.2423301122} {\bibfield  {journal} {\bibinfo  {journal} {Proceedings of the National Academy of Sciences}\ }\textbf {\bibinfo {volume} {122}},\ \bibinfo {pages} {e2423301122} (\bibinfo {year} {2025})}\BibitemShut {NoStop}%
\bibitem [{\citenamefont {Lubensky}\ \emph {et~al.}(2015)\citenamefont {Lubensky}, \citenamefont {Kane}, \citenamefont {Mao}, \citenamefont {Souslov},\ and\ \citenamefont {Sun}}]{lubenskyPhononsElasticityCritically2015}%
  \BibitemOpen
  \bibfield  {author} {\bibinfo {author} {\bibfnamefont {T.~C.}\ \bibnamefont {Lubensky}}, \bibinfo {author} {\bibfnamefont {C.~L.}\ \bibnamefont {Kane}}, \bibinfo {author} {\bibfnamefont {X.}~\bibnamefont {Mao}}, \bibinfo {author} {\bibfnamefont {A.}~\bibnamefont {Souslov}},\ and\ \bibinfo {author} {\bibfnamefont {K.}~\bibnamefont {Sun}},\ }\bibfield  {title} {\bibinfo {title} {Phonons and elasticity in critically coordinated lattices},\ }\href {https://doi.org/10.1088/0034-4885/78/7/073901} {\bibfield  {journal} {\bibinfo  {journal} {Reports on Progress in Physics}\ }\textbf {\bibinfo {volume} {78}},\ \bibinfo {pages} {073901} (\bibinfo {year} {2015})}\BibitemShut {NoStop}%
\end{thebibliography}
\end{document}